\documentclass[envcountsame,orivec,runningheads,11pt]{llncs}
\usepackage[utf8]{inputenc}
\usepackage{a4}
\usepackage{amsmath, amssymb, amsfonts,xspace,stmaryrd}
\usepackage{graphicx}
\usepackage{footmisc}
\makeatletter
\DeclareRobustCommand*{\bfseries}{%
  \not@math@alphabet\bfseries\mathbf
  \fontseries\bfdefault\selectfont
  \boldmath
}
\makeatother

\newtheorem{myexample}[theorem]{Example}
\newtheorem{fact}[theorem]{Fact}
\newtheorem{myremark}[theorem]{Remark}

\newtheorem{myquestion}[theorem]{Question}
\newtheorem{observation}[theorem]{Observation}
\newcommand{\keywords}[1]{\noindent {\bf Keywords:}\ #1}

\newcommand{\COMMENTED}[1]{}

\newcommand{\bin}{\mathrm{bin}}
\newcommand{\dom}{\operatorname{dom}}
\newcommand{\range}{\operatorname{range}}
\newcommand{\dC}{d_{\Cantor}}
\newcommand{\dD}{d_{\Cantor'}}
\newcommand{\dH}{d_{\Hilbert}}
\newcommand{\Reg}{\operatorname{Reg}}
\newcommand{\Card}{\operatorname{Card}}

\newcommand{\lin}{\operatorname{lin}}
\newcommand{\poly}{\operatorname{poly}}

\newcommand{\id}{\operatorname{id}}
\newcommand{\ball}{\operatorname{B}}
\newcommand{\cball}{\overline{\operatorname{B}}}
\newcommand{\Lip}{\operatorname{Lip}}

\newcommand{\preccurlyeqP}{\preccurlyeq_{\rm P}}
\newcommand{\preccurlyeqO}{\preccurlyeq_{\rm O}}
\newcommand{\preccurlyeqT}{\preccurlyeq_{\rm T}}
\newcommand{\calF}{\mathcal{F}}

\newcommand{\calH}{\mathcal{H}}
\newcommand{\calC}{\mathcal{C}}

\newcommand{\N}{\mathbb{N}}

\newcommand{\ID}{\mathbb{D}}
\newcommand{\IR}{\mathbb{R}}

\newcommand{\IZ}{\mathbb{Z}}
\newcommand{\IN}{\mathbb{N}}

\newcommand{\stateC}{\texttt{C}}

\newcommand{\sdzero}{\textup{\texttt{0}}\xspace}
\newcommand{\sdone}{\textup{\texttt{1}}\xspace}
\newcommand{\sdminus}{\textup{\texttt{\={1}}}\xspace}
\newcommand{\calK}{\mathcal{K}}
\newcommand{\calO}{\mathcal{O}}
\newcommand{\calP}{\mathcal{P}}
\newcommand{\calA}{\mathcal{S}}
\newcommand{\calo}{\mathcal{\scriptscriptstyle O}}
\newcommand{\calp}{\mathcal{\scriptscriptstyle P}}
\newcommand{\cala}{\mathcal{\scriptscriptstyle S}}
\newcommand{\Cantor}{\mathcal{C}}

\newcommand{\Hilbert}{\mathcal{H}}

\newcommand{\Machine}{\mathcal{M}}

\newcommand{\frakX}{\mathfrak{X}}
\newcommand{\Universe}{\frakX}

\newcommand{\exth}[1]{{#1}^*}
\newcommand{\extl}[1]{{#1}_*}
\newcommand{\extm}[1]{{#1}^*_*}
\renewcommand{\thefootnote}{\fnsymbol{footnote}}

\usepackage[usenames]{color}  %for internal comments from/to/among authors

\makeatletter
\newcommand*\bigdot{\mathpalette\bigdot@{.5}}
\newcommand*\bigdot@[2]{\mathbin{\vcenter{\hbox{\scalebox{#2}{$\m@th#1\bullet$\,\,}}}}}
\makeatother

\newcommand{\upinv}[1]{{#1}^{\overline{-1}}}
\newcommand{\loinv}[1]{{#1}^{\underline{-1}}}

\title{Representation Theory of Compact Metric Spaces \\ and Computational Complexity of Continuous Data\thanks{%
Supported by the National Research Foundation of Korea 
(grant NRF-2017R1E1A1A03071032)
and the International Research \& Development Program of
the Korean Ministry of Science and ICT (grant NRF-2016K1A3A7A03950702).
We thank Florian Steinberg for helpful discussions.
Example~\ref{x:Gleb} is due to Gleb Pogudin.}}
\titlerunning{Representation Theory and Computational Complexity of Compact Metric Spaces}
\institute{Dept. of Informatics, Kyushu University \and KAIST School of Computing}
\authorrunning{Akitoshi Kawamura, Donghyun Lim, Svetlana Selivanova, Martin Ziegler}
\author{Akitoshi Kawamura$^1$, 
	Donghyun Lim$^2$, 
	Svetlana Selivanova$^2$, 
	Martin Ziegler$^2$ \\
\small \texttt{kawamura@inf.kyushu-u.ac.jp},  \quad
\texttt{\{klimdhn,sseliv,ziegler\}@kaist.ac.kr}}
\date{\keywords{Computational Complexity of Continuous Data}}

\makeatletter
\renewcommand\maketitle{\newpage
  \refstepcounter{chapter}%
  \stepcounter{section}%
  \setcounter{section}{0}%
  \setcounter{subsection}{0}%
  \setcounter{figure}{0}
  \setcounter{table}{0}
  \setcounter{equation}{0}
  \setcounter{footnote}{0}%
  \begingroup
    \parindent=\z@
    \renewcommand\thefootnote{\@fnsymbol\c@footnote}%
    \if@twocolumn
      \ifnum \col@number=\@ne
        \@maketitle
      \else
        \twocolumn[\@maketitle]%
      \fi
    \else
      \newpage
      \global\@topnum\z@   % Prevents figures from going at top of page.
      \@maketitle
    \fi
    \thispagestyle{empty}\@thanks
    \def\\{\unskip\ \ignorespaces}\def\inst##1{\unskip{}}%
    \def\thanks##1{\unskip{}}\def\fnmsep{\unskip}%
    \instindent=\hsize
    \advance\instindent by-\headlineindent
    \if@runhead
       \if!\the\titlerunning!\else
         \edef\@title{\the\titlerunning}%
       \fi
       \global\setbox\titrun=\hbox{\small\rm\unboldmath\ignorespaces\@title}%
       \ifdim\wd\titrun>\instindent
          \typeout{Title too long for running head. Please supply}%
          \typeout{a shorter form with \string\titlerunning\space prior to
                   \string\maketitle}%
          \global\setbox\titrun=\hbox{\small\rm
          Title Suppressed Due to Excessive Length}%
       \fi
       \xdef\@title{\copy\titrun}%
    \fi
    \if!\the\tocauthor!\relax
      {\def\and{\noexpand\protect\noexpand\and}%
      \protected@xdef\toc@uthor{\@author}}%
    \else
      \def\\{\noexpand\protect\noexpand\newline}%
      \protected@xdef\scratch{\the\tocauthor}%
      \protected@xdef\toc@uthor{\scratch}%
    \fi
    \if@runhead
       \if!\the\authorrunning!
         \value{@inst}=\value{@auth}%
         \setcounter{@auth}{1}%
       \else
         \edef\@author{\the\authorrunning}%
       \fi
       \global\setbox\authrun=\hbox{\small\unboldmath\@author\unskip}%
       \ifdim\wd\authrun>\instindent
          \typeout{Names of authors too long for running head. Please supply}%
          \typeout{a shorter form with \string\authorrunning\space prior to
                   \string\maketitle}%
          \global\setbox\authrun=\hbox{\small\rm
          Authors Suppressed Due to Excessive Length}%
       \fi
       \xdef\@author{\copy\authrun}%
       \markboth{\@author}{\@title}%
     \fi
  \endgroup
  \setcounter{footnote}{\fnnstart}%
  \clearheadinfo}

\def\institutename{\par
 \begingroup
 \parskip=\z@
 \parindent=\z@
 \setcounter{@inst}{1}%
 \def\and{\qquad\stepcounter{@inst}%
 \noindent$^{\the@inst}$\enspace\ignorespaces}%
 \setbox0=\vbox{\def\thanks##1{}\@institute}%
 \ifnum\c@@inst=1\relax
   \gdef\fnnstart{0}%
 \else
   \xdef\fnnstart{\c@@inst}%
   \setcounter{@inst}{1}%
   \noindent$^{\the@inst}$\enspace
 \fi
 \ignorespaces
 \@institute\par
 \endgroup}

\makeatother

\begin{document}
\maketitle

\begin{abstract}
Choosing an encoding over binary strings for input/output to/by a Turing Machine
is usually straightforward and/or inessential for discrete data (like graphs),
but delicate --- heavily affecting computability and even more computational complexity ---
already regarding real numbers, not to mention more advanced (e.g. Sobolev)
spaces. %common in the analysis of, say, partial differential equations.
For a general theory of computational complexity over continuous data
we introduce and justify `quantitative admissibility' as requirement 
for sensible encodings of arbitrary compact metric spaces,
a refinement of qualitative `admissibility' due to [Kreitz\&Weihrauch'85]:
\par
An \emph{admissible} representation of a T$_0$ space $X$ is a (i) \emph{continuous} partial 
surjective mapping from the Cantor space of infinite binary sequences which is (ii) maximal 
w.r.t. \emph{continuous} reduction. By the Kreitz-Weihrauch (aka ``Main'') Theorem 
of computability theory for continuous data,
a function $f:X\to Y$ with admissible representations of co/domain
is \emph{continuous} ~iff~ it admits \emph{continuous} a 
code-translating mapping on Cantor space, a so-called \emph{realizer}.
We require a \emph{linearly/polynomially} admissible representation of a compact metric space
$(X,d)$ to have (i) asymptotically \emph{optimal} modulus of continuity, namely close to the entropy of $X$,
and (ii) be maximal w.r.t. reduction having \emph{optimal} modulus of continuity in a similar sense.
\par
Careful constructions show the category of such representations to be Cartesian closed,
and non-empty: every compact $(X,d)$ admits a linearly-admissible representation.
Moreover such representations give rise to a tight quantitative correspondence between the 
modulus of continuity of a function $f:X\to Y$ on the one hand and on the other hand
that of its realizer: a ``Main Theorem'' of computational \emph{complexity}.
\par
This suggests (how) to take into account the entropies of the spaces under consideration
when measuring/defining algorithmic cost over continuous data;
and to follow [Kawamura\&Cook'12] considering and classifying 
\emph{generalized} representations with domains `larger' than Cantor space
for (e.g. function) spaces of exponential entropy.
\end{abstract}

\setcounter{footnote}{1}

%\COMMENTED{
\setcounter{tocdepth}{3}
\renewcommand{\contentsname}{}
%\textbf{\small Table of Contents:} \nopagebreak
\section*{Table of Contents}
\begin{center}\small
\begin{minipage}[c]{0.98\textwidth}\vspace*{-8ex}%
\tableofcontents
\end{minipage}
\end{center}
%}

%%%%%%%%%%%%%%%%%%%%%%%%%%%%%%%%%%%%%%%%%%%%%%%%%%%%%%%%%%%%%%%%%%%
\section{Motivation, Background, and Summary of Contribution}
Arguably most computational problems in Science and Engineering
are concerned with continuous rather than with discrete data \cite{BC06,Bra13}.
Here the Theory of Computability exhibits new topological ---
and continuous complexity theory furthermore metric --- aspects
that trivialize, and are thus invisible, in the discrete realm.
In particular input and output require rather careful a choice
of the underlying encoding as sequences of bits
to be read, processed, and written by a Turing machine. For example,
\begin{itemize}
\item
encoding real numbers via their binary expansion 
$x=\sum_{n=0}^\infty b_n2^{-n-1}$, 
and thus operating on the Cantor space of infinite binary sequences $\bar b=(b_0,b_1,\ldots b_n,\ldots)$,
renders arithmetic averaging $[0;1]^2\ni(x,y)\mapsto (x+y)/2\in[0;1]$
\emph{dis}continuous and \emph{un}computable 
\cite{Turing37},\cite[Exercise~7.2.7]{Wei00}.
\item
Encoding real numbers via a sequence of (numerators and denominators, in binary, of)
rational approximations up to absolute error $\leq2^{-n}$ does render averaging computable
\cite[Theorem~4.3.2]{Wei00},
but admits no worst-case bound on computational cost \cite[Examples~7.2.1+7.2.3]{Wei00}.
\item
The \emph{dyadic} representation encodes
$x\in[0;1]$ as any integer sequence $a_n\in\{0,\ldots 2^n\}$ (in binary without leading 0) 
s.t. $|x-a_n/2^n|\leq2^{-n}$; and similarly encode $y\in[0;1]$ as $(b_n)$.
Then the/an integer $c_n$ closest to $(a_{n+1}+b_{n+1})/4$
satisfies $\big|(x+y)/2-c_n/2^n\big|\leq2^{-n}$,
and is easily computed --- but requires first reading/writing
$a_m,b_m,c_m$ for all $m<n$: a total of $\Theta(n^2)$ bits.
\item
Encoding $x\in[0;1]$ as \emph{signed} binary expansion
$x=\sum_{n\geq0} (2b_n+b_{n+1}-1)\cdot2^{-n-1}$ with $b_n\in\{\sdzero,\sdone\}$ s.t. $2b_n+b_{n+1}\in\{-1,0,1\}$,
and similarly $y$, renders averaging computable in linear time $\calO(n)$ \cite[Theorem~7.3.1]{Wei00}.
\end{itemize}
%
%Since there exist $\Theta(2^n)$ different approximations of $x,y,(x+y)/2\in[0;1]$
%up to absolute error $2^{-n}$, distinguishing them requires at least $\Omega(n)$ bits:
The signed binary expansion is thus asymptotically `optimal' up to a constant factor,
the dyadic representation is still optimal up to a quadratic polynomial,
the rational representation is `unbounded',
and the binary expansion is unsuitable.

But how to choose and quantitatively assess complexity-theoretically appropriate
encodings of spaces $X$ other than $[0;1]$,
such as those common in the analysis and solution theory of PDEs \cite{Triebel}?

The present work refines the existing classification of encodings from the 
computability theory of general continuous data while guided by and generalizing 
the well-established theory of computational complexity over real numbers.
There, the binary expansion is known to violate the technical condition
of \emph{admissibility}; and we introduce and investigate quantitative strengthenings
\emph{linear} admissibility (satisfied by the signed binary,
but neither by the dyadic nor by the rational representation) and
\emph{polynomial} admissibility (satisfied by the signed binary
and by the dyadic, but not by the rational representation).

%%%%%%%%%%%%%%%%%%%%%%%%%%%%%%%%%%%%%%%%%
\subsection{Computability over Continuous Data, Complexity in Real Computation}

Here we review established notions and properties
of computability and complexity theory over the reals,
as well as notions and properties of
computability theory over more general abstract spaces:
as guideline to the sensible complexity theory
of more general abstract spaces developed in the sequel.

\begin{figure}[htb]
\begin{center}\includegraphics[width=0.5\textwidth]{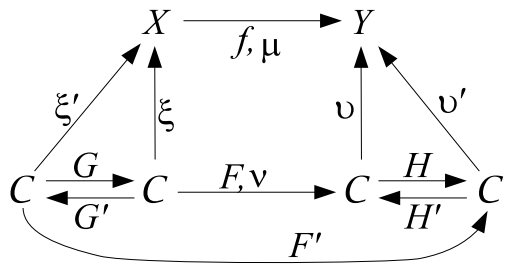}\end{center}
\caption{\label{f:commdiag}Realizers of $f:X\to Y$ with respect to different representations.}
\end{figure}

\begin{definition}
\label{d:Type2}
A \emph{Type-2 Machine} $\Machine$ is a Turing machine with 
dedicated one-way output tape and infinite read-only input tape
{\rm\cite[Definitions~2.1.1+2.1.2]{Wei00}}.

Naturally operating on infinite sequences of bits,
$\Machine$ \emph{computes}
a partial function $F:\subseteq\Cantor\to\Cantor$ 
on the Cantor space $\Cantor=\{\sdzero,\sdone\}^\omega$ of infinite binary sequences if, 
when run with any input $\bar b\in\dom(F)$ on its tape,
$\Machine$ keeps printing the symbols of $F(\bar b)$ one by one;
while its behaviour on other inputs may be arbitrary.

$\Machine$ computes $F$ in \emph{time} $t:\IN\to\IN$ if 
it prints
the $n$-th symbol of $F(\bar b)$ after at most $t(n)$ steps
regardless of $\bar b\in\dom(F)$.

For a fixed predicate $\varphi:\Cantor\to\{\sdzero,\sdone\}$, 
a Type-2 Machine with \emph{oracle} $\varphi$ 
can repeatedly query $\varphi(\vec z)\in\{\sdzero,\sdone\}$
for finite strings $\vec z$ during its computation.
\end{definition}
Concerning topological spaces $X$ of continuum cardinality beyond real numbers,
the \emph{Type-2 Computability Theory} systematically studies and compares
encodings, formalized as follows \cite[\S3]{Wei00}:

\begin{definition}
\label{d:TTE}
\begin{enumerate}
\item[a)]
A \emph{representation} of a set $X$ is a partial surjective
mapping $\xi:\subseteq\Cantor:=\{\sdzero,\sdone\}^{\IN}\twoheadrightarrow X$
on the Cantor space of infinite streams of bits.
\item[b)]
The \emph{product} of representations $\xi:\subseteq\Cantor\twoheadrightarrow X$
and $\upsilon:\subseteq\Cantor\twoheadrightarrow Y$ is
$\xi\times\upsilon:\subseteq\Cantor\ni (b_0,b_1,\ldots b_n,\ldots)\mapsto
\big(\xi(b_0,b_2,b_4,\ldots),\upsilon(b_1,b_3,\ldots)\big)\in X\times Y$.
\item[c)]
For representations $\xi:\subseteq\Cantor\twoheadrightarrow X$
and $\upsilon:\subseteq\Cantor\twoheadrightarrow Y$,
a $(\xi,\upsilon)$-\emph{realizer} of a function $f:X\to Y$
is a partial function $F:\dom(\xi)\to\dom(\upsilon)\subseteq\Cantor$ on Cantor space
such that $f\circ\xi=\upsilon\circ F$ holds; see Figure~\ref{f:commdiag}.
\item[d)]
$(\xi,\upsilon)$-\emph{computing} $f$ means to compute some $(\xi,\upsilon)$-realizer $F$ of $f$
in the sense of Definition~\ref{d:Type2}.
\item[e)]
A \emph{reduction} from representation $\xi\twoheadrightarrow X$ to $\xi'\twoheadrightarrow X$
is a $(\xi,\xi')$-realizer of the identity $\id:X\to X$; that is,
a partial function $F:\dom(\xi)\to\dom(\xi')$ on Cantor space such that $\xi=\xi'\circ F$.
We write $\xi\preccurlyeqT\xi'$ to express that a \emph{continuous} reduction exists,
where $\Cantor$ is equipped with the Cantor space metric $\dC(\bar b,\bar a)=2^{-\min\{n:b_n\neq a_n\}}$.
\end{enumerate}
\end{definition}
Examples~\ref{x:Binary}, \ref{x:Rational}, \ref{x:Dyadic}, and \ref{x:SignedDigit} below
formalize the above binary, rational, dyadic, and signed encodings of the reals 
as representations $\beta$, $\rho$, $\delta$, and $\sigma$, respectively.
It is well-known that the latter three, but not $\beta$, 
are pairwise continuously reducible \cite[Theorem~7.2.5]{Wei00}
and thus equivalent with respect to the notions of computability they induce on reals;
but only $\delta$ and $\sigma$ 
admit mutual reductions with \emph{polynomial} modulus of continuity.

\begin{myremark}
Recall {\rm\cite[\S6]{Wei03}} that a \emph{modulus\footnote{%
From the general logical perspective this constitutes a 
\emph{Skolemization} as canonical quantitative refinement of qualitative continuity \cite{Koh08a}.
We consider a modulus as an integer function due to its close connection
with asymptotic computational cost: Fact~\ref{f:Proper}a+b).
For the continuous conception of a modulus common in Analysis see Lemma~\ref{l:Modulus} below.} continuity}
of a function $f:X\to Y$ between compact metric spaces $(X,d)$ and $(Y,e)$ 
is a non-decreasing mapping $\mu:\IN\to\IN$ such that
$d(x,x')\leq2^{-\mu(n)}$ implies $e\big(f(x),f(x')\big)\leq2^{-n}$.
\\
Every uniformly continuous function has a (pointwise minimal) modulus of continuity;
Lipschitz-continuity corresponds to moduli $\mu(n)=n+\calO(1)$,
and H\"{o}lder-continuity to linear moduli $\mu(n)=\calO(n)$: 
see Fact~\ref{f:Topology}c). 
\end{myremark}
According to the sometimes so-called \emph{Main Theorem} of Computable Analysis,
a real function $f$ is continuous  iff $f$ is computable by some oracle Type-2 Machine 
w.r.t. $\rho$ and/or $\delta$ and/or $\sigma$ \cite[Definitions~2.1.1+2.1.2]{Wei00}.
For spaces beyond the reals, Kreitz and Weihrauch \cite{KW85} have identified 
\emph{admissibility} as central condition on `sensible' representations
in that these %(constitutes the weakest hypothesis that)
make the Main Theorem generalize \cite[Theorem~3.2.11]{Wei00}:

\begin{fact} \label{f:Main}
Let $X$ and $Y$ denote second-countable T$_0$ spaces
equipped with \emph{admissible} representations $\xi$ and $\upsilon$, respectively.
A function $f:X\to Y$ is continuous ~iff~
it admits a continuous $(\xi,\upsilon)$-\emph{realizer}.

Recall {\rm\cite[Theorem~3.2.9.2]{Wei00}} that
a representation $\xi:\subseteq\Cantor\twoheadrightarrow X$ is \emph{admissible} iff 
(i) it is continuous and (ii) every continuous representation $\zeta:\subseteq\Cantor\twoheadrightarrow X$
satisfies $\zeta\preccurlyeqT\xi$.
\end{fact}
Computability-theoretically `sensible' representations $\xi$ are thus those
maximal, among the continuous ones, with respect to continuous reducibility.
The present work refines these considerations and notions from qualitative computability to 
complexity. For representations satisfying our proposed strengthening of \emph{admissibility},
Theorem~\ref{t:Main2} below asymptotically optimally translates 
quantitative continuity between functions $f:X\to Y$ and their realizers $F$.
Such translations heavily depend on the co/domains $X,Y$ under consideration:

\begin{myexample}
\label{x:Max}
\begin{enumerate}
\item[a)]
A function $f:[0;1]\to[0;1]$ has polynomial
modulus of continuity ~iff~ it has a $(\delta,\delta)$-realizer 
w.r.t. the dyadic and/or signed binary expansion
(itself having a polynomial modulus of continuity)
~iff~ $f$ has a $(\sigma,\sigma)$-realizer 
of polynomial modulus of continuity;
cmp. {\rm\cite[Theorem~2.19]{Ko91},\cite[Exercise~7.1.7]{Wei00},\cite[Theorem~14]{DBLP:conf/lics/KawamuraS016}}.
\item[b)]
For the compact space $[0;1]'_1:=\Lip_1([0;1],[0;1])$ of non-expansive (=Lipschitz-continuous with constant 1) %functions 
$f:[0;1]\to[0;1]$ equipped with the supremum norm,
the application function\emph{al} $[0;1]'_1\times[0;1]\ni (f,r)\mapsto f(r)\in[0;1]$
is non-expansive,
yet admits \emph{no} realizer with \emph{sub-}exponential modulus of continuity
for \emph{any} product representation (Definition~\ref{d:TTE}b)
of $[0;1]'_1\times[0;1]$ {\rm\cite[Example~6g]{KMRZ15}}.
\end{enumerate}
\end{myexample}
It is naturally desirable that, like in the discrete setting, 
every computation of a total function $f:X\to Y$ admit some (possibly fast growing, but pointwise finite) 
worst-case complexity bound $t=t(n)$; however already for the real unit interval $X=[0;1]=Y$ 
this requires the representation $\xi$ of $X$ to be chosen with care,
avoiding both binary $\beta$ and rational $\rho$ \cite[Example~7.2.3]{Wei00}.
Specifically, one wants its domain $\dom(\xi)\subseteq\Cantor$ to be compact \cite{Sch95,Wei03,Sch04}:
cmp. Example~\ref{x:Rational}.

\begin{fact} \label{f:Proper}
\begin{enumerate}
\item[a)]
Suppose Type-2 Machine $\Machine$ (with/out some fixed oracle) computes a function
$F:\subseteq\Cantor\to\Cantor$ with compact $\dom(F)$.
Then $\Machine$ admits a bound $t(n)\in\IN=\{0,1,2,\ldots\}$ 
on the number of steps it takes to
print the first $n$ output symbols of the value $F(\bar b)$
\emph{regardless} of the argument $\bar b\in\dom(F)$;
see {\rm\cite[Exercise~7.1.2]{Wei00}}.
\item[b)]
If $F:\subseteq\Cantor\to\Cantor$ is computable (with/out some fixed oracle) in time $t$,
then $n\mapsto t(n)$ constitutes a modulus of continuity of $F$.
\item[c)]
Conversely to every continuous $F:\subseteq\Cantor\to\Cantor$ 
with modulus of continuity $\mu$ there exists an oracle\footnote{%
In our model, query tape and head are not altered by an oracle query.}
$\varphi$ and Type-2 Machine $\Machine^\varphi$ 
computing $F$ in time $\calO\big(n+\mu(n)\big)$;
cmp. {\rm\cite[Theorem~2.3.7.2]{Wei00}}.
\end{enumerate}
\end{fact}
Item~b) expresses that, on Cantor space, quantitative continuity
basically coincides with time-bounded relativized computability.
Item~a) requires a continuous representation $\xi$ to map closed subsets
of Cantor space to closed subsets of $X$ --- hence one cannot expect 
$\xi$ to be an \emph{open} mapping, as popularly posited
in computability \cite[Lemma~3.2.5.(2+3+5)]{Wei00}
and ingredient to (the proof of) the aforementioned Main Theorem;
cmp. \cite[Theorem~3.2.9]{Wei00}. 

%%%%%%%%%%%%%%%%%%%%%%%%%%%%%
\subsection{Summary of Contribution}

We establish a quantitative refinement of the Main Theorem
for arbitrary compact metric spaces,
tightly relating moduli of continuity of functions $f:X\to Y$ to those of their 
realizers $F$ \emph{relative} to the entropies of co/domains $X$ and $Y$:

Recall {\cite{Kolmogorov},\rm\cite[\S6]{Wei03}} that the \emph{entropy}\footnote{%
From the general logical perspective, entropy constitutes a 
\emph{Skolemization} as canonical quantitative refinement of qualitative precompactness \cite{Koh08a}.} 
of a compact metric space $(X,d)$ is the mapping $\eta:\IN\to\IN$ such that
$X$ can by covered by $2^{\eta(n)}$, but not by $2^{\eta(n)-1}$,
closed balls of radius $2^{-n}$. 
The real unit interval $[0;1]$ has entropy $\eta(n)=n-1$;
whereas $[0;1]'_1=\Lip_1([0;1],[0;1])$ has entropy $\eta'_1(n)=\Theta(2^n)$;
see Example~\ref{x:Entropy} for further spaces. 

\begin{myremark}
\label{r:Admissible} 
By Example~\ref{x:Entropy}f), for any modulus of continuity $\kappa$ of a representation 
$\xi:\subseteq\Cantor\twoheadrightarrow X$, the space $X$ has entropy $\eta\leq\kappa$;
and we require a \emph{linearly admissible} $\xi$ to (i) have modulus of continuity
$\kappa(n)\leq\calO\Big(\eta\big(n+\calO(1)\big)\Big)$ almost optimal:
permitting asymptotic `slack' a constant factor in value and constant shift in argument.

Moreover a linearly admissible $\xi$ must satisfy that,
(ii) to every representation $\zeta:\subseteq\Cantor\twoheadrightarrow X$
with modulus of continuity $\nu$ there exists a mapping $F:\dom(\zeta)\to\dom(\xi)$
with modulus of continuity $\mu$ such that $\zeta=\xi\circ F$
\emph{and} $\big(\mu\circ\kappa\big)(n)\leq \nu\big(\calO(n)\big)$.
\end{myremark}
This new Condition~(ii) strengthens previous qualitative continuous reducibility ``$\zeta\preccurlyeqT\xi$''
to what we call \emph{linear} metric reducibility ``$\zeta\preccurlyeqO\xi$'',
requiring a $(\zeta,\xi)$-realizer $F$ with almost optimal modulus of continuity:
For functions $\varphi:X\to Y$ and $\psi:Y\to Z$ with respective moduli of continuity $\mu$ and $\kappa$,
their composition $\psi\circ\varphi:X\to Z$ is easily seen to have modulus of continuity $\mu\circ\kappa$.

Abbreviating  $\lin(\nu):=\calO\Big(\nu\big(\calO(n)\big)\Big)$ 
and with the semi-inverse $\loinv{\nu}(n):=\min\{ m\::\: \nu(m)\geq n\}$,
our results are summarized as follows:
\smallskip
%\begin{theorem}\label{t:Main}
\begin{enumerate}
\item[a)]
Let $(X,d)$ and $(Y,e)$ denote infinite compact metric spaces %of diameter $\leq1$
with entropies $\eta$ and $\theta$
and equipped with \emph{linearly} admissible representations $\xi$ and $\upsilon$.
If $f:X\to Y$ has modulus of continuity $\mu$,
it admits a realizer $F$ with modulus of continuity
$\lin(\eta)\circ \mu\circ\lin\big(\loinv{\theta}\big)$.
Conversely if $F$ is a realizer of $f$ with modulus $\nu$,
then $f$ has modulus 
$\lin\big(\loinv{\eta}\big)\circ\nu\circ\lin(\theta)$. 
\item[b)]
Every compact metric space $(X,d)$ admits a linearly admissible representation $\xi$.
For `popular' spaces $X,Y$ having linear/polynomial entropy $\eta,\theta$,
the moduli of continuity of functions and their realizers are thus
linearly/polynomially related; yet according to Examples~\ref{x:Entropy}d+e)
there exist both spaces of entropy growing arbitrarily slow and arbitrarily fast.
Still, estimates (a) are asymptotically tight in a sense explained in Remark~\ref{r:Tight}.
\item[c)]
The category of quantitatively admissible representations is Cartesian closed:
Given linearly admissible representations for spaces $X_j$ ($j\in\IN$),
we construct one for the product space $\prod_j X_j$ 
w.r.t. the `Hilbert Cube' metric $\dH=\sup_j d_j/2^j$
(cmp. Example~\ref{x:Entropy}a)
such that the canonical projections $(x_0,\ldots x_j,\ldots)\mapsto x_j$
and embeddings $x_j\mapsto (x_0,\ldots x_j,\ldots)$
admit realizers with optimal modulus of continuity in the sense of (a).
For the compact space $\calK(X)$ of non-empty compact subsets w.r.t. Hausdorff Distance
we construct a canonical \emph{polynomially} admissible representation;
and for the compact space $X'_\mu:=\calC_\mu(X,[0;1])$ of functions
$f:X\to[0;1]$ having modulus of continuity $\mu$, equipped with the supremum norm,
one such that the application functional $X'_\mu\times X\ni (f,x)\mapsto f(x)$ 
has a realizer with optimal modulus of continuity in the sense of (a).
\end{enumerate}
%\end{theorem}
%
\smallskip\noindent
See Theorems~\ref{t:Main2} and \ref{t:Admissible} and \ref{t:Cartesian} and \ref{t:Functions}
for the precise statements. 
Example~\ref{x:SignedDigit} verifies that the signed digit expansion 
of the interval $[0;1]$ is linearly admissible;
hence our quantitative ``Main Theorem'' (a) indeed generalizes the real case
as well as quantitatively refining the qualitative Fact~\ref{f:Main}.

\medskip
Note in (a) the typical form of transition maps,
similar for instance to change-of-basis in linear algebra or change-of-chart in differential geometry.
It thus captures the information-theoretic `external' influence of the co/domain
according to Fact~\ref{f:Proper}b), and allows to separate that from the 
`inherent' recursion-theoretic complexity of a computational problem:
Informally speaking, an algorithm operating on continuous data is not 
to be `blamed' for incurring large cost if the underlying domain has 
large entropy, as in Example~\ref{x:Max}b): 
see Remark~\ref{r:SecondOrder} below.

\COMMENTED{
\begin{definition}
\label{d:Adapted}
Fix compact metric spaces $(X,d)$ and $(Y,e)$ 
with entropies $\eta$ and $\theta$ and
with linearly/polynomially admissible representations
$\xi$ and $\upsilon$, respectively.
\emph{Computing} a function $f:X\to Y$ 
with modulus of continuity $\mu$
in linear/polynomial time
means to compute a $(\xi,\upsilon)$-realizer $F$ of $f$
within time linear/polynomial in 
$\lin(\id+\log\eta)\circ \mu\circ\lin\big(\loinv{\id+\log\theta}\big)$.
\end{definition}
This straightforwardly generalizes to \emph{multi}functions \cite{PZ13}.
It consistently `overloads' Definition~\ref{d:Type2} regarding Cantor space,
and recovers the real case $X,Y\in\big\{\Cantor,[0;1]\big\}$ having 
linear entropies $\eta,\theta$.
}

%%%%%%%%%%%%%%%%%%%%%%%%%%%%%%%%%%%%%
\subsection{Previous and Related Work, Current Ideas and Methods}
\label{ss:Related}
Computability theory of real numbers was initiated by Alan Turing (1936),
then generalized to real functions by Grzegorzcyk (1957),
to Euclidean subsets by Kreisel and Lacombe (1957),
to Banach Spaces by Pour-El and Richards (1989),
to topological T$_0$ spaces by Weihrauch \cite[\S3.2]{Wei00},
and furthermore to so-called \emph{QCB} spaces by Schr\"{o}der \cite{Sch02}:
These works had introduced the fundamental notions
which in turn have enabled the abundance of investigations 
that constitute the contemporary computability theory over continuous data.

\medskip
Computational (bit-)complexity theory of real numbers and functions
was introduced by Harvey Friedman and Ker-I Ko \cite{KF82,Bra05e}.
It differs from the discrete setting for instance by measuring 
computational cost in dependence on the output approximation error $2^{-n}$.
Some effort, a careful choice of representation, and the hypothesis
of a compact domain is needed to prove that any total computable real function
actually admits a finite runtime bound depending only on $n$ \cite{Sch04}.
It took even more effort, as well as guidance from discrete 
Implicit Complexity Theory \cite{DBLP:journals/siamcomp/KapronC96,Lam06}, 
to proceed from this Complexity Theory of real functions \cite{Ko91}
to a suitable definition of computational complexity for real operators \cite{KC12}.

The latter involves a modified model of computation discussed in Subsection~\ref{ss:Hyper}.
Again, only the notions introduced in the above works
have enabled the present plethora of investigations 
and rigorous classifications of common numerical problems,
such as \cite{Ko91,RW02,BBY06,BBY07a,BGP11,KMRZ15,Poisson}.
And their sensible further generalization to abstract function 
(e.g. Sobolev) spaces common in analysis is still in under 
development and debate \cite{KSZ16,DBLP:journals/lmcs/Steinberg17}.
Indeed the real co/domain is special in that it has linear entropy;
hence the impact of co/domain on the computational complexity of
problems had been hidden before our quantitative ``Main Theorem'' (a).

In \cite{DBLP:conf/lics/KawamuraS016} we had picked up from \cite{Wei03}
towards a general theory of computational complexity for compact
metric spaces $(X,d)$: exhibiting its entropy $\eta$ as a lower bound 
on the bit-cost of real 1-Lipschitz functions $f:X\to[0;1]$,
and constructing a generic representation 
with modulus of continuity $\kappa(n)\leq\calO\big(n\cdot\eta(n)\big)$
that allows an appropriate (oracle) Type-2 Machine 
to compute any fixed such $f$ in time polynomial in $\eta$.

The present work generalizes and extends this as follows:
\begin{itemize}
\item 
  Theorem~\ref{t:Linear} constructs a generic representation $\xi$
  with (i) modulus of continuity $\kappa\leq\lin(\eta)$ \emph{linear} in the entropy $\eta$
\item 
  and (ii) establishes said $\xi$ \emph{maximal}/complete w.r.t. linear metric reduction
   ``$\zeta\preccurlyeqO\xi$'' among all uniformly continuous representations $\zeta$ of $X$.
\item
  We propose (i) and (ii) as axioms
  and quantitative strengthening of classical, qualitative admissibility
  for \emph{complexity-}theoretically sensible representations.
\item
  Theorem~\ref{t:Main2} strengthens the classical, qualitative Main Theorem
  by \emph{quantitatively} characterizing (moduli of) continuous functions $f:X\to Y$
  in terms of (moduli of) their $(\xi,\upsilon)$-realizers and the entropies of co/domain $X,Y$.
\item 
  Theorem~\ref{t:Admissible}d) confirms the classical categorical binary Cartesian product
  of representations \cite[Definition~3.3.3.1]{Wei00} to maintain \emph{linear admissibility} (i) and (ii).
\item
  The classical categorical countable Cartesian product of representations \cite[Definition~3.3.3.2]{Wei00} 
  does \emph{not} maintain linear (nor polynomial) admissibility; but our modified construction
  exhibited in Theorem~\ref{t:Cartesian} does. 
  In particular (ii) it is \emph{maximal}/complete w.r.t. linear metric reductions.
\item
  Moreover, as opposed to some linearly admissible representation constructed from `scratch'
  by invoking Theorem~\ref{t:Linear}, the representation from Theorem~\ref{t:Cartesian} additionally
  guarantees the canonical componentwise projections and embeddings 
  to admit continuous realizers with optimal moduli of continuity.
\item 
  Theorem~\ref{t:Functions} constructs from any linearly admissible representation $\xi$
  of compact metric $(X,d)$ a \emph{polynomially} admissible representation $\xi'_\mu$
  of $X'_\mu=\calC_\mu(X,[0;1])$. 
  In particular (ii) it is \emph{maximal}/complete w.r.t. polynomial metric reductions.
\item
  Moreover %, as opposed to some linearly admissible representation constructed from `scratch'
  %by invoking Theorem~\ref{t:Linear}, the representation from Theorem~\ref{t:Functions} 
  said representation $\xi'_\mu$
  guarantees the application functional $X'_\mu\times X\ni (f,x)\mapsto f(x)\in[0;1]$
  to have a continuous realizer with optimal modulus of continuity.
\end{itemize}
Revolving around notions like entropy and modulus of continuity,
our considerations and methods are mostly information-theoretic:
carefully constructing representations and realizers,
analyzing the dependence of their value on the argument,
and comparing thus obtained bounds on their modulus of continuity
to bounds on the entropy of the space under consideration,
estimated from above by constructing coverings with `few' balls of given radius $2^{-n}$
as well as bounded from below by constructing subsets of points 
of pairwise distance $>2^{-n}$.

%%%%%%%%%%%%%%%%%%%%%%%%%%%%
\subsection{Overview}
Section~\ref{s:Admissible} formally introduces our 
conception of quantitatively (polynomially/linearly) admissible representations.
Subsection~\ref{ss:Real} re-analyzes the above three representations of $[0;1]$
from this new perspective.
And Subsection~\ref{ss:Topology} collects metric properties of other popular spaces;
including new ones constructed
via binary and countable Cartesian products (w.r.t. Hilbert Cube metric), 
the hyperspace of non-empty compact subsets w.r.t. Hausdorff metric,
and function spaces.
Section~\ref{s:Standard} recalls and rephrases 
our previous results \cite[\S3]{DBLP:conf/lics/KawamuraS016} 
from this new perspective: 
constructing a generic polynomially-admissible representation
for any compact metric space $(X,d)$.
And Subsection~\ref{ss:Linear} improves that to linear admissibility.
Section~\ref{s:Category} finally formally states our 
complexity-theoretic Main Theorem~\ref{t:Main2} in quantitative detail;
and presents categorical constructions for new 
quantitatively admissible representations from given ones,
paralleling the considerations from Subsection~\ref{ss:Topology}:
binary and countable Cartesian products,
Hausdorff hyperspace of non-empty compact subsets, and function spaces.
Section~\ref{s:Conclusion} collects some questions
about possible refinements/strengthenings,
such as improving/dropping constant factors in our results.
Subsection~\ref{ss:Hyper} finally extends our considerations
to generalized representations for higher types in the sense of \cite{KC12}; 
and Subsection~\ref{ss:Future} puts them into the larger context
of quantitatively-universal compact metric spaces.

%%%%%%%%%%%%%%%%%%%%%%%%%%%%%%%%%%%%%
\section{Intuition and Definition of Quantitative Admissibility}
\label{s:Admissible}
In order to refine computability to a sensible theory of computational complexity
we propose in this section
two quantitative refinements of qualitative \emph{admissibility}
formalized in Definition~\ref{d:Admissible} below.
But first let us briefly illustrate how a reasonable representation 
can be turned into an unreasonable one,
and how that affects the computational complexity of a function:
to get an impression of what quantitative admissibility should prohibit.

\medskip
Consider `padding' a given representation $\xi$ with some
fixed strictly increasing $\varphi:\IN\to\IN$ in order 
to obtain a new representation $\xi_\varphi$ defined by 
$\xi_\varphi(\bar b):=\xi\big((b_{\varphi(n)})_{_n}\big)$
for $\bar b=(b_n)_{_n}\in\Cantor$.
Then $\dom(\xi_\varphi)$ is compact whenever $\dom(\xi)$ is;
but computing some $(\xi_\varphi,\upsilon)$-realizer
now may require `skipping' over $\varphi(n)$ bits of any 
given $\xi_\varphi$-name before reaching/collecting
the same information as contained in only the first $n$ bits
of a given $\xi$-name when computing a $(\xi,\upsilon)$-realizer:
possibly increasing the time complexity from $t$ to $t\circ\varphi$,
definitely increasing its optimal modulus of continuity.
On the other end computing a $(\xi,\upsilon_\varphi)$-realizer
might become easier, as now as many as $\varphi(n)$ bits of the 
padded output can be produced from only $n$ bits of the unpadded one:
possibly decreasing the time complexity from $t$ to $t\circ\upinv{\varphi}$,
see Definition~\ref{d:Admissible}c) below.

\begin{definition}
\label{d:Discrete}
\begin{enumerate}
\item[a)]
We abbreviate $\bar x|_{<n}:=(x_0,\ldots x_{n-1})$ and
\[ (x_0,\ldots x_{n-1})\circ Z^\IN \;=\; \big\{(x_0,\ldots x_{n_1},z_n,z_{n+1} \ldots):z_n,z_{n+1} \ldots\in Z\big\}  \enspace . \]
\[ \{\sdzero,\sdone\}^n \ni \vec x = (x_0,\ldots x_{n-1}) \mapsto 
\langle \vec x\rangle := (\sdzero\,x_0\,\sdzero\,x_1\,\sdzero x_2\ldots\,\sdzero\,x_{n-1}\,\sdone)\in\{\sdzero,\sdone\}^{2n+1} \]
denotes a self-delimiting encoding of finite binary strings.

Let $\lfloor r\rceil\in\IZ$ mean the integer closest to $r\in\IR$
with ties broken towards 0:
$\big\lfloor \pm n+\tfrac{1}{2}\big\rceil=\pm n$ for $n\in\IN$.
\item[b)]
Let $\Reg$ denote the set of all non-decreasing 
unbounded mappings $\nu:\IN\to\IN$. % with $\nu(0)=0$.
The \emph{lower} and \emph{upper semi-inverse} of $\nu\in\Reg$ are
\[ %\begin{eqnarray*}
\loinv{\nu}(n) \;:=\; \min\{ m\::\: \nu(m)\geq n\} %wrong: \max\{m \::\: \nu(m-1)<n \}  \\
, \quad \upinv{\nu}(n) \;:=\; %\max\{ m\::\: \nu(m)\leq n\} \;=\; 
\min\{m \::\: \nu(m+1)>n \}  
\enspace . \]
\item[c)]
Extend Landau's class of asymptotic growth 
\begin{eqnarray*} 
\nu \leq \calO(\mu) &\Leftrightarrow& 
  \exists C\in\IN \; \forall n\in\IN : \; \nu(n)\leq C\cdot\mu(n)+C \\
\nu \leq \calP(\mu) &:=&  \calO\big(\mu(n)\big)^{\calO(1)}  
  \Leftrightarrow\; \exists C\;\forall n:\; \nu(n)\leq \big(C+C\cdot \mu(n)\big)^C  \\
\nu \leq \calA(\mu) &:=&  \mu+\calO(1) 
  \Leftrightarrow\; \exists C\;\forall n:\; \nu(n)\leq \mu(n)+C  \\
%\nu \leq \calL(\mu) &:=&  \mu+\calO\big(\log^{\calO(1)}\mu\big) 
%  \Leftrightarrow\; \exists C\;\forall n:\; \nu(n)\leq \mu(n)+C+C\cdot\log^C\nu(n)  \\
%\cals(\mu) &:=$ 
% \big\{\nu\in\Reg \;|\; \exists C\in\IN \; \forall n\in\IN \; \nu(n)\leq \mu(n+C) \big\} ,  \\
\nu\leq \calo(\mu) &:=&  \mu\circ\calO(\id) \quad\Leftrightarrow\quad \exists C\;\forall n:\; \nu(n)\leq \mu(C\cdot n+C)\\
\nu \leq \calp(\mu) &:=&  \mu\circ\calP(\id) \quad\Leftrightarrow\quad \exists C\;\forall n:\; \nu(n)\leq \mu(C\cdot n^C+C)\\
\nu \leq \cala(\mu) &:=&  \mu\circ\calA(\id) \quad\Leftrightarrow\quad \exists C\;\forall n:\; \nu(n)\leq \mu(n+C)\\
%\nu \leq \call(\mu) &:=&  \mu\circ\calL(\id) \quad\Leftrightarrow\quad \exists C\;\forall n:\; \nu(n)\leq \mu(n+C+C\cdot\log^C n)\\[0.4ex]
\nu \leq \lin(\mu) &:=& \calO\big(\calo(\mu)\big) 
= \calo\big(\calO(\mu)\big) 
\;\Leftrightarrow\;\; \exists C\;\forall n:\; \nu(n)\leq C+C\cdot\mu(C+C\cdot n) \\
\nu \leq \poly(\mu) &:=& \calP\big(\calp(\mu)\big) \;\Leftrightarrow\; \exists C\;\forall n:\; \nu(n)\leq \big(n+C+C\cdot \mu(C\cdot n^C+C)\big)^C 
%\nu \leq \off(\mu) &:=& \calA\big(\cala(\mu)\big) \;\Leftrightarrow\; \exists C\;\forall n:\; \nu(n)\leq \mu(n+C)+C \\
\end{eqnarray*}
to denote sequences bounded linearly/polynomially/additively by, 
and/or after linearly/polynomially/additively growing the argument to, $\mu$.
Here $\id:\IN\to\IN$ is the identity mapping.
\end{enumerate}
\end{definition}
In Item~c), classes $\lin(\mu)\leq\poly(\mu)$ capture `relative' asymptotics
in increasing granularity. They are transitive and compositional in the following sense:

\begin{observation}
\label{o:Growth}
\begin{enumerate}
\item[a)]
$\mu\leq\calO(\nu)$ and $\nu\leq\calO(\kappa)$ implies $\mu\leq\calO(\kappa)$. \\
$\mu\leq\calP(\nu)$ and $\nu\leq\calP(\kappa)$ implies $\mu\leq\calP(\kappa)$. \\
$\mu\leq\calA(\nu)$ and $\nu\leq\calA(\kappa)$ implies $\mu\leq\calA(\kappa)$. 
%$\mu\leq\calL(\nu)$ and $\nu\leq\calL(\kappa)$ implies $\mu\leq\calL(\kappa)$. 
\item[b)]
$\mu\leq\calo(\nu)$ and $\nu\leq\calo(\kappa)$ implies $\mu\leq\calo(\kappa)$. \\
$\mu\leq\calp(\nu)$ and $\nu\leq\calp(\kappa)$ implies $\mu\leq\calp(\kappa)$. \\
$\mu\leq\cala(\nu)$ and $\nu\leq\cala(\kappa)$ implies $\mu\leq\cala(\kappa)$. 
%$\mu\leq\call(\nu)$ and $\nu\leq\call(\kappa)$ implies $\mu\leq\call(\kappa)$. 
\item[c)]
$\mu\leq\calO(\nu)$ implies $\mu\circ\kappa \leq \calO(\nu\circ\kappa)$. \\
$\mu\leq\calP(\nu)$ implies $\mu\circ\kappa \leq \calP(\nu\circ\kappa)$. \\
$\mu\leq\calA(\nu)$ implies $\mu\circ\kappa \leq \calA(\nu\circ\kappa)$. 
%$\mu\leq\calL(\nu)$ implies $\mu\circ\kappa \leq \calL(\nu\circ\kappa)$. 
\item[d)]
$\mu\leq\calo(\nu)$ implies $\kappa\circ\mu \leq \calo(\kappa\circ\nu)$. \\
$\mu\leq\calp(\nu)$ implies $\kappa\circ\mu \leq \calp(\kappa\circ\nu)$. \\
$\mu\leq\cala(\nu)$ implies $\kappa\circ\mu \leq \cala(\kappa\circ\nu)$. 
%$\mu\leq\call(\nu)$ implies $\kappa\circ\mu \leq \call(\kappa\circ\nu)$.
\item[e)]
$\calo(\mu)\circ\calO(\nu)
\;=\; \calo(\mu)\circ\nu
\;=\; \mu\circ\calO(\nu)$. \\
$\calp(\mu)\circ\calP(\nu) 
\;=\; \calp(\mu)\circ\nu
\;=\; \mu\circ\calP(\nu)$. \\
$\cala(\mu)\circ\calA(\nu)
\;=\; \cala(\mu)\circ\nu
\;=\; \mu\circ\calA(\nu)$. 
%$\call(\mu)\circ\calL(\nu)
%\;=\; \call(\mu)\circ\nu
%\;=\; \mu\circ\calL(\nu)$.
\end{enumerate}
\end{observation}
In particular polynomial `absolute' growth means $\poly(\id)=\calP(\id)=\calp(\id)$;
and linear means $\lin(\id)=\calO(\id)=\calo(\id)$.
We also collect some properties of the semi-inverses:

\begin{lemma}
\label{l:seminv}
\begin{enumerate}
\item[a)]
For $\mu\in\Reg$, $\loinv{\mu}$ and $\upinv{\mu}$ are again in $\Reg$
and $\loinv{\mu}\leq\upinv{\mu}$.
\item[b)]
Every $\mu\in\Reg$ satisfies
$\loinv{\mu}\circ\mu\;\leq\;\id\;\leq\;\upinv{\mu}\circ\mu$, \\
with equality in case $\mu$ is injective 
(necessarily growing at least linearly); \\[0.3ex]
and every $\mu\in\Reg$ satisfies
$\mu\circ\upinv{\mu}\;\leq\;\id\;\leq\mu\circ\loinv{\mu}$, \\
with equality in case $\mu$ is surjective
(necessarily growing at most linearly).
\item[c)]
For $\nu,\kappa\in\Reg$ it holds
\[
\mu\circ\nu\;\leq\;\kappa 		\;\;\Leftrightarrow\;\; 
\nu\;\leq\;\upinv{\mu}\circ\kappa 	\;\;\Leftrightarrow\;\; 
\mu\;\leq\;\kappa\circ\loinv{\nu}  \]
\item[d)]
Suppose $a,b,c,d\in\IN$ with $b,d\geq1$ and $\mu,\nu\in\Reg$ satisfy
$\forall n: \;\nu(n)\leq a+b\cdot \mu(c+d\cdot n)$.
Then it holds $\forall m: \;\loinv{\mu}(m)\leq c+d\cdot\loinv{\nu}(a+b\cdot m)$
and $\upinv{\mu}(m)\leq d\cdot\upinv{\nu}(a+b\cdot m)+c+d-1$.
\qquad\qquad\qquad
In particular, \\
$\calO\big(\upinv{\mu}\big)=\upinv{\calo(\mu)}$
and $\calo\big(\upinv{\mu}\big)=\upinv{\calO(\mu)}$
and $\lin\big(\upinv{\mu}\big)=\upinv{\lin(\mu)}$;
similarly
$\calO\big(\loinv{\mu}\big)=\loinv{\calo(\mu)}$
and $\calo\big(\loinv{\mu}\big)=\loinv{\calO(\mu)}$
and $\lin\big(\loinv{\mu}\big)=\loinv{\lin(\mu)}$.
\item[e)]
There are $\mu,\nu\in\Reg$ s.t. $\mu\leq\poly(\nu)$ 
but $\loinv{\nu}\not\leq\poly\big(\loinv{\mu}\big)$
and $\upinv{\nu}\not\leq\poly\big(\upinv{\mu}\big)$.
%\\
There are $\mu,\nu\in\Reg$ s.t.
$\mu\circ\loinv{\mu}\not\in\poly(\id)$ 
and $\upinv{\nu}\circ\nu\not\in\poly(\id)$ 
and $\loinv{\nu}\circ\calO(\nu)\not\in\poly(\id)$ 
and $\mu\circ\calO\big(\upinv{\mu}\big)\not\in\poly(\id)$.
\item[f)]
$\max\big\{\upinv{\mu},\upinv{\nu}\big\} =\upinv{\min\{\mu,\nu\}}$ and
$\min\big\{\upinv{\mu},\upinv{\nu}\big\} =\upinv{\max\{\mu,\nu\}}$ and
$\max\big\{\loinv{\mu},\loinv{\nu}\big\} =\loinv{\min\{\mu,\nu\}}$ and
$\min\big\{\loinv{\mu},\loinv{\nu}\big\} =\loinv{\max\{\mu,\nu\}}$.
%$\mu,\nu\in\Reg$.
\end{enumerate}
\end{lemma}
Here finally comes our formal definition of quantitative admissibility:

\begin{definition}
\label{d:Admissible}
\begin{enumerate}
\item[a)]
Consider a compact subset $K$ of a metric space $(X,d)$.
Its \emph{relative entropy} 
is the non-decreasing integer mapping
$\eta=\eta_{X,K}:\IN\to\IN$ such that some $2^{\eta(n)}$,
but no $2^{\eta(n)-1}$, closed balls $\cball(x,r)$ 
of radius $r:=2^{-n}$ with centers $x\in X$ can cover $K$. 
If $X$ itself is compact, we write $\eta_X:=\eta_{X,X}$
for its (intrinsic) entropy.
\item[b)]
Consider a uniformly continuous representation $\xi$ of the compact metric space $(X,d)$
and uniformly continuous mapping $\zeta:\subseteq\Cantor\to X$.
Refining Definition~\ref{d:TTE}e), call a reduction $F:\dom(\zeta)\to\dom(\xi)$ 
\emph{polynomial} (``$\zeta\preccurlyeqP\xi$'') if it has a modulus of continuity $\mu$
and $\xi$ has a modulus $\kappa$ satisfying 
$\mu\circ\kappa\leq\calp(\nu)$ for every modulus $\nu$ of $\zeta$.
\\
$F$ is \emph{linear} (``$\zeta\preccurlyeqO\xi$'') if it holds $\mu\circ\kappa\leq\calo(\nu)$.
%$F$ is \emph{polylogarithmic} (``$\zeta\preccurlyeqL\xi$'') if it holds $\mu\circ\kappa\leq\call(\nu)$:
%again for every every modulus of continuity 
%$\kappa$ of $\xi$ and $\nu$ of $\zeta$ and $\mu$ of $F$.
\item[c)]
A representation $\xi$ of the compact metric space $(X,d)$ is 
\emph{polynomially admissible} iff (i) it has a modulus of continuity
$\kappa\leq\calP\calo(\eta)$, i.e., 
bounded polynomially in the entropy with linearly transformed argument, 
and (ii) every uniformly continuous representation $\zeta:\subseteq\Cantor\twoheadrightarrow X$
satisfies $\zeta\preccurlyeqP\xi$ in the sense of (b).
\item[d)]
Representation $\xi$ of $(X,d)$ is \emph{linearly admissible} iff (i) it has a modulus of continuity
$\kappa\leq\calO\cala(\eta)$, i.e., not exceeding
the entropy by more than a constant factor in value and constant shift in argument, 
and (ii) every uniformly continuous representation $\zeta:\subseteq\Cantor\twoheadrightarrow X$
satisfies $\zeta\preccurlyeqO\xi$.
%\item[e)]
%Representation $\xi$ of $(X,d)$ is 
%\emph{polylogarithmically admissible} iff (i) it has a modulus of continuity
%$\kappa\leq\calL\cala(\eta)$,
%and (ii) every uniformly continuous representation $\zeta:\subseteq\Cantor\twoheadrightarrow X$
%satisfies $\zeta\preccurlyeqL\xi$.
\end{enumerate}
\end{definition}
According to Example~\ref{x:Entropy}f) below,
any representation's modulus of continuity satisfies $\kappa\geq\eta$,
i.e., is bounded from below by the entropy;
and Condition~(i) in Definition~\ref{d:Admissible}c+d)
requires a complexity-theoretically appropriate representation 
to be close to that optimum ---
which itself can be arbitrarily small/large according to Example~\ref{x:Entropy}d+e).
The converse Condition~(ii) in Definition~\ref{d:Admissible}c+d)
similarly requires that $\mu\circ\kappa$, a modulus of continuity of $\xi\circ F=\zeta$
be `close' to that of $\zeta$.
Note that linear admissible representations may (i) exceed the entropy by 
a constant factor in value and by an additive constant in the argument
while (ii) linear reduction only allows for the latter:
because (i) is what we can achieve in Theorem~\ref{t:Linear} 
while (ii) guarantees transitivity; similarly for the polynomial case.

\begin{myremark}
\label{r:Transitive}
We record that relations
``$\preccurlyeqP$'' and ``$\preccurlyeqO$'' are transitive:
Fix $\alpha$ with modulus of continuity $\lambda$,
$\beta$ with modulus $\mu$, and
$\gamma$ with $\nu$; 
and linear reduction $F:\dom(\alpha)\to\dom(\beta)$ with modulus of continuity $\iota$
such that $\alpha=\beta\circ F$ and $\iota\circ\mu\leq\calo(\lambda)$;
as well as linear reduction $G:\dom(\beta)\to\dom(\gamma)$ with modulus of continuity $\kappa$
such that $\beta=\gamma\circ G$ and $\kappa\circ\nu\leq\calo(\mu)$.
Then $\alpha=\gamma\circ G\circ F$, where reduction $G\circ F:\dom(\alpha)\to\dom(\gamma)$
has modulus $\iota\circ\kappa$ satisfying
$(\iota\circ\kappa)\circ \nu\leq \iota\circ \calo(\mu)\leq\calo(\lambda)$ 
by Observation~\ref{o:Growth}e).
\\
Also note that, according to Lemma~\ref{l:seminv}d), 
condition $\mu\circ\kappa\leq\calo(\nu)$ is equivalent to 
$\mu\leq\nu\circ\calO\big(\loinv{\kappa}\big)=\calo(\nu)\circ\loinv{\kappa}$;
similarly for %$\mu\circ\kappa\leq\call(\nu)$ and 
$\mu\circ\kappa\leq\calp(\nu)$.
\end{myremark}
Definition~\ref{d:Admissible}c) coincides with \cite[Definition~18]{DBLP:conf/lics/KawamuraS016}
and strengthens the computability-theoretically common qualitative notion of \emph{admissibility}
from \cite[Definitions~2.1.1+2.1.2]{Wei00};
while Definition~\ref{d:Admissible}d) in turn strengthens (c) from polynomial to linear.

\begin{proof}[Lemma~\ref{l:seminv}]
Most claims are immediate, we thus only expand on Item~d):
\begin{enumerate}
\item[d)]
First suppose $c=0$ and $d=1$. Then 
\begin{eqnarray*}
\loinv{\mu}(m) &=& \min\underbrace{\{n:a+b\cdot \mu(n)\geq a+b\cdot m\}}_{
\rotatebox[origin=c]{270}{$\supseteq$}}
\\ &\leq& \min\overbrace{\{n:\nu(n)\geq a+b\cdot m\}} \;=\; \loinv{\nu}(a+b\cdot m) \quad\text{ and} \\[1ex]
\upinv{\mu}(m) &=& \min\underbrace{\{n:a+b\cdot \mu(n+1)> a+b\cdot m\}}_{
\rotatebox[origin=c]{270}{$\subseteq$}}
\\ &\leq& \min\{n:\nu(n+1)> a+b\cdot m\} \;=\; \upinv{\nu}(a+b\cdot m) \enspace . \\[1ex]
\loinv{\mu}(m) &=& \min\underbrace{\{n:\mu(n)\geq m\}}_{
\rotatebox[origin=c]{270}{$\supseteq$}}
\\ &\leq& \min\overbrace{\{c+d\cdot n':\mu(c+d\cdot n')\geq m\}} 
\;\leq\; 
\min\{c+d\cdot n':\nu(n')\geq m\} 
\end{eqnarray*}
$=\; c+d\cdot\loinv{\nu}(m)$ in case $a=0$ and $b=1$. If additionally $d=1$, then
\begin{eqnarray*}
\upinv{\mu}(m)&=&
 \min\big\{n:\mu(n+1)> m\big\} 
\;\leq\; \min\big\{c+n:\mu(c+n+1)> m\} \\
&\leq& \min\big\{c+n:\nu(n+1)> m\}
\;=\; c+\upinv{\nu}(m) 
\end{eqnarray*}
Finally in case $a=0=c$ and $b=1\neq d$,
\begin{eqnarray*}
\upinv{\mu}(m)&=&
\min\big\{n:\mu(n+1)> m\big\} 
\;\leq\; \min\big\{d\cdot \lceil n/d\rceil :\mu(n+1)> m\big\}  \\
&\leq& \min\big\{d\cdot n' + (d-1) :\mu(c\cdot n'+1)> m\big\}  \\
&\leq& \min\big\{d\cdot n' + (d-1) :\nu(n'+1)> m\big\} 
\end{eqnarray*}
$=\; d\cdot\upinv{\nu}(m)+(d-1)$.
\qed\end{enumerate}\end{proof}
%
%%%%%%%%%%%%%%%%%%%%%%%%%%%%%%%%%%%%%%%%%
\subsection{Real Examples}
\label{ss:Real}

Here we formally recall, and analyze from the perspective of admissibility,
the three representations of the real unit interval mentioned in the introduction:
binary, dyadic and signed binary.
Let us record that the real unit interval $[0;1]$ 
has entropy $\eta_{[0;1]}(n)=n-1$ for all integers $n\geq1$.

\begin{myexample}[Binary Representation] \label{x:Binary}
The binary representation of the real unit interval 
\[ 
\beta: \; \Cantor \;\ni\; \bar b \;\mapsto\; \sum\nolimits_n b_n2^{-n+1} \;\in\; [0;1] 
\] 
is surjective and 1-Lipschitz, i.e., has the identity $\id:\IN\ni n\mapsto n\in\IN$ as modulus of continuity:
coinciding with the entropy up to shift 1, i.e., optimal!
However it is not (even qualitatively) admissible {\rm\cite[Theorem~4.1.13.6]{Wei00}},
does not admit a continuous realizer of, e.g.,
the continuous mapping $[0;1/3]\ni x\mapsto 3x\in[0;1]$;
cmp. {\rm\cite[Example~2.1.4.7]{Wei00}}.
\end{myexample}

\begin{myexample}[Rational Representation] \label{x:Rational}
Consider the binary encoding of non-negative integers without leading $\sdzero$:
\begin{equation}
\label{e:Binary}
 \bin \;:\; \IN \;\ni\; 2^n-1+\sum\nolimits_{0\leq j<n} b_j2^j \;\mapsto\; (b_0,\ldots b_{n-1})\; \in\; \{\sdzero,\sdone\}^n  \enspace .
\end{equation}
The \emph{rational} representation of $[0;1]$ is the mapping
$\rho:\subseteq\Cantor\twoheadrightarrow[0;1]$ with
\begin{multline*}
\big(
\langle\bin(a_0)\rangle\: \langle\bin(c_0)\rangle\;
\langle\bin(a_1)\rangle\: \langle\bin(c_1)\rangle\;
%\langle\bin(a_2)\rangle\: \langle\bin(c_2)\rangle\;
\ldots
\langle\bin(a_n)\rangle\: \langle\bin(c_n)\rangle\;
\ldots \big)
\;\mapsto\; \lim\nolimits_j a_j/c_j , \\[0.5ex]
\dom(\rho) \:=\: \big\{
\big(  %\langle\bin(a_0)\rangle\:
\ldots\:
\langle\bin(a_n)\:\langle\bin(c_n)\;\rangle\:\ldots\big) \::\:
\exists x\in[0;1] \: |a_n/c_n-x|\leq 2^{-n}\big\} \enspace .
\end{multline*}
Representation $\rho$ is continuous, but not uniformly continuous
(its domain is not compact) and thus has no modulus of continuity.
\end{myexample}
\begin{proof}
Consider a $\rho$-name of $r=1/2$ 
starting with any $a_0\in\IN$ and $b_0:=2a_0$.
Increasingly long $a_0$ thus give rise to a sequence of $\rho$-names of $r$
with no converging subsequence.
Moreover $a_0/b_0=1/2$ fixes $r$ up to error $2^{-n}$ only for $n:=0$;
but requires `knowing' the first $\mu(0)\geq\log_2(a_0)\to\infty$ bits of its $\rho$-name.
\qed\end{proof}

\begin{myexample}[Dyadic Representation] \label{x:Dyadic}
The \emph{dyadic representation} of the real unit interval $[0;1]$
\begin{multline*}
\delta:\subseteq\Cantor\;\ni\;\big(
\langle\bin(a_0)\rangle\:
\langle\bin(a_1)\rangle\:
\langle\bin(a_2)\rangle\:
\ldots
\langle\bin(a_n)\rangle\:
\ldots \big)
\;\mapsto\; \lim\nolimits_j a_j/2^j , \\[0.5ex]
\dom(\delta) \:=\: \big\{
\big(  %\langle\bin(a_0)\rangle\:
\ldots\:
\langle\bin(a_n)\rangle\:\ldots\big) \::\:
2^n \geq a_n\in\IN, \;
|a_n/2^n-a_m/2^m|\leq 2^{-n}+2^{-m}\big\}
\end{multline*}
\begin{enumerate}
\item[i)] has a quadratic modulus of continuity $\kappa(n):=2\cdot(n+1)\cdot(n+2)$
but no sub-quadratic one and in particular is not H\"{o}lder-continuous. 
\item[ii)] To every partial function $\zeta:\subseteq\Cantor\to [0;1]$
with modulus of continuity $\nu$
there exists a mapping $F:\dom(\zeta)\to\dom(\delta)$ 
with modulus of continuity $\nu$
such that $\zeta=\delta\circ F$ holds. \\
In particular $\delta$ is polynomially admissible. 
\item[iii)]
To every $m\in\IN$ and every $r,r'\in[0;1]$ with $|r-r'|\leq2^{-m-1}$,
there exist $\delta$-names $\bar y_r$ and $\bar y'_{r'}$ 
of $r=\delta(\bar y_r)$ and $r'=\delta(\bar y'_{r'})$
with $\dC(\bar y_r,\bar y'_{r'})\leq2^{-m-1}$.
\item[iv)]
If $(Y,e)$ is a compact metric space and 
$f:[0;1]\to Y$ such that $f\circ\delta:\dom(\delta)\subseteq\Cantor\to Y$
has modulus of continuity $\nu$, 
then $f$ has modulus of continuity $\nu$.
\end{enumerate}
\end{myexample}
Here and as opposed to Definition~\ref{d:Admissible},
(ii) applies also to non-surjective $\zeta$.

\begin{proof}
\begin{enumerate}
\item[i)]
Record %(Definition~\ref{d:TTE}d) 
that $0\leq a_n \leq 2^n$ implies $\langle\bin(a_n)\rangle\in\{\sdzero,\sdone\}^*$
to have length between 1 and $2n+1$; and
$\langle\bin(a_0)\rangle\: \ldots \langle\bin(a_{n})\rangle$
has binary length between $n+1$ and $\kappa(n)$.
Therefore perturbing a $\delta$-name $\bar y$ of some $r\in[0;1]$
to $\bar y'$ with $\dC(\bar y,\bar y')\leq2^{-\kappa(n)}$ 
will keep (the binary expansions of) $a_0,\ldots a_{n}$ unmodified;
and thus satisfies $|\delta(\bar y)-\delta(\bar y')|\leq 2^{-n}$.
Hence $\kappa(n)$ is a modulus of continuity.
On the other hand consider the $\delta$-name $\bar y$
of $r:=3/4$ with $a_n:=3\cdot 2^{n-2}$ for $n\neq2$
has $\langle\bin(a_n)\rangle$ of length $2n-1$
and starts at bit position $\sum_{m<n} (2m-1)\geq\Omega(n^2)$;
yet changing $a'_{m}:\equiv3\cdot 2^{m-2}+2^{m-n}$ for all $m>n$
turns it into a $\delta$-name $\bar y'$ of $r':=r+2^{-n}$
with $\dC(\bar y,\bar y')\leq2^{\calO(n^2)}$.
So $\delta$ has no sub-quadratic modulus of continuity.
\item[ii)]
We construct $F:\dom(\zeta)\to\dom(\delta)$
as limit $F(\bar x)=\lim_n F_n\big(\bar x|_{<\nu(n+1)}\big)$ 
of a sequence of partial functions
$F_n:\subseteq\{\sdzero,\sdone\}^{\nu(n+1)}\to\{\sdzero,\sdone\}^{<\kappa(n)}$
which is monotone in that $F_{n+1}\big(\bar x|_{<\nu(n+2)}\big)$
contains $F_{n}\big(\bar x|_{<\nu(n+1)}\big)$ as initial segment.
To every $n$ and each (of the finitely many) $\bar x|_{<\nu(n+1)}$ with $\bar x\in\dom(\zeta)$,
fix some $r_n=r_n\big(\bar x|_{<\nu(n+1)}\big)\in\zeta\big[\bar x|_{<\nu(n+1)}\circ\Cantor]\subseteq[0;1]$.
Then given $\bar x\in\dom(\delta)$ and iteratively for $n=0,1,\ldots$ let 
\[ F_{n}\big(\bar x|_{<\nu(n+1)}\big) \;:=\; F_{n-1}\big(\bar x|_{<\nu(n)}\big) \:\circ\: \langle\bin(a_n)\rangle,
\quad a_n:=\lfloor r_n\cdot2^n\rceil\in\{0,\ldots 2^n\} \enspace . \]
Since $\nu$ is a modulus of continuity of $\zeta$, it follows 
\[ \zeta\big[\bar x|_{<\nu(n+1)}\circ\Cantor\big] \;\subseteq\; \big[r_n-2^{-n-1};r_n+2^{-n-1}\big] 
\;\subseteq\; \big[\tfrac{a_n}{2^n}-2^{-n};\tfrac{a_n}{2^n}+2^{-n}\big] \]
as $|r_n-a_n/2^n|\leq2^{-n-1}$. Thus it holds
$|a_n/2^n-r|\leq2^{-n}$ for $r:=\zeta(\bar x)$ since $|r-r_n|\leq2^{-n-1}$;
$F(\bar x) = \lim\nolimits_n F_n(\bar x_n) = 
\big(\langle\bin(a_0)\rangle\:\ldots\: \langle\bin(a_n)\rangle\: \ldots \big)$
is a $\delta$-name of $r$; 
and fixing the first $\nu(n+1)$ symbols of $\bar x\in\dom(\zeta)$
fixes $\vec y:=F_n\big(\bar x|_{<\nu(n+1)}\big)$ as well as $(a_0,\ldots a_n)$ 
and therefore also (at least) the first $n+1$ symbols of $\bar y:=F(\bar x)$:
hence $F$ has modulus of continuity $\nu(n)$.
\\
As recorded above, $[0;1]$ has entropy $\eta(n)=n-1$;
hence (i) and (ii) imply polynomial admissibility
according to Definition~\ref{d:Admissible}c).
\item[iii)]
xTo $r\in [0;1]$ consider the $\delta$-name $\bar y_r:=\big(\ldots \langle\bin(a_n)\rangle\:\ldots \big)\in\Cantor$ of $r$
with $a_n:=\lfloor r\cdot2^n\rceil$.
For every $m\in\IN$, its initial segment
$\big( \langle\bin(a_0)\rangle\: \ldots \langle\bin(a_m)\rangle\big)$
has binary length between $m+1$ and $\kappa(m)$; 
and, for every $r'\in[0;1]$ with $|r-r'|\leq2^{-m-1}$,
can be extended to a $\delta$-name $\bar y'_{r'}$ 
via $a'_{m'}:=\lfloor r'\cdot2^{m'}\rceil$ for all $m'>m$.
\item[iv)]
Applying (iii) to $m:=\nu(n)$, the hypothesis implies
$e\big(f(r),f(r')\big)=e\big(f\circ\delta(\bar y_r),f\circ\delta(\bar y'_{r'})\big)\leq 2^{-n}$.
\qed\end{enumerate}\end{proof}
The dyadic representation in Example~\ref{x:Dyadic} has modulus of continuity
quadratic (i.e. polynomial), but not linear, in the entropy. This overhead comes
from the `redundancy' of the precision-$n$ approximation $a_n/2^n$ of binary length
$\calO(n)$ superseding all previous $a_m/2^m$, $m<n$. The signed binary representation 
on the other hand achieves precision $2^{-n}$ by appending one `signed' digit 
$\tilde b_{n-2}\in\{-1,0,1\}$, encoded as two binary digits $(b_{2n-4},b_{2n-3})\in\{\sdzero\sdzero,\sdzero\sdone,\sdone\sdzero\}$
via $\tilde b_{n-2}=2b_{2n-4}+b_{2n-3}-1$, to the previous 
approximation up to error $2^{-n+1}$, yielding a modulus of continuity linear in the entropy:

\begin{myexample}[Signed Binary Represent.] \label{x:SignedDigit}
The signed binary representation, considered as total mapping
\begin{equation}
\label{e:SignedDigit}
\sigma:\subseteq\{\sdzero\sdzero,\sdzero\sdone,\sdone\sdzero\}^\IN\subseteq\Cantor 
\;\ni \bar b \mapsto\; 
\tfrac{1}{2}+\sum\limits_{m=0}^{\infty} (2b_{2m}+b_{2m+1}-1) \cdot 2^{-m-2} 
%\;=\; \tfrac{1}{2}+\sum\limits_{m=2}^{\infty} (2b_{2m-4}+b_{2m-3}-1) \cdot 2^{-m} 
\;{\in}\; [0;1] 
\end{equation}
\begin{enumerate}
\item[i)]
is surjective and has modulus of continuity $\kappa(n)=2n$, 
i.e., is H\"{o}lder-continuous.
\item[ii)]
To every partial function $\zeta:\subseteq\Cantor\to [0;1]$
with modulus of continuity $\nu$
there exists a mapping $F:\dom(\zeta)\to\dom(\sigma)$
with modulus of continuity $\kappa:2m\mapsto\nu(m+1)$
such that $\zeta=\sigma\circ F$ holds. \\
In particular $\sigma$ is linearly admissible.
\item[iii)]
To every $n\in\IN$ and every $r,r'\in[0;1]$ with $|r-r'|\leq2^{-n}$,
there exist $\sigma$-names $\bar y_r$ and $\bar y'_{r'}$ 
of $r=\sigma(\bar y_r)$ and $r'=\sigma(\bar y'_{r'})$
with $\dC(\bar y_r,\bar y'_{r'})\leq2^{-2n}$.
\item[iv)]
If $(Y,e)$ is a compact metric space and 
$f:[0;1]\to Y$ such that $f\circ\sigma:\dom(\sigma)\subseteq\Cantor\to Y$
has modulus of continuity $2\nu$, then $f$ has modulus of continuity $\nu$.
\end{enumerate}
\end{myexample}
\begin{proof}[Example~\ref{x:SignedDigit}]
\begin{enumerate}
\item[i)]
Note that appropriate choice of $\tilde b_{0},\tilde b_{1},\ldots\in\{-1,0,1\}$
yields precisely any possible value $[-1/2;+1/2]\ni\sum_{m=0}^\infty \tilde b_{m-2} \cdot 2^{-m}$;
hence $\sigma$ is total and surjective. In one worst case,
changing $(b_{2m},b_{2m+1})=\sdzero\sdzero$ (encoding the signed digit $-1$)
to $(b'_{2m},b'_{2m+1})=\sdone\sdzero$ (encoding $+1$) for all $m\geq n$
changes the real number $r$ from Equation~(\ref{e:SignedDigit})
to $r'=r+\sum\nolimits_{m\geq n} (2) \cdot 2^{-m-2}=r+2^{-n-1}$;
while $\bar b'$ agrees with $\bar b$ up to position $2n-1$:
This asserts $\kappa(n)=2n$ to be a modulus of continuity and,
in view of $[0;1]$ having entropy $\eta(n)=n-1$, 
establishes Condition~(i) of Definition~\ref{d:Admissible}b).
\item[ii)]
Similarly to the proof of Example~\ref{x:Dyadic},
for every $n$ and every $\bar x|_{<\nu(n+1)}$ with $\bar x\in\dom(\zeta)$,
consider the compact set $\zeta\big[\bar x|_{<\nu(n+1)}\circ\Cantor\big]\subseteq[0;1]$:
Having diameter $\leq2^{-(n+1)}$ by the definition of $\nu$, 
it is contained in $[r_n-2^{-n-2};r_n+2^{-n-2}]$ for 
$r_n:=\big(\min\zeta\big[\bar x|_{<\nu(n+1)}\circ\Cantor\big]+\max\zeta\big[\bar x|_{<\nu(n+1)}\circ\Cantor\big]\big)/2\in[0;1]$;
hence $|r_{n+1}-r_n|\leq2^{-n-2}$.
Now let $r'_1:=\tfrac{1}{2}$ capture the constant term $\tilde b_{-1}:=-1$ in Equation~(\ref{e:SignedDigit})
such that $|r_1-r'_1|\leq\tfrac{1}{4}$ with $r_1\in\big(\tfrac{1}{4};\tfrac{3}{4}\big)$;
and for $n=2,3,\ldots$ inductively append one additional signed digit 
\[
2b_{2n-4}+b_{2n-3}-1\;=\;\tilde b_{n-1}\;:=\;\lfloor 2^{n}\cdot(\underbrace{r_n-r'_{n-1}}_{\leq\pm3\cdot2^{-n-1}}
)\rceil\in\{-1,0,+1\} \]
such that $r'_n\;:=\;\tfrac{1}{2}+\sum\nolimits_{m=1}^{n} \tilde b_{m-2} \cdot 2^{-m}$
again satisfies $|r_n-r'_n|\leq2^{-n-1}$ and $|r_{n+1}-r'_n|\leq|r_{n+1}-r_n|+|r_n-r'_n|\leq3\cdot2^{-n-2}$
and $|r-r'_n|\leq|r-r_n|+|r_n-r'_n|\leq2^{-n}$ for $r=\zeta(\bar x)\in\zeta\big[\bar x|_{<\nu(n+1)}\circ\Cantor\big]$:
Hence $F(\bar x):=(b_0,\ldots b_{2n-4},b_{2n-3},\ldots)$ is a $\sigma$-name of $r$,
and the thus defined function $F$ has modulus of continuity $2n\mapsto\nu(n+1)$.
\item[iii)]
To $r\in [0;1]$ and $n\in\IN$ consider signed digits
$\tilde b_0,\ldots \tilde b_{n-2}\in\{-1,0,1\}$
and $r'_n\;:=\;\tfrac{1}{2}+\sum\nolimits_{m=1}^{n} \tilde b_{m-2} \cdot 2^{-m}$
with $|r-r'_n|\leq2^{-n-1}$ as in (ii).
As in (i), appropriate choice of $\tilde b_{n-1},\tilde b_{n},\ldots\in\{-1,0,1\}$
yields any possible value $[-2^{-n};+2^{-n}]\ni\sum_{m=n-1}^\infty \tilde b_{m-2} \cdot 2^{-m}$;
hence every $r'\in[0;1]$ with $|r-r'|\leq2^{-n}$ admits a signed binary expansion
$r'=\tfrac{1}{2}+\sum_{m=1}^{\infty} \tilde b_{m-2}\cdot2^{-m}$
extending $(\tilde b_0,\ldots \tilde b_{n-2})$,
and $\sigma$-name $\bar y_{r'}$ coinciding on the first $2n$ binary symbols.
\item[iv)]
Applying (iii) to $n:=\nu(m)$, the hypothesis implies
$e\big(f(r),f(r')\big)=e\big(f\circ\sigma(\bar y_r),f\circ\sigma(\bar y'_r)\big)\leq 2^{-m}$.
\qed\end{enumerate}\end{proof}
The signed binary representation renders real addition
computable by a finite-state transducer:

\begin{center}
\includegraphics[width=0.9\textwidth]{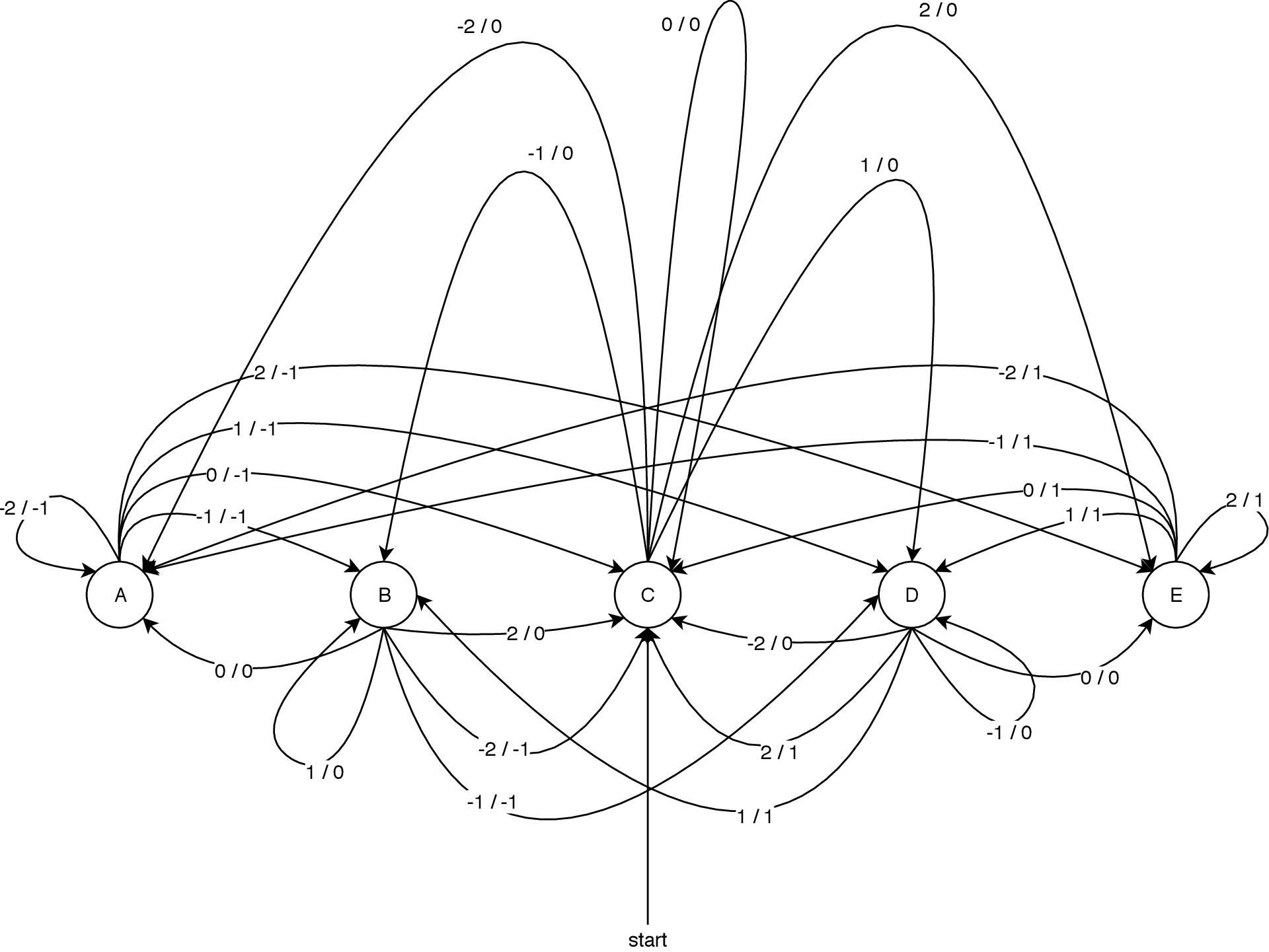}
\end{center}
Starting off in state \stateC,
in each round $\#n=0,1,\ldots$ it reads the next signed digits $a_{n},b_{n}\in\{-1,0,1\}$ 
in the respective expansions of real arguments $x=\sum_n a_n2^{-n}$ and $y=\sum_n b_n2^{-n}$,
and follows that edge whose \emph{first} label agrees with $a_n+b_n$
while outputting the \emph{second} label $c_{n-2}\in\{-1,0,1\}$  
of said edge such that $x+y=\sum_n c_n2^{-n}$. 
The transducer works by storing for each state the accumulated value from previous input except those already output.

\medskip
The signed-digit expansion's modulus of continuity 
leaves a constant-factor gap to the entropy, attained by the binary expansion.
One can trade between both, namely permit $\sdminus$ only at asymptotically fewer positions (i)
while incurring possible `carry ripples' between them over asymptotically longer ranges (ii):

\begin{myexample} \label{x:Gleb}
Fix a strictly increasing function $\varphi:\IN\to\IN$ with $\varphi(0)=0$.
Representation $\sigma_\varphi$
`interpolates' between Examples~\ref{x:Binary} and \ref{x:SignedDigit} 
by considering signed binary expansions
$\sum\limits_{m=1}^{\infty} \tilde c_{m-1} \cdot 2^{-m}$
with $\tilde c_{m}\in\{\sdminus,\sdzero,\sdone\}$ for every $m\in\range\varphi=\varphi[\IN]$ 
but $\tilde c_m\in\{\sdzero,\sdone\}$ for all $m\in\IN\setminus\varphi[\IN]$.
Each $\tilde c_m$ with $m\in\IN\setminus\varphi[\IN]$ is encoded as one bit,
each $\tilde c_m$ with $m\in\varphi[\IN]$ as two bits.
Thus $\sigma_{\id}=\sigma$ recovers Example~\ref{x:SignedDigit}.
\begin{enumerate}
\item[i)]
$\sigma_\varphi$ is surjective and has modulus of continuity $n\mapsto n+\upinv{\varphi}(n)$. 
\item[ii)]
There exists a mapping $F_\varphi:\dom(\sigma)\to\dom(\sigma_\varphi)$
with modulus of continuity $2\varphi\circ\upinv{(\id+\varphi)}$ 
such that $\sigma=\sigma_\varphi\circ F_\varphi$ holds.
\end{enumerate}
In particular $\sigma_{n^2}:=\sigma_{n\mapsto n^2}$ has modulus of continuity $n+\calO(\sqrt{n})$;
and to every partial function $\zeta:\subseteq\Cantor\to [0;1]$
with modulus of continuity $\nu$
there exists a mapping $F_{n^2}:\dom(\zeta)\to\dom(\sigma_{n^2})$
with modulus of continuity $n\mapsto\nu(n+1)+\calO\big(\sqrt{\nu(n+1)}\big)$
such that $\zeta=\sigma_{n^2}\circ F_{n^2}$.
\end{myexample}
Note that $\varphi(n):=2^n$ has 
$\upinv{(\id+\varphi)}(n)\geq \log_2(n)+1$ infinitely often
and therefore $\varphi\circ\upinv{(\id+\varphi)}(n)\geq 2\cdot n$:
yielding a linear reduction $\zeta\preccurlyeqO\sigma_\varphi$, 
but no better.

\begin{proof}[Example~\ref{x:Gleb}]
\begin{enumerate}
\item[i)]
Similarly to the proof of Example~\ref{x:SignedDigit}i),
the first $n$ digits $\tilde c_0,\ldots \tilde c_{n-1}$ of an expansion
fix the value up to absolute error $<2^{-n}$.
Differing from Example~\ref{x:SignedDigit},
this initial segment of the expansion occupies not $2n$ 
but $n+\upinv{\varphi}(n)$ bits since `signed' digits (permitted)
only at the $\upinv{\varphi}(n)$ positions $\varphi[\IN]\cap\{0,\ldots n-1\}$.
\item[ii)]
We describe a transformation $F_\varphi$ converting a given signed-digit expansion
$r=\tfrac{1}{2}+\sum\limits_{m=1}^{\infty} \tilde b_{m-2} \cdot 2^{-m}$
with $\tilde b_m\in\{\sdminus,\sdzero,\sdone\}$ to the required form
$r=\sum\limits_{m=1}^{\infty} \tilde c_{m-1} \cdot 2^{-m}$
with $\tilde c_m\neq\sdminus$ except for positions $m\in\varphi[\IN]$. 
Reflecting the constant term in Equation~(\ref{e:SignedDigit}), 
initially let $c_0:=\sdone$, tentatively.
Now iteratively for $k=1,2,\ldots$ re-code the signed integer
\begin{eqnarray*}
-2^{\varphi(k)}+1 &=& 
0 \:- 2^{\varphi(k)-1} \:- 2^{\varphi(k)-2} \:-\ldots\: -2 \:-1 \\
&\leq&
2^{\varphi(k)}\cdot c_{\varphi(k-1)}\:+\:
2^{\varphi(k)-1}\cdot \tilde b_{\varphi(k-1)} \:+\:
2^{\varphi(k)-2}\cdot \tilde b_{\varphi(k-1)+1} \:+\ldots \\
&& \ldots+\: 2 \cdot \tilde b_{\varphi(k)-2} \:+\: \tilde b_{\varphi(k)-1}  \\
&=:&
2^{\varphi(k)}\cdot \tilde c_{\varphi(k-1)} \:+\:
2^{\varphi(k)-1}\cdot \tilde c_{\varphi(k-1)+1} \:+\: 2^{\varphi(k)-2}\cdot \tilde c_{\varphi(k-1)+2} \:+\ldots \\
&& \ldots+\: 2 \cdot \tilde c_{\varphi(k)-1} \:+\: c_{\varphi(k)}
\end{eqnarray*}
uniquely with $\tilde c_{\varphi(k-1)}\in\{\sdminus,\sdzero,\sdone\}$ and
$\tilde c_{\varphi(k-1)+1},\ldots \tilde c_{\varphi(k)-1},c_{\varphi(k)}\in\{\sdzero,\sdone\}$:
the latter again only tentatively.
Thus the $\varphi(k)+k$ bits of $(\tilde c_0,\ldots \tilde c_{\varphi(k)-1})$
depend precisely on the $2\varphi(k)$ bits of $(\tilde b_0,\ldots \tilde b_{\varphi(k)-1})$:
the transformation $F_\varphi$ on Cantor space thus 
has modulus of continuity $\varphi(k)+k\mapsto2\varphi(k)$ for all $k\in\IN$, 
and $2\varphi\circ\upinv{(\id+\varphi)}$ in general.
\qed\end{enumerate}\end{proof}

%%%%%%%%%%%%%%%%%%%%%%%%%%%%%%%%%%%%%
\subsection{Abstract Examples}
\label{ss:Topology}

This subsection collects some properties, relations, and examples of
moduli of continuity and entropies of spaces.

\begin{fact}
\label{f:Topology}
\begin{enumerate}
\item[a)]
Every compact metric space $(X,d)$ can be covered by
finitely many open balls $\ball(x,r)$; therefore its entropy $\eta$ is well-defined.
If $X$ is infinite %of diameter $\diam(X):=\max\{d(x,x'):x,x'\in X\}\leq1$, 
then $\eta\in\Reg$.
\item[b)]
Every continuous function $f:X\to Y$ between compact metric spaces $(X,d)$ and $(Y,e)$
is uniformly continuous and therefore has a modulus $\mu$ of continuity. 
%If $\diam(Y)\leq1$ one may w.l.o.g. suppose $\mu\in\Reg$.
%If $f:X\to Y$ has modulus of continuity $\mu$
%and $g:Y\to Z$ has $\nu$, then $g\circ f$ has $\mu\circ\nu$.
\item[c)]
Lipschitz-continuous functions have moduli of continuity
$\mu(n)=n+\calO(1)$, and vice versa.
H\"{o}lder-continuous functions have linear moduli of continuity
$\mu(n)=\calO(n)$, and vice versa.
\item[d)]
%Proceeding from metric $d$ on $X$ to H\"{o}lder-equivalent $d'$ 
%with $k,\ell$ according to Definition~\ref{d:Admissible}c)
%and similarly proceeding from $e$ on $Y$ to $e'$ with $K,L$,
%turns a modulus of continuity $\mu$ of $f:X\to Y$ into $\mu'$ with
%\[ \mu(n) \;\leq\; k\cdot\big(\ell+\mu'(L+K\cdot n)\big) 
%\quad\wedge\quad
%\mu'(n) \;\leq\; \ell+k\cdot\mu\big((n+L)\cdot K\big) \enspace . \]
Proceeding from metric $d$ on $X$ to a metric $d'$ with 
$d'\leq2^{-n}$ whenever $d\leq2^{-\nu(n)}$  
changes the entropy $\eta$ to $\eta'\leq\eta\circ\nu$.
%$d\leq 2^{-\nu(n)}$ implies $d'\leq 2^{-n}$;
%and $2^{\eta'(n)-1}$ balls of $d'$-radius $2^{-n}$ do not suffice to cover,
%so \eta(\nu(n)) > \eta'(n) - 1
Additionally proceeding from $e$ on $Y$ to $e'$ satisfying $e'\leq2^{-\kappa(n)}\Rightarrow e\leq2^{-n}$
turns a modulus of continuity $\mu$ of $f:X\to Y$ into $\mu'$ 
with $\mu\leq\nu\circ\mu'\circ\kappa$.
\end{enumerate}
\end{fact}

\begin{myexample}[Entropy]
\label{x:Entropy}
\begin{enumerate}
\item[a)]
The real unit interval $[0;1]$ 
has entropy $\eta_{[0;1]}(n)=n-1$ for all integers $n\geq1$.
Cantor space has entropy $\eta_{\Cantor}=\id$.
The Hilbert Cube $\Hilbert=\prod_{j\geq0}[0;1]$ with metric $\dH(\bar x,\bar y)=\sup_j |x_j-y_j|/2^j$
has entropy $\eta_{\Hilbert}(n)=\Theta(n^2)$. 
\item[b)]
Let compact $(X,d)$ and $(Y,e)$ have entropies $\eta_X$ and $\eta_Y$, respectively.
Then the entropy $\eta_{X\times Y}$ of compact $\big(X\times Y,\max\{d,e\}\big)$ satisfies
\[ \forall n: \quad \eta_X(n)+\eta_Y(n) \;\leq\; \eta_{X\times Y}(n+1)+1 \;\leq\; \eta_X(n+1)+\eta_Y(n+1)+1 \enspace . \]
\item[c)]
Let compact $(X_j,d_j)$ all have diameter $\leq1$ and entropy $\eta_j$, $j\in\IN$.
Then $\big(\prod_jX_j,\sup_j d_j/2^j\big)$ is compact and has entropy $\eta$ satisfying
\[ \forall n: \quad \sum\nolimits_{j\leq n} \eta_j(n-j) \;\leq\; 
\eta(n+1)+\lceil n/2\rceil \;\leq\; \sum\nolimits_{j\leq n} \eta_j(n+1-j) +\lceil n/2\rceil \enspace .  \]
For $X_j\equiv[0;1]$ this recovers the Hilbert Cube $\prod_j[0;2^{-j}]$.
\item[d)]
Let $X$ be a compact space with metric $d\leq1$ and entropy $\eta$.
Then $D(x, y) := 1/\big(\log_2 2/d(x,y)\big)\leq1$
constitutes a topologically equivalent metric yet
inducing entropy $H(n) = \eta(2^n - 1)$.
\item[e)] 
Fix an arbitrary non-decreasing unbounded $\varphi:\IN\to\IN$
and re-consider Cantor space, now equipped with 
$d_\varphi:(\bar a,\bar b)\mapsto 2^{-\varphi(\min\{m:a_m\neq b_m\})}\in[0;1]$.
This constitutes a metric, topologically equivalent to $\dC=d_{\id}$
but with entropy $\eta_\varphi=\loinv{\varphi}$.
\item[f)]
For $K$ a closed subset of compact $(X,d)$, 
it holds $\eta_{X,K}\leq\eta_{X,X}$
but not necessarily $\eta_K\leq\eta_X$.
The image $Z:=f[X]\subseteq Y$ has entropy
$\eta_{Z}\leq \eta_X\circ\mu$,
where $\mu$ denotes a modulus of continuity of $f$.
Every connected compact metric space $X$ has entropy at least linear $\eta(n)\geq n+\Omega(1)$.
\item[g)]
Fix a compact metric space $(X,d)$ with entropy $\eta$.
Let $\calK(X)$ denote the set of non-empty closed subsets of $X$
and equip it with the Hausdorff metric
$D(V,W)=\max\big\{\sup\{ d_V(w) :w\in W\},\sup\{d_W(v):v\in V\}\big\}$,
where  $d_V:X\ni x\mapsto\inf\{ d(x,v) : v\in V\}$ denotes the distance function.
Then $\big(\calK(X),D\big)$ constitutes a compact metric space \cite[Exercise~8.1.10]{Wei00}.
It has entropy $H\leq2^{\eta}$ with $2^{\eta(n)-1}<H(n+1)$.
%If $X$ is connected, then so is $\calK(X)$; if $X$ is convex, then so is $\calK(X)$.
\item[h)]
Fix a connected compact metric space $(X,d)$ %of diameter $\leq 1$ 
with entropy $\eta$,
and consider the convex metric space $X':=\calC(X,[0;1])$ of continuous real functions 
equipped with the supremum norm $|f|=\sup_{x\in X} |f(x)|$.
Its subset $X'_1:=\Lip_1(X,[0;1])$ of non-expansive functions $f:X\to[0;1]$
is compact by Arzel\'a-Ascoli; it has relative entropy 
$\eta'_1(n):=\eta_{X',X'_1}(n)=\Theta\big(2^{\eta(n\pm\calO(1))}\big)$; 
more precisely: $2^{\eta(n-1)-1} \;<\;
\eta'_1(n) \;\leq\;\calO\big(2^{\eta(n+2)}\big)$.
\end{enumerate}
\end{myexample}
Item~(d) yields spaces with asymptotically large entropy;
and Item~(e) does similarly for small entropy ---
of a totally disconnected space in view of Example~\ref{x:Entropy}f).
The connection between modulus of continuity and entropy
(Item~f) had been observed in \cite[Lemma~3.1.13]{SteinbergDisse}.
We emphasize that Item~h) refers to $X'_1$ as 
subset of $X'$, not as metric space of its own:
see also Question~\ref{q:Future}e) below.
According to Item~(d) of the following Lemma,
analyses of function spaces $\calC_\mu(X,[0;1])$
may indeed w.l.o.g. suppose $\mu=\id$, i.e.,
the consider the non-expansive case. %ZZZ

\begin{lemma}
\label{l:Modulus}
For non-decreasing unbounded $\mu:\IN\to\IN$ let  
\[ %\begin{equation} \label{e:Modulus}
\omega_\mu \;:\; [0;\infty) \;\ni\;
t \;\mapsto\; \inf\Big\{ \sum\limits_{j=1}^J 2^{-n_j} \::\: J,n_1,\ldots n_j\in\IN, \: t\leq \sum\limits_{j=1}^J 2^{-\mu(n_j)} \Big\}
\;\in\; [0;\infty) 
\] %\end{equation}
\begin{enumerate}
\item[a)]
It holds $\omega_\mu(0)=0$ and $\omega_\mu(t)>0$ for $t>0$.
$\omega_\mu$ is subadditive: $\omega(s+t)\leq\omega(s)+\omega(t)$.
$\omega_\mu$ has modulus of continuity $\mu$.
\item[b)]
If $\mu$ is strictly increasing, 
then $\mu(n)=\min\big\{ m\in\IN \::\: \omega_\mu(2^{-m})\leq2^{-n}\big\}$.
\item[c)]
For a compact \emph{convex} metric space $(X,d)$ and
any $x,y\in X$, there exists an isometry $\imath:[0;d(x,y)]\to X$
with $\imath(0)=x$ and $\imath\big(d(x,y)\big)=y$.
\item[d)]
If $(X,d)$ is compact convex and $\mu$ a modulus of continuity of $f:X\to\IR$,
then $|f(x)-f(x')|\leq \omega_\mu\big(d(x,x')\big)$ for all $x,x'\in X$. \\
In particular $\calC_\mu\big((X,d),\IR\big)=\Lip_1\big((X,\omega_\mu\circ d),\IR\big)$
holds for every strictly increasing $\mu$.
\end{enumerate}
\end{lemma}
Recall that a (not necessarily linear) metric space $X$ is called
\emph{convex} if, to any distinct $x,y\in X$, 
there exists a $z\in X\setminus\{x,y\}$ with $d(x,y)=d(x,z)+d(z,y)$.
Examples include compact convex subsets of Euclidean space with its inherited metric,
but also connected compact subsets when equipped with the intrinsic (=shortest-path) distance,
while Cantor space is not convex.

\begin{myexample}[Modulus of Continuity]
\label{x:Topology}
\begin{enumerate}
\item[a)]
The function $(0;1]\ni t\mapsto 1/\ln(e/t)\in(0;1]$
extends uniquely continuously to $0$
and has an exponential, but no polynomial,
modulus of continuity. 
\item[b)]
Picking up on Example~\ref{x:Entropy}c),
let $\xi_j:\subseteq\Cantor\twoheadrightarrow X_j$ have modulus of continuity $\kappa_j$
and fix some injective `pairing' function
\[ \IN\times\IN \;\ni\;  (n,m) \;\mapsto\;\langle n,m\rangle %\;:=\;(n+m)\cdot(n+m+1)/2+n\;\leq\;\calO(n^2+m^2) \enspace .\]
\;\in\; \IN \]
such as Cantor's $\langle n,m\rangle\;=\;(n+m)\cdot(n+m+1)/2+n$. Then 
\[ \prod\nolimits_j x_j:\subseteq\Cantor\;\ni\;\bar b\;\mapsto\;
\Big(\xi_j\big(b_{\langle j,0\rangle}, %b_{\langle j,1\rangle},
\ldots b_{\langle j,n\rangle},\ldots\big)_j\Big) 
\;\in\;\prod\nolimits_j X_j  \]
has modulus of continuity $n\mapsto\sup_{j<n} \langle j,\kappa_j(n-j)\rangle$.
\item[c)]
If $\xi$ is a representation of $X$ with modulus of continuity $\mu$,
then the following $2^\xi$ is a representation of $\calK(X)$ with modulus 
of continuity $m\mapsto 2^{\mu(m)+1}-1$:
$(b_0,b_1,\ldots b_m,\ldots)\in\Cantor$ is a $2^\xi$-name of $A\in\calK(X)$ iff,
for every $n\in\IN$ and every $\vec v\in\{\sdzero,\sdone\}^{<\mu(n)}$ it holds:
%\forall n\in\IN \;\; \forall \vec v\in\{\sdzero,\sdone\}^{<\mu(n)}\; : \qquad 
%\begin{equation} \label{e:Hausdorff}
%\left\{ \begin{array}{rl} 
\begin{eqnarray} \label{e:Hausdorff}
(\vec v\circ\Cantor)\cap\dom(\xi)\neq\emptyset
\;\wedge\; b_{\bin(\vec v)}=\sdone &\Rightarrow& 
\xi[\vec v\circ\Cantor]\cap\cball(A,2^{-n})\neq\emptyset  \\  \nonumber
(\vec v\circ\Cantor)\cap\dom(\xi)\neq\emptyset
\;\wedge\; b_{\bin(\vec v)}=\sdzero &\Rightarrow& 
\xi[\vec v\circ\Cantor]\cap\cball(A,2^{-n-1})=\emptyset 
%\end{array} \right. \end{equation}
\end{eqnarray}
where $\vec v\circ\Cantor \;:=\; \{\vec v\bar w: \bar w\in\Cantor\} \;\subseteq\; \Cantor$
and $\bin(v_0,\ldots v_{n-1})=v_0+2v_1+4v_2+\cdots+2^{n-1}v_n+2^n-1$
and $\cball(A,r):=\bigcup_{a\in A}\cball(a,r)$.
\end{enumerate}
\end{myexample}
Note that $\mu\leq\calO(\eta)$ implies $2^{\mu}\leq\calP(2^\eta)$:
reflected in Theorem~\ref{t:Admissible}d) and Theorem~\ref{t:Functions} below.
According to Example~\ref{x:Entropy}c),
Example~\ref{x:Topology}b) does \emph{not} preserve linear admissibility
already in case of spaces $X_j$ with quadratic entropy:
a more sophisticated construction is needed in Theorem~\ref{t:Cartesian}.

%%%%%%%%%%%%%%%%%%%%%%%%%%%%%%%%%%%%%
\subsection{Proofs}

\begin{proof}[Example~\ref{x:Topology}]
\begin{enumerate}
\item[c)]
$2^\xi$ is a representation, as $A\in\calK(X)$ can be recovered from any name $\bar b$:
On the one hand, for every $\xi$-name $\bar v$ of $a\in A$
and every $n\in\IN$, $\displaystyle b_{\bin(v_0,\ldots v_{\mu(n)-1})}=\sdone$;
on the other hand, for every $\xi$-name $\bar v$ of $a\not\in A$,
$\ball(a,2^{-n})\cap A=\emptyset$ implies
$\displaystyle b_{\bin(v_0,\ldots v_{\mu(n)-1})}=\sdzero$.
Since $\bin(v_0,\ldots v_{\mu(n)-1})<2^{\mu(n)+1}-1$,
this also establishes $2^{\mu+1}-1$ as modulus of continuity of $2^\xi$.
\qed\end{enumerate}\end{proof}

\begin{proof}[Lemma~\ref{l:Modulus}]
\begin{enumerate}
\item[a)] Regarding subadditivity, $\omega_\mu(s+t)=$
\begin{eqnarray*}
&=&
\inf \Big\{ \sum\limits_{j=1}^J 2^{-n_j} + \sum\limits_{k=1}^K 2^{-m_j} 
\::\: J,K,n_1,\ldots n_j,m_1,\ldots m_K\in\IN, \\
&& \phantom{\inf } \underbrace{\phantom{\Big\{}
\qquad\qquad\qquad s+t\leq \sum\limits_{j=1}^J 2^{-\mu(n_j)} +
\sum\limits_{k=1}^K 2^{-\mu(m_k)} \Big\}}_{\rotatebox[origin=c]{270}{$\supseteq$}} \\
&\leq&
\inf\overbrace{\Big\{ \sum\limits_{j=1}^J 2^{-n_j} +\sum\limits_{k=1}^K 2^{-m_j} 
\::\: J,K,n_1,\ldots n_j,m_1,\ldots m_K\in\IN, } \\
&& \phantom{\inf \Big\{ } \qquad s\leq \sum\nolimits_{j=1}^J 2^{-\mu(n_j)} \;\wedge\;
t\leq \sum\nolimits_{k=1}^K 2^{-\mu(m_k)} \Big\} \\
&=& \omega_\mu(s)\;+\; \omega_\mu(t)
\end{eqnarray*}
By definition ($J:=1$) it holds $0\leq\omega_\mu(t)\leq 2^{-n}$ for $t\leq2^{-n}$,
and in particular $\omega_\mu(0)=0$. By subadditivity
and whenever $0\leq\delta\leq2^{-\mu(n)}$, we have both
$\omega_\mu(t)\leq\omega_\mu(t+\delta)\leq\omega_\mu(t)+\omega_\mu(\delta)\leq\omega_\mu(t)+2^{-n}$ and 
$\omega_\mu(t)-2^{-n}\leq\omega_\mu(t)-\omega_\mu(\delta)=\omega_\mu(t-\delta+\delta)-\omega_\mu(\delta)
\leq\omega_\mu(t-\delta)\leq\omega_\mu(t)$.
\item[b)]
For every $t\leq2^{-\mu(n)}$ it holds $\omega_\mu(t)\leq2^{-n}$ by definition, and hence 
$\tilde\mu(n):=\min\big\{ m\in\IN \::\: \omega_\mu(2^{-m})\leq2^{-n}\big\}\leq \mu(n)$.
Conversely for $m\leq n_1,\ldots,n_j\in\IN$,
\[ \sum\nolimits_j 2^{-\mu(n_j)} \;\leq\;
\sum\nolimits_j 2^{-\mu(m)-n_j+m} 
\;=\; 2^{-\mu(m)}\cdot 2^{m}\cdot\sum\nolimits_j 2^{-n_j} \]
from strict monotonicity $\mu(n_j)=\mu(m+n_j-m)\geq\mu(m)+(n_j-m)$ by induction.
So $\sum_j 2^{-\mu(n_j)}\geq 2^{-\mu(m)}$ implies $\sum_j 2^{-n_j}\geq2^{-m}$
and $\omega_\mu\big(2^{-\mu(m)}\big)\geq2^{-m}$ 
and $\tilde\mu(n)\geq\mu(n)$.
\item[c)]
Using transfinite induction and completeness,
\cite[Exercise~5.1.17]{Kaplansky}
constructs a $z\in X$ with $d(x,z)=d(z,y)=d(x,y)/2$.
Now iterating with both $(x,z)$ and $(z,y)$ in place of $(x,y)$
yields a sequence of refinements $z_n\in X$, $n=0,\ldots,N=2^k$
$z_0=x$ and $z_N=y$ and $d(z_n,z_{n+1})=d(x,y)/N$.
Again by completeness,
$\imath_k(t):=z_{\min\{n: n/2^k\geq t\}}$ thus converges
uniformly to the claimed isometry.
\item[d)]
For $t:=d(x,x')$, and to any $J\in\IN$ and $n_1,\ldots,n_j\in\IN$
with $t\leq \sum\nolimits_{j=1}^J 2^{-\mu(n_j)}$,
c) yields $x=:x_0,x_1,\ldots,x_J=x'\in X$ with $d\big(x_{j-1},x_j\big)\leq2^{-\mu(n_j)}$.
Hence
\[ %\begin{eqnarray*}
\big|f(x)-f(x')\big|
\;\leq\; \sum\nolimits_{j=1}^J \big|f(x_{j-1})-f(x_j)\big| 
\;\leq\; \sum\nolimits_{j=1}^J 2^{-n_j} \;\leq\; \omega_\mu(t) 
\] %\end{eqnarray*}
\qed\end{enumerate}\end{proof}

\begin{fact}
\label{f:Extension}
Fix a compact metric space $(X,d)$,
%\begin{enumerate}\item[a)]
non-empty $Z\subseteq X$ and $L>0$. 
For $L$-Lipschitz $f:Z\to\IR$, the functions
\begin{equation}
\label{e:Extension}
\extl{f}:x\;\mapsto\; \sup\nolimits_{z\in Z}  f(z)-L\cdot d(z,x),
\quad \exth{f}:x\;\mapsto\;\inf\nolimits_{z\in Z}  f(z)+L\cdot d(z,x)
\end{equation}
extend $f$ to $X$ while preserving $L$-Lipschitz continuity.
Moreover every $L$-Lipschitz extension $\tilde f:X\to\IR$ of $f$ to $X$ 
satisfies $\extl{f}\leq\tilde f\leq\exth{f}$,
where $\exth{f}-\extl{f}\leq 2L |d_Z|=2L\sup_x d_Z(x)$.
The extension operator 
$\Lip_L(Z,\IR)\ni f\mapsto \extm{f}:=(\extl{f}+\exth{f})/2\in\Lip_L(X,\IR)$
is a well-defined isometry of compact metric spaces w.r.t. the supremum norm.
%\item[b)]
%Let $\lfloor y\rceil$ denote the integer closest to $y\in\IR\setminus(\IZ+\tfrac{1}{2})$
%with tie broken downwards $\lfloor y+\tfrac{1}{2}\rceil:=y$ for $y\in\IZ$. %well-defined!
%For $I,J\subseteq\IR$ (possibly unbounded) closed intervals with integer endpoints (or infinity)
%and $f:I\to J$ a (possibly partial) 1-Lipschitz function, 
%its rounded restriction $g:=\lfloor f|_\IZ\rceil:(I\cap \IZ)\to (J\cap\IZ)$ 
%is still 1-Lipschitz;
%and its piecewise linear 1-Lipschitz extension 
%$\extm{g}:[\min(I);\max(I)]\subseteq\IR \to [\min(J);\max(J)]$
%is well-defined and satisfies $\big|f(t)-\extm{g}(t)\big|\leq1$ for all $t\in[\min(I);\max(I)]$.
%\item[c)] 
%If $X$ is convex, then to any $x,y\in X$
%there exists an isometry $\imath:[0;d(x,y)]\to X$
%with $\imath(0)=x$ and $\imath\big(d(x,y)\big)=y$.
%\end{enumerate}
\end{fact}
$\big(\extl{f},\exth{f}\big)$ 
is known as \emph{McShane-Whitney pair} \cite{Petrakis}.
For the purpose of self-containment, we include a proof:

\begin{proof}[Fact~\ref{f:Extension}]
%\begin{enumerate}\item[a)]
W.l.o.g. $L=1$.
For $x\in Z$, choosing $z:=x$ shows $\extl{f}(x)\geq f(x)$ 
while $f(z)-d(z,x)\geq f(z)-|f(z)-f(x)|\geq f(x)$
implies $\extl{f}(x)\leq f(x)$.
Moreover, for every $z\in Z$, we have
\[ -\sup\big\{f(z')-d(z',x'):z'\in Z\big\}
\;\leq\; -f(z)+d(z,x') \;\leq\; -f(z)+d(z,x)+d(x,x') \]
and hence %$\extl{f}(x)-\extl{f}(x') \leq$
\[ \extl{f}(x)-\extl{f}(x') \;\leq\; 
\sup\big\{ f(z)+d(z,x) - f(z)+d(z,x)+d(x,x'): z\in Z\big\} 
\;=\; d(x,x') . \]
The estimates for $\exth{f}$ proceed similarly.
%Dually, for every $z\in Z$ and $x,x'\in X$ it holds
%\begin{eqnarray*}
%\exth{f}(x')-\exth{f}(x) &=&
%\underbrace{\inf\big\{f(z')+d(z',x'):z'\in Z\big\}}_{
%\leq f(z)+\omega(d(z,x'))}
%\underbrace{-\inf\big\{f(z)+d(z,x):z\in Z\big\}}_{
%=\sup\{-f(z)-d(z,x):z\in Z\}} \\
%&\leq& 
%\sup\big\{ f(z)+d(z,x)+d(x,x')
%-f(z)-d(z,x):z\in Z\big\} \;=\; d(x,x')
%\end{eqnarray*}
For $x\in X$ and with $z,z',w,w'$ ranging over $Z$, 
$\displaystyle
\big(\extl{f}(x)+\exth{f}(x)\big)-\big(\extl{g}(x)+\exth{g}(x)\big) \;=$
\begin{eqnarray*}
&=&\sup\limits_z f(z)\!-\!d(z,x)
\:-\: \sup\limits_w g(w)\!-\!d(w,x) 
\:-\: \inf\limits_{w'} g(w')\!+\!d(w',x)
\:+\: \inf\limits_{z'} f(z')\!+\!d(z',x)
\\
&=& \sup_z f(z)\!-\!d(z,x) 
\:+ \inf_w -g(w)+d(w,x)
\:+ \sup_{w'} -g(w')-d(w',x) 
\:+ \inf_{z'} f(z')+d(z',x) \\
&\leq& \sup\limits_z f(z)-d(z,x) 
\:-\: g(z)+d(z,x) 
\:+\: \sup\limits_{w'} -g(w')-d(w',x) 
\:+\: f(w')+d(w',x) \\
&=& 2\cdot \sup_z f(z)-g(z)\;\leq\; 2\cdot\sup\nolimits_z |g(z)-f(z)| \enspace .
\qed \end{eqnarray*}
%\item[b)]
%Let us first record that
%\begin{equation} \label{e:Rounding}
%|y-y'|\leq k\in\IN \quad\Rightarrow\quad
%\big|\lfloor y\rceil - \lfloor y'\rceil\big|, \;
%\big|\lfloor y\rfloor - \lfloor y'\rfloor\big|, \;
%\big|\lceil y\rceil - \lceil y'\rceil\big| \;
%\leq\; k \enspace . \end{equation}
%Now consider w.l.o.g. $I=[0;1]$ and the two cases
%\[ f(0)\in(-1/2;1/2] \;\wedge\;f(1)\in(-1/2;1/2] , \quad
%f(0)\in(-1/2;1/2] \;\wedge\;f(1)\in(1/2;3/2] \]
%since all others follow by symmetry.
%In the first case $g(0)=0=g(1)$, hence $\extm{g}(t)\equiv0$ for $0\leq t\leq 1$;
%while $|f(t)|\leq1$ for $t\in[0;1/2]$
%as well as for $t\in[1/2;1]$ since $f$ is 1-Lipschitz, so the claim holds.
%In the second case $g(0)=1$ and $g(1)=1$, hence 
%$\extm{g}(t)=x$ for $0\leq t\leq 1$;
%and 1-Lipschitz similarly implies $|f(t)-\extm{g}(t)|\leq1$.
%\item[c)]
%Using transfinite induction and metric completeness,
%\mycite{Exercise~5.1.17}{Kaplansky}
%constructs a $u\in X$ with $d(x,u)=d(u,y)=d(x,y)/2$.
%Now iterating with both $(x,u)$ and $(u,y)$ in place of $(x,y)$
%yields a sequence of refinements $u_n\in X$, $n=0,\ldots N=2^k$
%$u_0=x$ and $u_N=y$ and $d(u_n,u_{n+1})=d(x,y)/N$.
%Again by metric completeness,
%$\imath_k(t):=u_{\min\{n: n/2^k\geq t\}}$ thus converges
%uniformly to the claimed isometry.
%\qed\end{enumerate}\end{proof}
\end{proof}

\begin{proof}[Example~\ref{x:Entropy}]
Let $\calH_X(n)$ denote the least number of closed balls of radius $2^{-n}$ covering $X$,
so that $\eta_X(n)=\lceil\log_2\calH(n)\rceil$.
Let $\calC_X(n)$ denote the largest number of points in $X$
of pairwise distance $>2^{-n}$, also known as \emph{capacity}.
(Since $X$ is not an integer function, there is not danger of 
confusion this notation with that of a space of continuous functions\ldots)
Then $\calC_X(n)\leq\calH_X(n+1)$:
To cover $X$ requires covering the $\calC_X(n)$ points as above;
but any closed ball of radius $2^{-(n+1)}$ can contain at most one 
of those points having distance $>2^{-n}$.
On the other hand $\calH_X(n)\leq\calC_X(n)$,
since balls of radius $2^{-n}$ whose centers form a maximal set $X_n$ of pairwise distance $>2^{-n}$
cover $X$: if they missed a point, that had distance $>2^{-n}$ to all centers in $X_n$
and thus could be added to $X_n$: contradicting its maximality.
\begin{enumerate}
\item[a)]
Cover $[0;1]$ by $2^{n-1}$ closed intervals $I_{n,j}:=\big[j\cdot2^{-(n-1)};(j+1)\cdot2^{-(n-1)}\big]$,
$j=0,\ldots 2^{n-1}-1$, of radius $2^{-n}$ around centers $(2j+1)2^{-n}$: optimally. \\
Cover $\Cantor$ by $2^n$ closed balls $\vec x\circ\Cantor$
of radius $2^{-n}$ around centers $\vec x\in\{\sdzero,\sdone\}^n$: optimally. 
Cover $\Hilbert$ by $2^{n-1}\cdot2^{n-2}\cdots 2\cdot 1=2^{n(n+1)/2}$ closed balls
\[ I_{n,j_0}\times I_{n-1,j_1} \times \cdots \times I_{1,j_{n-1}} \times \prod\nolimits_{j\geq n} [0;1] \]
of radius $2^{-n}$ with indices ranging as follows:
$0\leq j_0<2^{n-1}, \quad 0\leq j_1<2^{n-2}, \quad \cdots \quad j_{n-1}=0$.
\item[b)]
Obviously $\calH_{X\times Y}\leq \calH_X\cdot \calH_Y$
and $\calC_{X\times Y}\geq\calC_X\cdot\calC_Y$:
$\calH_X(n)\cdot\calH_Y(n) \leq$
\[ \leq\; \calC_X(n)\cdot\calC_Y(n) \;\leq\; \calC_{X\times Y}(n)
\;\leq\; \calH_{X\times Y}(n+1) \;\leq\; \calH_X(n+1)\cdot\calH_Y(n+1) \]
%\;\leq\; \calC_X(n+1)\cdot\calC_Y(n+1) \;\leq\; \calC_{X\times Y}(n+1) \]
Also record $\lceil s\rceil+\lceil t\rceil \geq \lceil s+t\rceil\geq \lceil s\rceil+\lceil t\rceil-1$
for all $s,t>0$.
%\[ \forall n: \quad \eta_X(n)+\eta_Y(n) \;\leq\; \eta_{X\times Y}(n+1)+1 \;\leq\; \eta_X(n+1)+\eta_Y(n+1)+1 \enspace . \]
\item[c)]
Abbreviating $\calH_j:=\calH_{X_j}$ and $\calC_j:=\calC_{X_j}$,
we have $\calH(n)\leq \prod_{j<n} \calH_j(n-j)$ 
and $\calC(n)\geq \prod_{j<n} \calC_j(n-j)$:
note that $\calH_j(0)=1=\calC_j(0)$ as $X_j$ has diameter $\leq1$. 
Finally $\sum_{j=0}^{n-1} \lceil t_j\rceil \geq \lceil\sum_{j=0}^{n-1} t_j\rceil \geq \sum_{j=0}^{n-1} \lceil t_j\rceil -\lfloor n/2\rfloor$.
\item[f)]
For a counterexample to $\eta_K\leq\eta_X$
consider a circle/hyper-/sphere with and without center.
\\
Regarding the lower bound for connected compact metric spaces,
consider $N:=2^{\eta(n)}$
and $x_1,\ldots x_N\in X$ such that balls with centers $x_j$ and radius $2^{-n}$ cover $X$:
$X\subseteq\bigcup_{n=1}^N \cball(x_n,2^{-n})$.
Consider the finite undirected graph $G_n=(V_n,E_n)$
with vertices $V_n=\{1,\ldots N\}$ and edges
$\{i,j\}\in E_n\Leftrightarrow\ball(x_i,2^{-n+1})\cap\ball(x_j,2^{-n+1})\neq\emptyset$
whenever the two open balls with centers $x_i,x_j$ and radius twice $2^{-n}$ intersect.
This graph is connected: If $I,J\subseteq V_n$ were distinct connected components,
then $\bigcup_{n\in I}\ball(x_n,2^{-n+1})$ and $\bigcup_{n\not\in I}\ball(x_n,2^{-n+1})$
were two disjoint open subsets covering $X$.
Therefore any two vertices $i,j\in V_n$ are connected via $\leq N-1$ edges;
and for every edge $\{a,b\}$, it holds $d(x_a,x_b)<2^{-n+2}$ by definition of $E_n$:
Hence $x_i$ and $x_j$ have metric distance $d(x_i,x_j)$ at most $(N-1)\cdot 2^{-n+2}$;
and any two $x,y\in X$ have $d(x,y)\leq N\cdot2^{-n+2}$:
requiring $2^{\eta(n)}=N\geq d(x,y)\cdot 2^{n-2}$.
\item[g)]
Obviously $\calH_{\calK(X)}\leq2^{\calH_X}$
and $\calC_{\calK(X)}\geq2^{\calC_X}$.
\item[h)]
Fix $n\in\IN$ and consider a maximal set
$X_n\subseteq X$ of $N:=\calC_X(n)$ points of pairwise distance $>2^{-n}$.
There are $2^{\calC_X(n)}$ different $f:X_n\to\{0,2^{-n}\}$;
each is 1-Lipschitz, and extends to $\extm{f}:X\to[0;1]$;
and, according to Fact~\ref{f:Extension},
different such $f$ give rise to $\extm{f}$ of 
mutual supremum distance $\geq2^{-n}$:
Hence $\calC_{X'_1}(n)\geq 2^{\calC_X(n)}$,
and $X'_1\subseteq X'$ 
has intrinsic entropy $\eta'_1(n)\geq
\log_2 \calC_{X'_1}(n-1)
\geq \calC_X(n-1)\geq \calH_X(n-1)>2^{\eta(n-1)-1}$.
\\
Conversely, for any 1-Lipschitz $f:X\to[0;1]$, 
consider $f'_n:=\lfloor 2^n\cdot f\big|_{X_n}\rceil/2^n$:
still $(1+1/2)$-Lipschitz since rounding affects the value by at most $2^{-n-1}$
on arguments of distance $>2^{-n}$.
As argued before, maximality of $X_n$ implies
that the closed balls around centers $x\in X_n$
of radius $2^{-n}$ cover $X$ (hence $d_{X_n}\leq 2^{-n}$);
consequently so do the open balls with radius $2^{-n+1}$.
Similarly to the proof of (f),
consider the finite undirected and connected graph $G_n=(X_n,E_n)$ with edge 
$\{x,y\}\in E_n\;:\Leftrightarrow\;\ball(x,2^{-n+1})\cap\ball(y,2^{-n+1})\neq\emptyset$.
%present iff the two open balls with centers $x,y$ and radius twice $2^{-n}$ intersect.
Any vertex $y$ of $G_n$ adjacent to some $x$ has distance $d(x,y)<2^{-n+2}$;
and since $f'_n:X_n\to\ID_n$ is $\tfrac{3}{2}$-Lipschitz,
this implies $\ID_n\ni|f'_n(x)-f'_n(y)|\leq \tfrac{3}{2}\cdot2^{-n+2}$
leaving no more than 13 possible values for $f'_n(x)-f'_n(y)\in\big\{-6\cdot2^{-n},\ldots 0,\ldots +6\cdot2^{-n}\big\}$.
Connectedness of $G_n$ with $N$ vertices thus limits
the number of different $\tfrac{3}{2}$-Lipschitz $f'_n:X_n\to\ID_n$
to $\leq(1+2^n)\cdot 13^{N-1}\leq 2^{\calO(\calC_X(n))}$
in view of (f). And by Fact~\ref{f:Extension}
each such $f'_n$ extends to some $\tfrac{3}{2}$-Lipschitz $\extm{f'_n}:X\to[0;1]$.
Moreover $|d_{X_n}|\leq 2^{-n}$ implies
$\big|f-\extm{f_n}\big|\leq \big|\exth{f_n}-\extl{f_n}\big|/2 \leq \tfrac{3}{2}\cdot 2^{-n}$
for the $\tfrac{3}{2}$-Lipschitz (!) extension of the restriction $f_n:=f\big|_{X_n}$.
Since $g\mapsto\extm{g}$ is an isometry,
this implies $\big|f-\extm{f'_n}\big|
\leq \big|f-\extm{f_n}\big|+\big|f_n-f'_n\big|
\leq \tfrac{3}{2}\cdot2^{-n}+2^{-n-1}=2^{-n+1}$.
The $2^{\calO(\calC_X(n))}$ closed balls of radius $2^{-n+1}$ 
around centers $\extm{f'_n}\in\Lip_{3/2}(X,[0;1])\subseteq X'$ thus cover $\Lip_1(X,[0;1])=X'_1$:
$\eta'_1(n-1)=\eta_{X',X'_1}(n-1)\leq \calO\big(\calC_X(n)\big) \leq \calO\big(\calH_X(n+1)\big) \leq \calO\big(2^{\eta(n+1)}\big)$.
\qed\end{enumerate}\end{proof}

%%%%%%%%%%%%%%%%%%%%%%%%%%%%%%%%%%%%%%%%%
\section{Concise Standard Representations}
\label{s:Standard}

\cite[Definitions~3.2.2+3.2.7]{Wei00} introduce qualitative 
admissibility in terms of a \emph{standard} representation
which \cite[Lemma~3.2.5]{Wei00} then shows to satisfy properties (i) and (ii) from Fact~\ref{f:Main}.
Here we first recall from \cite[Definition~15]{DBLP:conf/lics/KawamuraS016}
the construction of a \emph{concise} standard representation 
of any fixed compact metric space $(X,d)$
that generalizes Example~\ref{x:Dyadic}:
For each $n$, fix a covering of $X$ by $\leq 2^{\eta(n)}$ balls of radius $2^{-n}$ according to the entropy;
assign to each ball a binary string $\vec a_n$ of length $\eta(n)$; 
then every $x\in X$ can be approximated by the center of some of these balls;
finally define a name of $x$ to be such a sequence $(\vec a_n)_{_n}$ of binary strings.
Theorem~\ref{t:Polynomial} establishes that this representation is polynomially admissible,
provided the balls' radius is reduced to $2^{-n-1}$.
Subsection~\ref{ss:Linear} improves the construction to yield a linearly admissible standard representation.

\begin{definition}
\label{d:Standard}
Let $(X,d)$ denote a compact metric space 
%of diameter $\diam(X)=\max\{d(x,x'):x,x'\in X\}\geq1$
with entropy $\eta$.
For each $n\in\IN$ fix some partial mapping 
$\xi_n:\subseteq\{\sdzero,\sdone\}^{\eta(n+1)}\to X$ 
such that $X=\bigcup_{\vec a\in\dom(\xi_n)} \cball\big(\xi_n(\vec a),2^{-n\pmb{-1}}\big)$,
where $\cball(x,r)=\{x'\in X:d(x,x')\leq r\}$ 
denotes the closed ball around $x$ of radius $r$.
The \emph{standard} representation of $X$ 
(with respect to the family $\xi_n$ of partial dense enumerations)
is the mapping
\begin{gather}
\label{e:Dyadic}
\xi \; :\subseteq \Cantor \;\ni\;
\big( (\vec a_0) \:  (\vec a_1) \:
\ldots
(\vec a_n) \:
\ldots \big)
\;\mapsto\;
\lim\nolimits_n \xi_n(\vec a_n) \;\in\; X ,  
\\[0.5ex]  \nonumber
\dom(\xi) :=\:  \big\{  \big(  \:\ldots\:
(\vec a_n) \: \ldots\big) \::\: \vec a_n\in\dom(\xi_n), \;
d\big(\xi_n(\vec a_n),\xi_m(\vec a_m)\big)\:\leq\: 2^{-n}\!+\!2^{-m}\big\}
\end{gather}
\end{definition}
Fact~\ref{f:Topology}a) asserts such $\xi_n$ to exist.
The real Example~\ref{x:Dyadic} is\footnote{%
Well, almost: $\bin(a_n)$ has length between $1$ and $2n+1$ while
here we make all strings in $\dom(\xi_n)$ have the same length $=\eta(n+1)$:
Using strings of varying length $<\eta(n+1)$ would additionally require encoding delimiters.}
a special case of this definition
with $\eta_{[0;1]}(n+1)=n$ according to Example~\ref{x:Entropy}a) and
\[ \delta'_n \;:\; \{\sdzero,\sdone\}^n \;\ni\; \vec a\;\mapsto\; 
 \big(\tfrac{1}{2}+a_0+2a_1+4a_2+\cdots+2^{n-1}a_{n-1}\big)/2^n %\;\in\; \ID_{n}+2^{-n-1} 
\enspace . \]
The covering balls' radius being $2^{-n\pmb{-1}}$ instead of $2^{-n}$ is exploited in the following theorem:

\begin{theorem}
\label{t:Polynomial}
\begin{enumerate}
\item[i)]
The standard representation $\xi$ of $(X,d)$ w.r.t. $(\xi_n)$ according to Definition~\ref{d:Standard} 
has modulus of continuity $\kappa(n):=\sum_{m=0}^{n} \eta(m+1)$.
\item[ii)]
To every partial function $\zeta:\subseteq\Cantor\to X$
with modulus of continuity $\nu$ 
there exists a mapping $F:\dom(\zeta)\to\dom(\xi)$ 
with modulus of continuity $\mu=\nu\circ\big(1+\loinv{\kappa}\big):\kappa(n)\mapsto\nu(n+1)$ 
such that $\zeta=\xi\circ F$ holds. \\
In particular $\xi$ is polynomially admissible, 
provided that the entropy grows at least with some positive power
$\eta(n)\geq\Omega(n^{\epsilon})$, $\epsilon>0$.
\item[iii)]
To every $m\in\IN$ and every $x,x'\in X$ with $d(x,x')\leq2^{-m-1}$,
there exist $\xi$-names $\bar y_x$ and $\bar y'_{x'}$ 
of $x=\xi(\bar y_x)$ and $x'=\xi(\bar y'_{x'})$
with $\dC(\bar y_x,\bar y'_{x'})\leq2^{-\kappa(m)}$.
\item[iv)]
If $(Y,e)$ is a compact metric space and 
$f:X\to Y$ such that $f\circ\xi:\dom(\xi)\subseteq\Cantor\to Y$
has modulus of continuity $\kappa\circ\nu$,
then $f$ has modulus of continuity $\nu+1$.
\end{enumerate}
\end{theorem}
Again (ii) strengthens Definition~\ref{d:Admissible}c)
in applying to not necessarily surjective $\zeta$.
In view of Lemma~\ref{l:seminv}c), (iv) can be rephrased as follows:
$f\circ\xi$ with modulus of continuity $\nu$
implies $f$ to have modulus of continuity $1+\loinv{\kappa}\circ\nu$.
However (ii) is \emph{not} saying that
$\zeta$ with modulus of continuity $\nu\circ\mu$
yields $F$ with modulus of continuity $n\mapsto\nu(n+1)$.

\begin{proof}[Theorem~\ref{t:Polynomial}]
\begin{enumerate}
\item[i)]
First observe that $\xi$ is well-defined:
as compact metric space, $X$ is complete 
and the dyadic sequence $\xi_n(\vec a_n)\in X$ therefore converges.
Moreover $\xi$ is surjective:
To every $x\in X$ and $n\in\IN$ there exists by hypothesis some
$\vec a_n\in\dom(\xi_n)$ with $x\in\cball\big(\xi_n(\vec a_n),2^{-n-1}\big)$;
hence $d\big(\xi_n(\vec a_n),\xi_m(\vec a_m)\big)\leq 2^{-n}+2^{-m}$ 
and $\lim_n\xi_n(\vec a_n)=x$.
Furthermore, $\vec a_n$ has binary length $\eta(n+1)$; hence 
$\big(\vec a_0 \: \ldots  \vec a_{n}\big)$
has length $\kappa(n)$ as above; and fixing this initial segment of a $\xi$-name $\bar x$
implies $d\big(\xi(\bar x),\xi_n(\vec a_n)\big)\leq2^{-n}$ by Equation~(\ref{e:Dyadic}).
\item[ii)]
To every $n$ and each (of the finitely many) $\bar x|_{<\nu(n+1)}$ with $\bar x\in\dom(\zeta)$,
fix some $\vec a_n=\vec a_n\big(\bar x|_{<\nu(n+1)}\big)\in\dom(\xi_n)$ such that
\[ \xi_n(\vec a_n)\;\in\;\zeta\big[\bar x|_{<\nu(n+1)}\circ\Cantor\big]\;\subseteq\;\bigcup\nolimits_{\vec a\in\dom(\xi_n)}
\cball\big(\xi_n(\vec a),2^{-n-1}\big) \enspace . \]
Then iteratively for $n=0,1,\ldots$ let
similarly to the proof of Example~\ref{x:Dyadic},
\[ F_{n}\big(\bar x|_{<\nu(n+1)}\big) \;:=\; F_{n-1}\big(\bar x|_{<\nu(n)}\big) \:\circ\: (\vec a_n)
\;\in\; \{\sdzero,\sdone\}^{\kappa(n-1)+\eta(n+1)=\kappa(n)} \enspace . \]
This makes $F(\bar x):=\lim_n F_n\big(\bar x|_{<\nu(n+1)}\big)\in\Cantor$
well-defined with modulus of continuity $\kappa:\kappa(n)\mapsto\nu(n+1)$.
Moreover it holds $F(\bar x)\in\dom(\xi)$ and $\xi\big(F(\bar x)\big)=\zeta(\bar x)$
since $\xi_n(\vec a_n),\xi(\bar x)\in\zeta\big[\bar x|_{<\nu(n+1)}\circ\Cantor\big]\subseteq\cball\big(\xi_n(\vec a_n),2^{-n}\big)$
because $\nu$ is a modulus of continuity of $\zeta$.
\\
Finally observe $\eta(n+1)\leq\kappa(n)\leq(n+1)\cdot\eta(n+1)\in\poly(\eta)$;
hence (i) and (ii) of Definition~\ref{d:Admissible}c) hold.
\item[iii)]
To $x\in X$ consider the $\xi$-name $\bar y_x:=\big( \ldots \: (\vec a_m) \: \ldots \big)$ of $x$
with $d\big(\xi_m(\vec a_m),x\big)\leq2^{-m-1}$.
For $m\in\IN$ its initial segment $\big( (\vec a_0)\: \ldots \: (\vec a_m)\big)$
has binary length $\kappa(m)$; 
and, for every $x'\in X$ with $d(x,x')\leq2^{-m-1}$,
can be extended to a $\xi$-name $\bar y'_{x'}$.
\item[iv)]
Applying (iii) to $m:=\nu(n)$, the hypothesis implies
$e\big(f(x),f(x')\big)=e\big(f\circ\xi(\bar y_x),f\circ\xi(\bar y'_{x'})\big)\leq 2^{-n}$.
\qed\end{enumerate}\end{proof}
%

%%%%%%%%%%%%%%%%%%%%%%%%%%%%%%%%%%%%%%%%%
\subsection{Improvement to Linear Admissibility}
\label{ss:Linear}

The generic representation $\xi$ 
of a compact metric space $(X,d)$ according to Definition~\ref{d:Standard}
being `only' polynomially admissible,
this subsection improves the construction to achieve linear admissibility.
Note that $\kappa(n)=\sum_{m=0}^{n} \eta(m+1)$ according to Theorem~\ref{t:Polynomial}a)
already is in $\calO\big(\eta(m+1)\big)$ whenever $\eta(m)\geq 2^{\Omega(m)}$ grows at least exponentially;
hence we focus on spaces with sub-exponential entropy.
To this end fix some unbounded non-decreasing $\varphi:\IN\to\IN$
and define a representation $\xi^\varphi$ of $X$ 
(with respect to the family $\xi_n$ of partial dense enumerations)
based on the \emph{sub}sequence $\xi_{\varphi(n)}$ of $\xi_n$:
\begin{gather}
\label{e:Dyadic2}
\xi^\varphi \; :\subseteq \Cantor \;\ni\;
\big( \vec a_0 \:
\ldots
\vec a_n \:
\ldots \big)
\;\mapsto\;
\lim\nolimits_n \xi_{\varphi(n)}\big(\vec a_n\big) \;\in\; X , 
 \qquad \dom(\xi^\varphi):=  \\[0.5ex]  
\big\{  \big( \vec a_0\: \ldots\:
\vec a_n\:\ldots\big) \::\: \vec a_n\in\dom\big(\xi_{\varphi(n)}\big), \;
d\big(\xi_{\varphi(n)}(\vec a_n),\xi_{\varphi(m)}(\vec a_m)\big)\leq 2^{-n}\!+\!2^{-m}\big\}
\nonumber \end{gather}
Intuitively, proceeding to a subsequence $\xi_{\varphi(n)}$ 
amounts to `skipping' intermediate precisions/error bounds 
and `jumping' directly from $2^{-\varphi(n-1)}$ to $2^{-\varphi(n)}$.
It formalizes a strategy implemented for instance by the \texttt{iRRAM C++} 
library for Exact Real Computation \cite{Mue01} which %(overridable by command-line parameters) 
starts with $\varphi(0)=50$ bits \texttt{double} precision and 
in phases $n=\#1,\#2,\ldots$
increases to $\varphi(n)=\lfloor \tfrac{6}{5}\cdot \varphi(n-1)\rceil+20$.

The proof of Theorem~\ref{t:Polynomial} carries over literally to see:
\begin{enumerate}
\item[i)]
Representation $\xi^\varphi$ 
has modulus of continuity 
\[ \kappa^\varphi(n)\;:=\;\sum_{m=0}^{\loinv{\varphi}(n)} \eta\big(\varphi(m)+1\big) \]
%\overset{(*)}{\geq\;} \max\big\{\eta(n+1),\loinv{\varphi}(n)\big\} \]
%with (*) since $\eta(2)\geq1$ by our hypothesis of $\diam(X)\geq1$.
\item[ii)]
To every partial function $\zeta:\subseteq\Cantor\to X$
with modulus of continuity $\nu$
there exists a mapping $F:\dom(\zeta)\to\dom(\xi^\varphi)$
with modulus of continuity $\nu\circ\big(1+\loinv{\kappa^\varphi}\big)$
such that $\zeta=\xi^\varphi\circ F$ holds.
%In particular $\xi^\varphi$ is linearly admissible.
\item[iii)]
To every $m\in\IN$ and every $x,x'\in X$ with $d(x,x')\leq2^{-m-1}$,
there exist $\xi^\varphi$-names $\bar y_x$ and $\bar y'_{x'}$ 
of $x=\xi^\varphi(\bar y_x)$ and $x'=\xi^\varphi(\bar y'_{x'})$
with $\dC(\bar y_x,\bar y'_{x'})\leq2^{-\kappa^\varphi(m)}$.
\item[iv)]
If $(Y,e)$ is a compact metric space and 
$f:X\to Y$ such that $f\circ\xi^\varphi:\dom(\xi)\subseteq\Cantor\to Y$
has modulus of continuity $\kappa^\varphi\circ\nu$,
then $f$ has modulus of continuity $\nu+1$.
\end{enumerate}
\begin{theorem}
\label{t:Linear}
Let $(X,d)$ denote a compact metric space of entropy $\eta$,
equipped with partial mappings $\xi_n:\subseteq\{\sdzero,\sdone\}^{\eta(n+1)}\to X$ 
such that $X=\!\!\!\!\!\bigcup\limits_{\vec a\in\dom(\xi_n)} \!\!\!\!\!\cball\big(\xi_n(\vec a),2^{-n-1}\big)$.
There exists an unbounded non-decreasing $\varphi:\IN\to\IN$
such that the representation $\xi^\varphi$ from Equation~(\ref{e:Dyadic2})
has modulus of continuity $\kappa^\varphi(n)\leq\tfrac{27}{4}\cdot\eta(n+1)$
and $\kappa^\varphi(n)\geq \eta(n+1)$.
In particular $\xi^\varphi$ is linearly admissible.
\end{theorem}
The proof of Theorem~\ref{t:Linear} follows immediately from Item~d) of the following lemma,
applied to $c:=3/2$ with $\eta(n+1)$ in place of $\eta(n)$.

\begin{lemma}
\label{l:Donghyun}
Let $\eta:\IN\to\IN$ be unbounded and non-decreasing and fix $c>1$.
Then there exists a strictly increasing mapping $\varphi:\IN\to\IN$ such that it holds
\begin{enumerate}
\item[a)]
$\displaystyle\forall m\in\IN: \quad \eta\big(\varphi(m+1)\big)\;\leq\; c^2\cdot \eta\big(\varphi(m)+1\big)$.
\item[b)]
$\displaystyle\forall m\in\IN: \quad c\cdot\eta\big(\varphi(m)\big)\;\leq\; \eta\big(\varphi(m+1)\big)$.
\item[c)]
$\displaystyle\forall m\in\IN: \quad c\cdot\eta\big(\varphi(m)+1\big)\;\leq\; \eta\big(\varphi(m+1)+1\big)$.
\item[d)]
$\displaystyle \sum\nolimits_{m=0}^{\loinv{\varphi}(n)} \eta\big(\varphi(m)\big) \;\leq\; \tfrac{c^3}{c-1}\cdot\eta(n)$.
\end{enumerate}
\end{lemma}
%
%Regarding Lemma~\ref{l:Donghyun},
Think of an infinite roll of toilet papers with numbers $\eta(0)$, $\eta(1)$, \ldots printed on them.
We shall cut this roll into appropriate \emph{runs} from sheet $\#\varphi(m)$ to $\#\varphi(m+1)-1$.
Item~a) asserts that integers on sheets within the same run differ by no more than factor $c^2$.
Items~b) and c) formalizes that labels on consecutive runs grow at least exponentially.

%Note that the modulus of continuity $\mu^\varphi$ according to Theorem~\ref{t:Linear} and Lemma~\ref{l:Donghyun} 
%`jumps' %YYY can achieve linear differences

\begin{proof}[Lemma~\ref{l:Donghyun}]
	We will construct an infinite subset of $\N$ by picking elements one by one. Its elements in increasing order will then constitute the sequence $\varphi$.
	
	First, in case there exists $x \in \N$ such that $\varphi(x) = 0$, pick the largest such $x$. And pick all those $x \in \N$ satisfying
	$$0 < \varphi(x) \cdot c \le \varphi(x+1).$$
	Possibly we have picked only finitely many elements. Let $m$ be the largest number picked so far. Pick $x > m+1$ such that
	$$c \le \frac{\varphi(x)}{\varphi(m+1)} < c^2.$$
	Such $x$ is guaranteed to exist so that we can choose. Now take $m = x$ and repeat this process infinitely. We can mechanically check that conditions (b) and (c) are met now. What remains is to pick more numbers so that (a) be satisfied while maintaining (b) and (c).
	
	We will pick some more numbers for each $i \in \N$ that fails condition (a). Suppose that $i \in \N$ fails (a). Denote for convenience $a := f(i)+1$ and $b := f(i+1)$. There are two cases.
	
	\textbf{Case i)} Suppose that $\frac{\varphi(b)}{\varphi(a)} \in [c^{2k}, c^{2k+1})$ for some $k$. Pick $x_1, x_2, \cdots x_k$ such that the followings hold for $j=1, 2, \cdots, k$:
	$$\frac{\varphi(x_j)}{\varphi(a)} < c^{2j-1} \le \frac{\varphi(x_j + 1)}{\varphi(a)}.$$
	
	\textbf{Case ii)} Suppose that $\frac{\varphi(b)}{\varphi(a)} \in [c^{2k+1}, c^{2k+2})$ for some $k$. Pick $x_1, x_2, \cdots x_k$ such that the followings hold for $j=1, 2, \cdots, k$:
	$$\frac{\varphi(x_j)}{\varphi(a)} < c^{2j} \le \frac{\varphi(x_j + 1)}{\varphi(a)}.$$
	
	It is now mechanical to check that all conditions (a), (b), and (c) are fulfilled. 

%	Concerning (d), \martin{[Donghyun?]}
\qed\end{proof}
%

%%%%%%%%%%%%%%%%%%%%%%%%%%%%%%%%%%%%%%%%%
\section{Quantitative Main Theorem and Categorical Constructions}
\label{s:Category}
We can now establish the quantitative Main Theorem %~\ref{t:Main}
strengthening the classical qualitative one \cite[Theorem~3.2.11]{Wei00}.

\begin{theorem}[Main Theorem of Type-2 \emph{Complexity} Theory]
\label{t:Main2}
Let $(X,d)$ be compact with entropy $\eta$
and linearly admissible representation $\xi$
of modulus of continuity $\kappa$.
Let $(Y,e)$ be compact with entropy $\theta$
and linearly admissible representation $\upsilon$
of modulus of continuity $\lambda$.
\begin{enumerate}
\item[a)]
If $f:X\to Y$ has modulus of continuity $\mu$,
then it admits a $(\xi,\upsilon)$-realizer $F$
with modulus of continuity
\[ \nu \;=\; \kappa\circ(1+\mu)\circ\big(\loinv{\lambda}+\calO(1)\big)
\;\in\; \lin(\eta)\circ\mu\circ\lin\big(\loinv{\theta}\big) \]
\item[b)]
If $f:X\to Y$ has $(\xi,\upsilon)$-realizer $F$
with modulus of continuity $\nu$,
then $f$ has modulus 
\[ \mu \;=\; \loinv{\kappa}\circ\nu\circ\lambda(1+\id)+\calO(1)
\;\in\; \lin\big(\loinv{\eta}\big)\circ\nu\circ\lin(\theta) \]
\end{enumerate}
\end{theorem}
The estimated moduli of continuity are (almost) tight:

\begin{myremark}\label{r:Tight}
Applying first (a) and then (b) always recovers
$f$ to have modulus of continuity $\mu':n\mapsto\mu\big(n+\calO(1)\big)+\calO(1)$
in place of $\mu$, that is, optimal up to a constant shift; recall Lemma~\ref{l:seminv}c).
\\
On the other hand applying first (b) and then (a) in general recovers $F$ only to have modulus of continuity 
$\nu'=\kappa\circ\loinv{\kappa}\circ\nu\circ\lambda\circ\big(\loinv{\lambda}+\calO(1)\big)+\calO(1)$:
which simplifies to $m\mapsto \nu\big(m+\calO(1)\big)+\calO(1)$ under additional hypotheses such as
\begin{itemize}
\itemsep0pt
\item Both $\kappa$ and $\lambda$ being surjective (and hence growing at most linearly), or
\item $\nu$ being of the form $\kappa\circ\nu'\circ\loinv{\lambda}$.
\end{itemize}
Since the real unit cube $[0;1]^d$ 
has linear modulus of continuity (Example~\ref{x:Entropy}a+b)
and the signed binary representation is linearly admissible (Example~\ref{x:SignedDigit}), 
Theorem~\ref{t:Main2} yields the following strengthening of Example~\ref{x:Max}a): \\
For any fixed $d,e\in\IN$ and non-decreasing $\mu:\IN\to\IN$, 
a function $f:[0;1]^d\to[0;1]^e$ has modulus of continuity $\lin(\mu)$
~iff~ it admits a $(\sigma,\sigma)$-realizer with modulus of continuity $\lin(\mu)$.
\end{myremark}
Recall (Remark~\ref{r:Admissible}) that 
\emph{linear metric} reducibility ``$\zeta\preccurlyeqO\xi$'' 
refines continuous reducibility ``$\zeta\preccurlyeqT\xi$'' 
by requiring a reduction $F:\dom(\zeta)\to\dom(\xi)$ with $\zeta=\xi\circ F$ 
to have modulus of continuity $\mu\circ\calO\big(\loinv{\kappa}\big)=\calo(\nu)\circ\loinv{\kappa}$ 
for every modulus of continuity $\nu$ of $\zeta$ and some $\kappa$ of $\xi$.

\begin{theorem}
\label{t:Admissible}
\begin{enumerate}
\item[a)]
Every infinite compact metric space $(X,d)$ of diameter $\leq1$
admits a linearly admissible representation $\xi$ of $X$.
Linear metric reducibility is transitive.
A representation $\zeta$ is linearly admissible iff
(i) it has a modulus of continuity in $\lin(\eta)$, and
(ii) admits a linear metric reduction $\xi\preccurlyeqO\zeta$.
\item[b)]
If $\xi:\subseteq\Cantor\twoheadrightarrow X$ is a linearly admissible
representation of the same compact space $X$ with respect to two metrics $d$ and $d'$,
then it holds $d\leq_\nu\leq d'\leq_\nu d$ 
for some $\nu\in\lin\big(\loinv{\eta}\big)\circ\lin(\eta)\circ\lin(\eta)$.
\item[c)]
Let $\xi$ and $\upsilon$ be linearly admissible representations for compact $(X,d)$ and $(Y,e)$, respectively.
Then $\xi\times\upsilon:\subseteq\Cantor\ni\bar b\mapsto\big(\xi(b_0,b_2,b_4,\ldots),\upsilon(b_1,b_3,b_5,\ldots)\big)$ 
is linearly admissible for $\big(X\times Y,\max\{d,e\}\big)$.
\\
Moreover it satisfies the following universal properties:
The projections $\pi_1:X\times Y\ni(x,y)\mapsto x\in X$ 
has a $\big(\xi\times\upsilon,\xi\big)$-realizer
with linear modulus of continuity $n\mapsto2n$;
$\pi_2:X\times Y\ni(x,y)\mapsto y\in Y$
has a $\big(\xi\times\upsilon,\upsilon\big)$-realizer
with modulus of continuity $n\mapsto2n+1$.
Conversely for every fixed $y\in Y$ 
the embedding $\imath_{2,y}:X\ni x\mapsto (x,y)\in X\times Y$
has a $\big(\xi,\xi\times\upsilon\big)$-realizer
with modulus of continuity $2n\mapsto n$;
and for every fixed $x\in X$ 
the embedding $\imath_{1,x}:Y\ni y\mapsto (x,y)\in X\times Y$
has a $\big(\upsilon,\xi\times\upsilon\big)$-realizer
with modulus of continuity $2n+1\mapsto n$.
\item[d)]
Let $\xi$ be a linearly admissible representation of connected compact $(X,d)$.
Then the representation $2^\xi$ of $\calK(X)$ from Example~\ref{x:Topology}c) 
is \emph{polynomially} admissible. 
\end{enumerate}
\end{theorem}
Note that linear `slack' in a modulus of continuity of $\xi$
translates to polynomial one in $2^\xi$.
Item~(c) justifies \cite[Definition~3.3.3.1]{Wei00}
constructing a representation of $X\times Y$
\emph{from} such of $X$ and $Y$ that it
(as opposed one created one `from scratch' by invoking a)
is compatible with the canonical morphisms $X\times Y$.
However \cite[Definition~3.3.3.2]{Wei00} for countable products 
does \emph{not} preserve linear admissibility; and neither does
Example~\ref{x:Topology}b), already in case of 
spaces $X_j$ with quadratic entropy
according to Example~\ref{x:Entropy}c).
For that purposes a more careful construction is needed:

\begin{theorem}
\label{t:Cartesian}
Fix compact metric spaces $(X_j,d_j)$ of entropies $\eta_j$ and diameters between $1/2$ and $1$, $j\in\IN$.
Let $\xi_j:\subseteq\Cantor\twoheadrightarrow X_j$ be \emph{uniformly} linearly admissible 
in that (i) it has modulus of continuity $\kappa_j(n)\leq c+c\cdot\eta_j(c+c\cdot n)$
and (ii) to every representation $\zeta_j:\subseteq\Cantor\twoheadrightarrow X_j$
with modulus of continuity $\nu_j$
there exists a mapping $F_j:\dom(\zeta_j)\to\dom(\xi_j)$
with modulus of continuity $\mu_j$
with $\mu_j\big(\kappa_j(n)\big)\leq\nu_j(c+c\cdot n)$
for some $c\in\IN$ \emph{in}dependent of $j$.
\\
Let a name of $(x_0,x_1,\ldots x_j,\ldots)\in\prod_j X_j$ 
be any infinite binary sequence
\begin{multline} 
\label{e:Cartesian}
\bar b^{(0)}|_{\kappa_0(0):\kappa_0(1)}, \qquad
\bar b^{(0)}|_{\kappa_0(1):\kappa_0(2)}, \quad
\bar b^{(1)}|_{\kappa_1(0):\kappa_1(1)}, \\ 
\bar b^{(0)}|_{\kappa_0(2):\kappa_0(3)}, \quad
\bar b^{(1)}|_{\kappa_1(1):\kappa_1(2)}, \quad
\bar b^{(2)}|_{\kappa_2(0):\kappa_2(1)}, 
\qquad\ldots\ldots \\[1ex] \ldots\ldots\qquad
\bar b^{(0)}|_{\kappa_0(n-1):\kappa_0(n)}, \;
\bar b^{(1)}|_{\kappa_1(n-2):\kappa_1(n-1)}, \;
\ldots \\ \ldots \;
\bar b^{(j)}|_{\kappa_j(n-j-1):\kappa_j(n-j)}, \;
\ldots \;
\bar b^{(n-1)}|_{\kappa_{n-1}(0):\kappa_{n-1}(1)}, \qquad \ldots\ldots
\end{multline}
such that $\bar b^{(j)}$ is a $\xi_j$-name of $x_j$.
Here $\bar b|_{k:\ell}$ abbreviates the finite segment $\bar b_{k},\ldots b_{\ell-1}$ of $\bar b$.
The thus defined representation $\xi:=\prod_j\xi_j:\subseteq\Cantor\twoheadrightarrow\prod_j X_j=:X$ 
has modulus of continuity $\kappa:n\mapsto\sum\nolimits_{j<n} \kappa_j(n-j)$
and is linearly admissible for $\big(\prod_j X_j,\sup_j d_j/2^j\big)$.
\\
Moreover the projection $\pi_j:\prod_j X_j\ni (x_0,\ldots x_j,\ldots)\mapsto x_j\in X_j$ 
has a $\big(\prod_i\xi_i,\xi_j\big)$-realizer with modulus of continuity
$m\mapsto \kappa\big(\loinv{\kappa_j}(m)+j\big)$;
and for every fixed $\bar x\in X$ and $j\in\IN$, the embedding 
$\imath_{j,\bar x}:X_j\ni x_j\mapsto (x_0,\ldots x_j,\ldots)\in X$
has a $\big(\xi_j,\prod_i\xi_i\big)$-realizer 
with modulus of continuity 
$m\mapsto \kappa_j\big(\max\big\{0,\loinv{\kappa}(m)-j\big\}\big)$.
\end{theorem}
Note that the derived moduli of continuity of the canonical morphisms
$\pi_j$ and $\imath_{j,\bar x}$ agree linearly 
with those predicted by Theorem~\ref{t:Main2},
are thus optimal in the sense of Remark~\ref{r:Tight}.
For Cartesian closure we finally treat function spaces,
in view of Lemma~\ref{l:Modulus}c) w.l.o.g. the 1-Lipschitz case:

\begin{theorem}
\label{t:Functions}
Fix convex compact metric space $(X,d)$.
To any linearly admissible representation $\xi$ of $X$
there exists a polynomially admissible representation $\xi'_1$
of the convex compact space $X'_1=\Lip_1(X,[0;1])$ of 
non-expansive functions $f:X\to[0;1]$.
It is canonical in that it asserts the application functional
$X'_1\times X\ni (f,x)\mapsto f(x)\in[0;1]$
admit a $(\xi'_1\mu\times\xi,\sigma)$-realizer $F$
with asymptotically optimal modulus of continuity $\leq\lin(\eta'_1)$.
\end{theorem}
Recall that Condition~(ii) of polynomial admissibility means
polynomial metric reducibility ``$\zeta\preccurlyeqP\xi'_1$''
to $\xi'_1$ of any other continuous representation $\zeta$ of $X'_1$.

%%%%%%%%%%%%%%%%%%%%%%%%%%%%%%%%%%%%%%%%%%%%%%%%%%%%%%%
\subsection{Proofs} \label{s:Proofs}

\begin{proof}[Theorem~\ref{t:Main2}]
\begin{enumerate}
\item[a)]
First suppose $\xi=\xi^\varphi$ the representation from Theorem~\ref{t:Linear}
with modulus of continuity $\kappa(n)\leq\calO\big(\eta(n+1)\big)$
and similarly $\upsilon$ with $\lambda(n)\leq\calO\big(\theta(n+1)\big)$.
Applying (ii) to $\zeta:=f\circ\xi:\dom(\xi)\subseteq\Cantor\to Y$ 
with modulus of continuity $\kappa\circ\mu$ yields 
a $(\xi,\upsilon)$-realizer $F$ with 
modulus $\nu=\kappa\circ\mu\circ\big(1+\loinv{\lambda}\big)$.
\\
Next consider arbitrary linearly admissible $\xi'$ with modulus $\kappa'$
and $\upsilon'$ with $\lambda'$.
Applying (ii) to $\xi'$ yields $G:\dom(\xi')\to\dom(\xi)$
with modulus $\kappa'\circ\big(1+\loinv{\lambda'}\big)$
such that $\xi'=\tilde\xi\circ G$;
and applying (ii) of Definition~\ref{d:Admissible}b) to $\upsilon$ yields $H:\dom(\upsilon)\to\dom(\upsilon')$
with modulus $\lambda\circ\big(\loinv{\lambda'}+\calO(1)\big)$
such that $\upsilon=\upsilon'\circ H$.
Together, $F':=H\circ F\circ G$ constitutes a $(\xi',\upsilon')$-realizer of $f$
with modulus of continuity
\begin{eqnarray*}
 \nu' &=& \kappa'\circ\big(1+\loinv{\lambda'}\big)\;\circ\; 
\kappa\circ\mu\circ\big(1+\loinv{\lambda}\big) \;\circ\;
\lambda\circ\big(\loinv{\lambda'}+\calO(1)\big) \\
&\leq& \kappa' \circ (1+\mu) \circ \big(\loinv{\lambda'}+\calO(1)\big) 
\;\in\; \lin(\eta)\circ\mu\circ\lin\big(\loinv{\theta}\big) 
\end{eqnarray*}
by Lemma~\ref{l:seminv}c+e) 
since $\eta\leq\kappa'\in\lin(\eta)$ and $\theta\leq\lambda'\in\lin(\theta)$ 
according to Definition~\ref{d:Admissible}b\,i) and Example~\ref{x:Entropy}f).
\item[b)]
As in (a) first suppose $F$ is a $(\xi,\upsilon)$-realizer of $f$
with modulus $\nu$, for $\xi=\xi^\varphi$ from Theorem~\ref{t:Linear}
with modulus of continuity $\kappa(n)\leq\calO\big(\eta(n+1)\big)$
and similarly $\upsilon$ with $\lambda(n)\leq\calO\big(\theta(n+1)\big)$.
Then $f\circ\xi=\upsilon\circ F:\dom(\xi)\subseteq\Cantor\to Y$
has modulus of continuity $\nu\circ\lambda\leq \kappa\circ\loinv{\kappa}\circ\nu\circ\lambda$
by Lemma~\ref{l:seminv}c); and (iv) implies $f$ to have modulus 
$\mu=1+\loinv{\kappa}\circ\nu\circ\lambda$.
\\
Next consider arbitrary linearly admissible $\xi'$ with modulus $\kappa'$
and $\upsilon'$ with $\lambda'$;
and let $F'$ be a $(\xi',\upsilon')$-realizer of $f$ with modulus $\nu'$.
(ii) yields $H'$ with $\upsilon'=\upsilon\circ H'$ of modulus $\lambda'\circ\big(1+\loinv{\lambda}\big)$;
and $G'$ with $\xi=\xi'\circ G'$ of modulus $\kappa\circ\big(\loinv{\kappa'}+\calO(1)\big)$
according to Definition~\ref{d:Admissible}b).
Together, $F:=H'\circ F'\circ G'$ constitutes a $(\xi,\upsilon)$-realizer of $f$
with modulus $\nu=\kappa\circ\big(\loinv{\kappa'}+\calO(1)\big)\circ \nu'\circ \lambda'\circ\big(1+\loinv{\lambda}\big)$.
So our initial consideration implies $f$ to have modulus
\begin{eqnarray*}
\mu' &=& 1+\loinv{\kappa}\;\circ\; \kappa\circ\big(\loinv{\kappa'}+\calO(1)\big)\circ \nu'\circ \lambda'\circ\big(1+\loinv{\lambda}\big)
\;\circ\; \lambda \\
&\leq& \calO(1)+\loinv{\kappa'}\circ\nu'\circ \lambda'(1+\id) 
\; \in \; \lin\big(\loinv{\eta}\big)\circ\nu'\circ\lin(\theta) 
\end{eqnarray*}
by Lemma~\ref{l:seminv}c+e) 
since $\eta\leq\kappa'\in\lin(\eta)$ and $\theta\leq\lambda'\in\lin(\theta)$ 
according to Definition~\ref{d:Admissible}b\,i) and Example~\ref{x:Entropy}f).
\qed\end{enumerate}\end{proof}

\begin{proof}[Theorem~\ref{t:Admissible}]
\begin{enumerate}
\item[a)]
Theorem~\ref{t:Linear} asserts the first claim.
For the second 
let $\zeta$ have modulus $\nu$ and $\zeta'$ have modulus $\nu'$ and $\zeta''$ have modulus $\nu''$;
let $F:\dom(\zeta)\to\dom(\zeta')$ with $\zeta=\zeta'\circ F$ 
have modulus of continuity $\nu\circ\calO\big(\loinv{\nu'}\big)$
and $F':\dom(\zeta')\to\dom(\zeta'')$ with $\zeta'=\zeta''\circ F'$ 
have modulus of continuity $\nu'\circ\calO\big(\loinv{\nu''}\big)$.
Then $\zeta=\zeta''\circ F'\circ F$, where $F'\circ F$ has modulus
\[ \nu\circ\calO\big(\loinv{\nu'}\big)
\;\circ\;  \nu'\circ\calO\big(\loinv{\nu''}\big)
\;=\;  \nu\circ\calO\big(\loinv{\nu''}\big) \enspace . \]
Finally, $\tilde\xi\preccurlyeqO\zeta$ is necessary according to 
Definition~\ref{d:Admissible}b\,ii);
while any other representation $\zeta'\preccurlyeqO\tilde\xi\preccurlyeqO\zeta$
since $\tilde\xi$ is admissible.
\item[b)]
First consider $\tilde\xi=\xi^\varphi$ the representation from Theorem~\ref{t:Linear}
with modulus $\tilde\kappa$.
By (iii), to every $x,x'\in X$ with $d(x,x')\leq2^{-m-1}$,
there exist $\tilde\xi$-names $\bar y_x$ and $\bar y'_{x'}$ 
of $x=\tilde\xi(\bar y_x)$ and $x'=\tilde\xi(\bar y'_{x'})$
with $\dC(\bar y_x,\bar y'_{x'})\leq2^{-\tilde\kappa(m)}$.
Hence $d'(x,x')=d'\big(\tilde\xi(\bar y_x),\tilde\xi(\bar y'_{x'})\big)\leq2^{-n}$
by (i) for $\tilde\kappa'\in\lin(\eta)$ modulus of continuity of $\tilde\xi$ w.r.t. $d'$
and $\tilde\kappa(m)\geq\tilde\kappa'(n)$, that is
%$n=\tilde\kappa\circ\upinv{\tilde\kappa'}(m)$
for $m=\loinv{\tilde\kappa}\circ\tilde\kappa'(n)$
by Lemma~\ref{l:seminv}c):
$d'\leq_\nu d$ for 
$\nu:=1+\loinv{\tilde\kappa}\circ\tilde\kappa'\in\lin\big(\loinv{\eta}\big)\circ\lin(\eta)$.
\\
Now let $\xi$ be linearly admissible with modulus $\kappa$.
Then $\xi=F\circ\tilde\xi$ for some $F:\dom(\xi)\to\dom(\tilde\xi)$
with modulus $\mu\leq\kappa\circ\calO\big(\loinv{\tilde\kappa}\big)$ 
Hence $\bar z_x:=F(\bar y_x)$ and $\bar z'_{x'}=F(\bar y'_{x'})$
have $\dC(\bar z_x,\bar z'_{x'})\leq 2^{-m'}$
for $\tilde\kappa(m)\geq\mu(m')$; 
and again $d'(x,x')\leq2^{-n}$
for $m'\geq\tilde\kappa'(n)$ modulus of continuity of $\xi$ w.r.t. $d'$:
$d'\leq_\nu d$ for 
$\nu:=1+\loinv{\tilde\kappa}\circ\mu\circ\tilde\kappa'\in\lin\big(\loinv{\eta}\big)\circ\lin(\eta)\circ\lin(\eta)$.
\item[c)]
Let $X$ and $Y$ have entropies $\eta$ and $\theta$, respectively;
$\xi$ and $\upsilon$ moduli of continuity $\mu\leq\lin(\eta)$ and $\nu\leq\lin(\theta)$.
(i) $X\times Y$ has entropy at least $\eta(n-1)+\theta(n-1)-1$ by Example~\ref{x:Entropy}b);
and $\xi\times\upsilon$ has modulus of continuity 
$2\cdot\max\{\mu,\nu\}\leq\lin\big(n\mapsto \eta(n-1)+\theta(n-1)-1\big)$.
(ii) Let $\zeta:\subseteq\Cantor\twoheadrightarrow X\times Y$ 
be any representation with modulus $\kappa$.
Then its projection onto the first component $\zeta_1$ 
constitutes a continuous representation of $X$ with some modulus $\kappa$.
Hence there exist by hypothesis $F:\dom(\zeta)\to\dom(\xi)$
of modulus $\kappa\circ\calO\big(\loinv{\mu}\big)$
such that $\zeta_1=\xi\circ F$;
similarly $\zeta_2=\upsilon\circ G$ with $G:\dom(\zeta)\to\dom(\upsilon)$ 
of modulus $\kappa\circ\calO\big(\loinv{\nu}\big)$.
Then $F\times G:=(F_0,G_0,F_1,G_1,F_2,\ldots):\dom(\zeta)\to\dom(\xi\times\upsilon)$
satisfies $\zeta=(\xi\times\upsilon)\circ(F\times G)$ and has modulus 
\[ 2\cdot\max\Big\{\kappa\circ\calO\big(\loinv{\mu}\big),\kappa\circ\calO\big(\loinv{\nu}\big)\Big\}
\;=\; 2\cdot\kappa\circ \calO\big(\loinv{\min\{\mu,\nu\}}\big) \]
$\leq\; \kappa\circ\calO\big(\loinv{2\cdot\max\{\mu,\nu\}}\big)$
by Lemma~\ref{l:seminv}h).
\\
Finally, $(b_0,b_1,b_2,b_3,b_4,\ldots)\mapsto (b_0,b_2,b_4,\ldots)$ 
is a $\big(\xi\times\upsilon,\xi\big)$-realizer of $\pi_1$
with modulus of continuity $n\mapsto2n$;
and $(b_0,b_1,b_2,b_3,b_4,\ldots)\mapsto (b_1,b_3,\ldots)$ 
a $\big(\xi\times\upsilon,\upsilon\big)$-realizer of $\pi_2$
with modulus of continuity $n\mapsto2n+1$.
For any fixed $\upsilon$-name $(b_1,b_3,\ldots)$ of $y$,
$(b_0,b_2,\ldots)\mapsto (b_0,b_1,b_2,b_3,b_4,\ldots)$
is a realizer of $\imath_{2,y}$ with modulus of continuity $2n\mapsto n$;
and any fixed $\xi$-name $(b_0,b_2,\ldots)$ of $x$,
$(b_1,b_3,\ldots)\mapsto (b_0,b_1,b_2,b_3,b_4,\ldots)$
is a realizer of $\imath_{1,x}$ with modulus of continuity $2n+1\mapsto n$.
\item[d)]
Combine Example~\ref{x:Topology}c) with Example~\ref{x:Entropy}g).
\qed\end{enumerate}\end{proof}

\begin{proof}[Theorem~\ref{t:Cartesian}]
Since $X_j$ has diameter between $1/2$ and $1$, w.l.o.g. $\eta_j(0)=0$ and $\eta_j(n)\geq1$ for $n\geq2$ and w.l.o.g. $\kappa_j(0)=0$.
$X:=\prod_j X_j$ has entropy $\eta(n)\geq\sum_{j<n}\eta_j(n-1-j)-\lfloor n/2\rfloor$ by Example~\ref{x:Entropy}c).
On the other hand the initial segment $\bar b^{(j)}|_{0:\kappa_j(n-j)}$
of a $\xi_j$-name $\bar b^{(j)}$ determines $x_j=\xi_j\big(\bar b^{(j)}\big)$
up to error $2^{-(n-j)}$ w.r.t. $d:=d_j/2^j$;
and is located among the first $\kappa_0(n)+\kappa_1(n-1)+\cdots+\kappa_n(0)=\kappa(n)$
symbols of a $\xi$-name of $(x_j)_{_j}$ according to Equation~(\ref{e:Cartesian});
recall $\kappa_n(0)=0$.
Therefore (i) $\xi$ has modulus of continuity $\kappa(n)$,
which is $\leq\lin\big(\eta(n)\big)$ since 
$\kappa_j(n)\leq c+c\cdot\eta_j(c+c\cdot n)$
and $\eta_j(n)\geq1$ for $n\geq2$ `covers' the $\lfloor n/2\rfloor$.
\\
Regarding (ii), let $\zeta:\subseteq\Cantor\twoheadrightarrow X$
have modulus of continuity $\nu$.
The projection $\pi_j:X\ni (x_0,\ldots x_j,\ldots)\mapsto x_j\in X_j$
has modulus of continuity $n\mapsto n+j$ since $X$ is equipped with metric $d=\sup_j d_j/2^j$.
The representation $\zeta_j:=\pi_j\circ\zeta:\dom(\zeta)\subseteq\Cantor\twoheadrightarrow X_j$
thus has modulus $\nu_j:n\mapsto\nu(n+j)$.
By hypothesis (ii) on $\xi_j$, there exists a mapping $F_j:\dom(\zeta)\to\dom(\xi_j)$
whose modulus of continuity $\mu_j$ satifies $\mu_j\big(\kappa_j(n)\big)\leq\nu_j(c+c\cdot n)$
such that it holds $\pi_j\circ\zeta=\xi_j\circ F_j$.
Now let $F:=$
\begin{multline*} 
F_0|_{\kappa_0(0):\kappa_0(1)}, \qquad
F_0|_{\kappa_0(1):\kappa_0(2)}, \quad
F_1|_{\kappa_1(0):\kappa_1(1)}, \\ 
F_0|_{\kappa_0(2):\kappa_0(3)}, \quad
F_1|_{\kappa_1(1):\kappa_1(2)}, \quad
F_2|_{\kappa_2(0):\kappa_2(1)}, 
\qquad\ldots\ldots \\[1ex] \ldots\ldots\qquad
F_0|_{\kappa_0(n-1):\kappa_0(n)}, \;
F_1|_{\kappa_1(n-2):\kappa_1(n-1)}, \;
\ldots \\ \ldots \;
F_j|_{\kappa_j(n-j-1):\kappa_j(n-j)}, \;
\ldots \;
F_{n-1}|_{\kappa_{n-1}(0):\kappa_{n-1}(1)}, \qquad \ldots\ldots
\end{multline*}
so that $\zeta=\xi\circ F$.
Moreover $F_j|_{0:\kappa_j(n-j)}$ depends on 
the first $\mu_j\big(\kappa_j(n-j)\big)$ symbols of its argument;
hence $F$ has modulus of continuity $\mu$ with
\[ \mu\big(\kappa(n)\big) 
\;=\; \sup\nolimits_{j<n} \mu_j\big(\kappa_j(n-j)\big) 
\;\leq\; \sup\nolimits_{j<n} \nu_j\big(c+c\cdot(n-j)\big)
\;=\; \nu(c+c\cdot n). \]
Mapping an infinite binary sequence according to Equation~(\ref{e:Cartesian})
to $\bar b^{(j)}$ constitutes a $\big(\prod_i\xi_i,\xi_j\big)$-realizer
of $\pi_j$; and for $n\geq j$ the first $\kappa_j(n-j)$ bits of $\bar b^{(i)}$
are determined by the first $\kappa(n)$ bits of the given $\xi$-name:
hence $\kappa_j(n-j)\mapsto\kappa(n)$ a modulus of continuity.
Conversely mapping $\xi_j$-name $\bar b^{(j)}$ of $x_j$
to the infinite binary sequence from Equation~(\ref{e:Cartesian})
constitutes a $\big(\xi_j,\prod_i\xi_i\big)$-realizer of $\imath_{j,\bar x}$
with modulus of continuity $\kappa(n)\mapsto \kappa_j\big(\max\{0,n-j\}\big)$.
\qed\end{proof}
\begin{proof}[Theorem~\ref{t:Functions}]
Let $\eta$ denote the entropy of $X$
and $\kappa\leq\lin(\eta)$ a modulus of continuity of $\xi$.
Only for notational simplicity, consider the case $\mu=\id$
of 1-Lipschitz functions $X'_1:=\Lip_1(X,[0;1])$.
We pick up on, and refine, the entropy analysis from the proof of Example~\ref{x:Entropy}h).
Fix $n\in\IN$ and, for every $\vec w\in \{\sdzero,\sdone\}^{\kappa(n)}$
with $\xi\big[\vec w\,\Cantor\big]\neq\emptyset$,
choose\footnote{Different choices lead to different representations $\xi'_1$ of $X'_1$.\label{f:contingent}} 
some $x_{\vec w}\in \xi\big[\vec w\,\Cantor\big]$.
Record that, as centers of closed balls of radius $2^{-n}$, these cover $X$.
\\
Next choose\footref{f:contingent}
some subset $W_n\subseteq \{\sdzero,\sdone\}^{\kappa(n)}$
such that any two distinct $\vec w\in W_n$ 
satisfy $d\big(x_{\vec w},x_{\vec v}\big)>2^{-n}$
while the closed balls $\cball\big(x_{\vec w},2^{-n+1}\big)$
of double radius still cover $X$.
Such $W_n$ can be created greedily
by repeatedly and in arbitrary order weeding out
one of $(\vec w,\vec v)$ whenever $d\big(x_{\vec w},x_{\vec v}\big)\leq2^{-n}$:
observe $\cball\big(x_{\vec w},2^{-n+1}\big)\supseteq \cball\big(x_{\vec v},2^{-n}\big)$
and abbreviate $X_n:=\{ x_{\vec w}:\vec w\in W_n\}$.
\\
We now formalize the idea that a $\xi'_1$-name of $f\in X'_1$
encodes a sequence $f_n:X_n\to\ID_n$ of $\tfrac{3}{2}$-Lipschitz functions
whose $\tfrac{3}{2}$-Lipschitz extensions $\extm{f_n}$ approximate
$f$ up to error $2^{-n+1}$:
As in the proof of Example~\ref{x:Entropy}h), 
$f'_n:=\lfloor 2^n\cdot f\big|_{X_n}\rceil/2^n$
satisfies this condition, hence asserting every $f\in X'_1$ to have a $\xi'_1$-name,
i.e. $\xi'_1$ be surjective.
\\ 
In order to encode the $f_n$ succinctly, and make $\xi'_1$ have modulus
of continuity $\kappa'_1\leq\lin(\eta'_1)$ for the entropy $\eta'_1$ of $X'_1$,
recall the connected undirected graph $G_n=(X_n,E_n)$ from the proof of Example~\ref{x:Entropy}f+h)
with edge $(x,x')$ present iff the open balls 
of radius $2^{-n+\pmb{2}}$ around centers $x,x'$ intersect.
Choose\footref{f:contingent} some directed spanning tree $F_n\subseteq E_n$ of $G_n$
with root $x_{n,0}$ and remaining nodes $x_{n,1},\ldots x_{n,N_n-1}$
in some \footref{f:contingent} topological order, where $N_n:=\Card(W_n)\leq 2^{\kappa(n)}$.
As in the proof of Example~\ref{x:Entropy}h), 
every $\tfrac{3}{2}$-Lipschitz $f_n:X_n\to\ID_n$
is uniquely described by $f_n\big(x_{n,0}\big)\in\ID_n\cap[0;1]$ 
together with the sequence 
\[ f_n\big(x_{n,m}\big)\;-\;f_n\big(x_{n,m-1}\big) \;\in\; \big\{-6\cdot2^{-n},\ldots 0,\ldots +6\cdot2^{-n}\big\} \quad , 1\leq m<N_n \enspace . \]
The first takes $n+1$ bits to describe, the latter $(N_n-1)\times \log_2(13)$ bits:
in view of Example~\ref{x:Entropy}f) a total of $\calO\big(2^{\kappa(n)}\big)$ to encode $f_n$ in some $\vec u_n$.
So the initial segment $(\vec u_0,\ldots \vec u_{n+1})$ of a thus defined $\xi'_1$-name $\bar u=(\vec u_0,\vec u_1,\ldots)$ of $f$,
encoding $f_0,\ldots f_{n+1}$ and thus determining $f$ up to error $2^{-n}$,
has length $\kappa'_1(n)=\calO\big(2^{\kappa(0)}\big)+\cdots+\calO\big(2^{\kappa(n+1)}\big)\leq\calO\big(2^{\kappa(n+2)}\big)\leq\lin(\eta'_1)$
again by Example~\ref{x:Entropy}f+h): thus establishing Condition~(i).
\\ %ZZZ (ii)
Regarding the application functional $(f,x)\mapsto f(x)$,
consider the following `algorithm' and recall Theorem~\ref{t:Admissible}c):
Given a $(\xi'_1\times\xi)$-name $(u_0,v_0,u_1,v_1,\ldots u_m,v_m,\ldots)$ 
of $(f,x)$ and $n\in\IN$,
let $\vec v:=\bar v_{<\kappa(n)}$ and `find' some $\vec w\in W_n$
with $d\big(x_{\vec w},x_{\vec v}\big)\leq2^{-n+1}$;
then `trace' the path in spanning tree $(W_n,F_n)$ from
its root $x_{n,0}$ to $x_{\vec w}=x_{n,M}$: all information contained 
within the first $\kappa'_1(n)$ bits of $\bar u$ 
encoding $\tfrac{3}{2}$-Lipschitz $f_n:X_n\to\ID_n$ 
whose extension approximates $f$ up to error $2^{-n+1}$,
sufficient to recover the value 
\[ y_n \;:=\; f_n\big(x_{\vec w}\big) \;=\;  f_n\big(x_{n,0}\big)   \;+\;
\sum\nolimits_{m=1}^M  \Big(f_n\big(x_{n,m}\big)-f_n\big(x_{n,m-1}\big)\Big)
\;\in\;\ID_n \]
satisfying $|y_n-f(x)|\leq |y_n-f_n(x)|+|f_n(x)-f(x)|\leq \tfrac{3}{2}\cdot 2^{-n+1}+2^{-n+1}\leq 2^{-n+3}$.
The (initial segment of length $n+3$ of) sequence $y_0,\ldots y_n,\ldots$ in turn is easily converted
to (an initial segment of length $n$ of) a signed binary expansion of $f(x)$
\cite[Lemma~7.3.5]{Wei00}: yielding a $(\xi'_1\times\xi,\sigma)$-realizer of $(f,x)\mapsto f(x)$ with 
asymptotically optimal modulus of continuity $\mu(n)=\max\{2\kappa(n+3),2\kappa'_1(n+3)\}\leq\lin\big(\eta'_1(n+3)\big)$
by Example~\ref{x:Entropy}f+h).
\qed\end{proof}

%%%%%%%%%%%%%%%%%%%%%%%%%%%%%%%%%%%%%%%%%
\section{Conclusion and Perspective}
\label{s:Conclusion}

For an arbitrary compact metric space $(X,d)$ 
we have constructed a generic representation $\xi$ 
with optimal modulus of continuity, namely agreeing with the space's entropy 
up to a constant factor. And we have shown this representation to exhibit 
properties similar to the classical \emph{standard} representation of a topological T$_0$ space
underlying the definition of qualitative \emph{admissibility},
but now under the quantitative perspective crucial for a generic 
resource-bounded complexity theory for computing with continuous data:
$\xi$ is maximal with respect to optimal metric reduction
among all continuous representations.
The class of such metrically optimal representations 
is closed binary and countable Cartesian products,
and gives rise to metrically optimal representations 
of the Hausdorff space of compact subsets
and of the space of non-expansive real functions.
Moreover, with respect to such \emph{linearly} admissible representations 
$\xi$ and $\upsilon$ of compact metric spaces $X$ and $Y$,
optimal moduli of continuity of functions $f:X\to Y$
and their $(\xi,\upsilon)$-realizers $F:\dom(\xi)\to\dom(\upsilon)$
are linearly related up to
composition with (the lower semi-inverse) of the entropies
of $X$ and $Y$, respectively. 

All our notions (entropy, modulus of continuity) and arguments are information-theoretic:
according to Fact~\ref{f:Proper}b) these precede, and under suitable oracles coincide with, 
complexity questions. 
They thus serve as general guide to investigations over concrete advanced
spaces of continuous data, such as of integrable or weakly differentiable
functions employed in the theory of Partial Differential Equations
\cite{DBLP:journals/lmcs/Steinberg17}.
%And they allow to translate between the two flavours of complexity theory for continuous data:
%one measuring algorithmic cost in dependence on the number $m$ of bits
%of a realizer $F$ being output, and one in dependence on
%the absolute output approximation error $2^{-n}$. 

\COMMENTED{
\begin{definition}
\label{d:Adapted}
Fix compact metric spaces $(X,d)$ and $(Y,e)$ 
with entropies $\eta$ and $\theta$ and
with linearly/polynomially admissible representations
$\xi$ and $\upsilon$, respectively.
\emph{Computing} a function $f:X\to Y$ 
with modulus of continuity $\mu$
in linear/polynomial time
means to compute a $(\xi,\upsilon)$-realizer $F$ of $f$
within time linear/polynomial in 
$\lin(\id+\log\eta)\circ \mu\circ\lin\big(\loinv{\id+\log\theta}\big)$.
\end{definition}
This straightforwardly generalizes to \emph{multi}functions \cite{PZ13}.
It consistently `overloads' Definition~\ref{d:Type2} regarding Cantor space,
and recovers the real case $X,Y\in\big\{\Cantor,[0;1]\big\}$ having 
linear entropies $\eta,\theta$.
}

%\begin{myremark}
%Our hypothesis on the metric spaces to be \emph{compact}
%can be relaxed to \emph{uniformly bounded},
%but then the functions under consideration 
%must be required \emph{uniformly} continuous 
%in order to have some modulus of continuity.
%\end{myremark}
%
In order to strengthen our Main Theorem~\ref{t:Main2},
namely to further decrease the gap between (a) and (b)
according to Remark~\ref{r:Tight}, we wonder:

\begin{myquestion}
\label{q:Future}
Which infinite compact metric spaces $(X,d)$ with entropy $\eta$ admit%
\begin{enumerate}\itemsep0pt%
\item[a)]
a %(not necessarily admissible) 
representation 
with $\eta\big(n+\calO(1)\big)+\calO(1)$ as modulus of continuity?
\item[b)]
an \emph{admissible} representation
with modulus of continuity $\eta\big(n+\calO(1)\big)+\calO(1)$?
\item[c)]
a \emph{linearly} admissible representation
with modulus $\eta\big(n+\calO(1)\big)+\calO(1)$ ?
\smallskip
\item[d)]
If $\xi$ is a linearly admissible representation of $(X,d)$
and $Z\subseteq X$ closed, \\ is the restriction 
$\xi|^{Z}$ then again linearly admissible?
{\rm\cite[Lemma~3.3.2]{Wei00}}
\item[e)]
In view of Example~\ref{x:Entropy}f),
how large can be the asymptotic gap between
intrinsic $\eta_{K,K}$ and relative entropy $\eta_{X,K}$ ?
\item[f)]
(How) does Theorem~\ref{t:Functions} generalize
from real $[0;1]$ to other compact metric codomains $Y$?
\item[g)]
How do the above considerations carry over from the Type-2
setting of computation on streams to that of oracle arguments
{\rm\cite{Ko91,KC12}}?
%This avoids the application functional (Theorem~\ref{t:Functions})
%having exponential complexity\ldots
\end{enumerate}
\end{myquestion}
%

%%%%%%%%%%%%%%%%%%%%%%%%%%%%%%%%%%%%%%%%%
\subsection{Algorithmic Cost and Representations for Higher Types}
\label{ss:Hyper}

It seems counter-intuitive that the application functional
should be as hard to compute as Example~\ref{x:Max} predicts.
To `mend' this artefact, the Type-2 Machine model
from Definition~\ref{d:Type2} --- computing on `streams' 
of infinite sequences of bits
$\bar b=(b_0,b_1,\ldots b_n,\ldots)\in\{\sdzero,\sdone\}^\IN$ 
with sequential/linear-time access --- is commonly modified
for problems involving continuous subsets or functions as arguments
to allow for `random'/logarithmic-time access {\cite{KC12}}:

\begin{definition}
\label{d:Type3}
\begin{enumerate}
\item[a)]
Abbreviate with $\Cantor':=\{\sdzero,\sdone\}^{\{\sdzero,\sdone\}^*}$ 
the set of total finite string predicates $\varphi:\{\sdzero,\sdone\}^*\to\{\sdzero,\sdone\}$,
equipped with the metric $\dD(\varphi,\psi)=2^{-\min\{|\vec u|:\varphi(\vec u)\neq\psi(\vec u)\}}$,
where $|\vec u|\in\IN$ denotes the length $n$ of $(u_0,\ldots u_{n-1})\in\{\sdzero,\sdone\}^*$.
\item[b)]
A \emph{Type-3 Machine} $\Machine^?$ is an ordinary oracle Turing machine
with variable oracle. It \emph{computes} the partial function $\calF:\subseteq\Cantor'\to\Cantor'$
if, for every $\varphi\in\dom(\calF)\subseteq\Cantor'$,
$\Machine^\varphi$ computes $\calF(\varphi)\in\Cantor'$ in that it
accepts all inputs $\vec v\in\big(\calF(\varphi)\big)^{-1}[\sdone]$
and rejects all inputs $\vec v\in\big(\calF(\varphi)\big)^{-1}[\sdzero]$;
The behaviour of $\Machine^\varphi$ for $\varphi\not\in\dom(\calF)$ may be arbitrary. 
\item[c)]
For $\psi\in\Cantor'$,
$\Machine^{\psi,?}$ denotes a Type-3 Machine with fixed oracle $\psi$ and additional variable oracle
in that it operates, for every $\varphi\in\dom(\calF)\subseteq\Cantor'$,
like $\Machine^{\psi\otimes\varphi}$, where
\[ \psi\otimes\varphi: (\sdzero\,\vec u)\mapsto\psi(\vec u),
\quad (\sdone\,\vec u)\mapsto\varphi(\vec u) \enspace . \]
\item[d)]
$\Machine^?$ computes $\calF$ in \emph{time} $t:\IN\to\IN$ if 
$\Machine^\varphi$ on input $\vec v\in\{\sdzero,\sdone\}^n$
stops after at most $t(n)$ steps 
regardless of $\varphi\in\dom(\calF)$;
similarly for $\Machine^{\psi,?}$.
\end{enumerate}
\end{definition}
Encoding a continuous space using oracles $\psi\in\Cantor'$
rather than sequences $\bar b\in\Cantor$ gave rise to a
different, new notion of representation \cite[\S3.4]{KC12}.
Here we shall call it \emph{hyper-}representation,
and avoid confusion with the previous conception
now referred to as \emph{stream} representation.

\begin{definition}
\label{d:Hyper}
\begin{enumerate}
\item[a)]
A \emph{hyper-}representation of a space $X$ is 
a partial surjective mapping $\Xi:\subseteq\Cantor'\twoheadrightarrow X$.
\item[b)]
For hyper-representations $\Xi:\subseteq\Cantor'\twoheadrightarrow X$
and $\Upsilon:\subseteq\Cantor'\twoheadrightarrow Y$,
a $(\Xi,\Upsilon)$-\emph{realizer} of a function $f:X\to Y$
is a partial function $\calF:\dom(\Xi)\to\dom(\Upsilon)$
such that $f\circ\Xi=\Upsilon\circ\calF$ holds.
\item[c)]
A \emph{reduction} from $\Xi:\subseteq\Cantor'\twoheadrightarrow X$
to $\Xi':\subseteq\Cantor'\twoheadrightarrow X$ is a
$(\Xi,\Xi')$-realizer of the identity $\id:X\to X$.
\item[d)]
$(\Xi,\Upsilon)$-\emph{computing} $f$ means to compute some $(\Xi,\Upsilon)$-realizer $\calF$ of $f$
in the sense of Definition~\ref{d:Type3}b).
\item[e)]
The \emph{product} hyper-representation 
of $\Xi:\subseteq\Cantor'\twoheadrightarrow X$ 
and $\Upsilon:\subseteq\Cantor'\twoheadrightarrow Y$ is 
\[ \Xi\times\Upsilon:\subseteq\Cantor'
\;\ni\;
\varphi\;\mapsto\; \Big(\Xi\big(\vec v\mapsto \varphi(\sdzero\,\vec v)\big),\Upsilon\big(\vec v\mapsto\varphi(\sdone\,\vec v)\big)\Big)
\;\in\; X\times Y \]
\item[f)]
Consider the hyper-representation (sic!) 
$\imath_\Cantor: \Cantor' \;\ni\; \varphi \;\mapsto\;  \big( \varphi(\sdone^n)_{_n}\big) \;\in\; \Cantor$.
For stream representation $\xi:\subseteq\Cantor\twoheadrightarrow X$,
let $\xi\circ\imath_\Cantor:\subseteq\Cantor'\twoheadrightarrow X$ 
denote its induced \emph{unary} hyper-representation.
\item[g)]
Abusing notation, consider the hyper-representation
\[ \bin: \Cantor' \;\ni\; \varphi \;\mapsto\; \Big(\varphi\big(\bin(n)\big)_{_n}\Big) \;\in\; \Cantor \enspace . \]
For stream representation $\xi:\subseteq\Cantor\twoheadrightarrow X$,
let $\xi\circ\bin\subseteq\Cantor'\twoheadrightarrow X$ 
denote its induced \emph{binary} hyper-representation.
\end{enumerate}
\end{definition}
Indeed, according to \cite[\S4.3]{KC12},
appropriate hyper-representations now
allow to compute the application functional in
time more reasonable than in Example~\ref{x:Max}b);
cmp. Item~d) of the following Proposition.
Items~a) to c) correspond to Fact~\ref{f:Proper}.

\begin{proposition}
\label{p:Type3}
\begin{enumerate}
\item[a)]
$\Cantor'$ is compact of entropy $\eta_{\Cantor'}=2^{\id}-1$.
If $\calF:\subseteq\Cantor'\to\Cantor'$ is computed by 
$\Machine^{?}$ and if $\dom(\calF)$ is compact,
then this computation admits a time bound $t:\IN\to\IN$
in the sense of Definition~\ref{d:Type3}d);
similarly for $\Machine^{\psi,?}$\ldots
\item[b)]
If $\Machine^{\psi,?}$ computes $\calF:\subseteq\Cantor'\to\Cantor'$ in time $t(n)$,
then $\calF$ has modulus of continuity $t(n)$. 
\item[c)]
If $\calF:\subseteq\Cantor'\to\Cantor'$ has modulus
of continuity $\mu:\IN\to\IN$, then there exists an oracle $\psi\in\Cantor'$
and a Type-3 Machine $\Machine^{\psi,?}$ computing $\calF$
in time $\calO\big(n+2^{t(n)}\big)$.
\item[d)]
The compact space $[0;1]'_1$ from Example~\ref{x:Max}b) 
admits a hyper-representation $\Delta_1$ such that application
$[0;1]'_1\times[0;1]\ni(f,r)\mapsto f(r)\in[0;1]$ 
is $(\Delta_1\times\tilde\delta,\delta\circ\imath_\Cantor)$-computable in polynomial time
for the unary hyper-representation $\delta\circ\imath_\Cantor$
induced by the aforementioned dyadic stream representation $\delta$ of $[0;1]$.
\item[e)]
Hyper-representation (sic!) $\imath_\Cantor:\subseteq\Cantor'\twoheadrightarrow\Cantor$ is an isometry.
Stream representation $\xi:\subseteq\Cantor\twoheadrightarrow X$ has modulus of continuity $\kappa$
iff induced unary hyper-representation $\xi\circ\imath_\Cantor$ does.
\item[f)]
For $(\xi,\upsilon)$-realizer $F:\subseteq\Cantor\to\Cantor$ of $f:X\to Y$,
$\calF:=\imath_\Cantor^{-1}\circ F\circ\imath_\Cantor:\subseteq\Cantor'\to\Cantor'$ 
is a $\big(\xi\circ\imath_\Cantor,\upsilon\circ\imath_\Cantor\big)$-realizer 
of $f:X\to Y$. $F$ has modulus of continuity $\mu$ iff $\calF$ does.
\item[g)]
Hyper-representation $\bin:\subseteq\Cantor'\twoheadrightarrow\Cantor$ has logarithmic modulus of continuity.
Its inverse, the stream representation $\bin:\Cantor\twoheadrightarrow\Cantor'$
has exponential modulus of continuity. Both are optimal.
Stream representation $\xi:\subseteq\Cantor\twoheadrightarrow X$ has modulus of continuity $\calO\big(2^\kappa\big)$
iff induced binary hyper-representation $\xi\circ\bin:\subseteq\Cantor'\twoheadrightarrow X$
has modulus of continuity $\kappa$.
\end{enumerate}
\end{proposition}
Note the gap between Type-3 Proposition~\ref{p:Type3}b+c),
absent in the Type-2 Fact~\ref{f:Proper}b+c).

From a high-level perspective, 
the exponential lower complexity bound to the application functional
in Example~\ref{x:Max}b) is due to $[0;1]'$ having 
exponential entropy while stream representations'
domain $\Cantor$ has only linear entropy:
see Example~\ref{x:Entropy}a+h).
Proposition~\ref{p:Type3}d)
avoids that information-theoretic bottleneck 
by proceeding to hyper-representations 
with domain $\Cantor'$ also having exponential entropy.

In fact {\cite[\S4.3]{KC12}} extends the representation
$\Delta_1$ of $[0;1]'_1=\Lip_1([0;1],[0;1])$ from Proposition~\ref{p:Type3}d)
to entire $\calC\big([0;1],[0;1]\big)=\bigcup_{\mu} \calC_\mu\big([0;1],[0;1]\big)$,
where the union ranges over all strictly increasing $\mu:\IN\to\IN$. 
Lacking compactness, in view of Proposition~\ref{p:Type3}a)
one cannot expect a time bound 
%for the application functional
%$\calC\big([0;1],[0;1]\big)\times[0;1]\ni (f,x)\mapsto f(x)\in[0;1]$
on entire $\calC\big([0;1],[0;1]\big)$
depending only on the output precision $n$.
Instead {\cite[\S3.2]{KC12}} considers runtime \emph{polynomial} 
if bounded by some term $P=P(n,\mu)$ 
in both the integer output precision parameter $n$ and a 
modulus of continuity $\mu:\IN\to\IN$ of the given function argument $f$:
a higher-type parameter \cite{EikeFlorian}.
The considerations in the present work suggest
a natural 

\begin{myremark}
\label{r:SecondOrder}
According to Example~\ref{x:Entropy}h) 
$\calC_\mu([0;1],[0;1])$ has entropy $\eta=\Theta\big(2^{\mu}\big)$
such that $\log\eta=\Theta(\mu)$ `recovers' the modulus of continuity.
Moreover our complexity-theoretic ``Main'' Theorem~\ref{t:Main2}
confirms that even metrically well-behaved (e.g. 1-Lipschitz) functionals $\Lambda:X\to[0;1]$
%$\Lambda_\mu:X_\mu=\calC_\mu([0;1],[0;1])\to[0;1]=:Y$ like application $(f,x)\mapsto f(x)$
can only have realizers with modulus of continuity/time complexity
growing in $X$'s entropy $\eta$. This suggests generalizing
second-order polynomial runtime bounds $P=P(n,\mu)$ 
from spaces of continuous real functions {\cite[\S3.2]{KC12}}
to $P(n,\log\eta)$ for compact metric spaces $X$ beyond $\calC_\mu([0;1],[0;1])$.
The logarithm is consistent with the quantitative properties of hyper-representations 
expressed in Proposition~\ref{p:Type3}.
\end{myremark}

\begin{proof}[Proposition~\ref{p:Type3}]
\begin{enumerate}
\item[a)]
Compactness of $\Cantor'$ follows from K\"{o}nig's Lemma:
it is an infinite finitely (only, as opposed to $\Cantor$, 
increasingly) branching tree.
Cover $\Cantor'$ by $2^{2^n-1}$ closed balls 
$\displaystyle\big\{\varphi:\varphi\big|_{\{\sdzero,\sdone\}^{<n}}=\psi\big\}$,
$\psi:\{\sdzero,\sdone\}^n\to\{\sdzero,\sdone\}$ of radius $2^{-n}$: optimally.
\\
Consider the number $N:=t_\Machine(\varphi,\vec u)\in\IN$ 
$\Machine^{\varphi}$ makes on input $\vec u\in\{\sdzero,\sdone\}^n$ 
for $\varphi\in\dom(\calF)\subseteq\Cantor'$.
During this execution, $\Machine^{\varphi}$ 
can construct and query oracle $\varphi$ 
only on strings $\vec v$ of length $|\vec v|<N$:
Replacing $\varphi$ with some $\psi\in\Cantor'$
of distance $\dD(\varphi,\psi)\leq2^{-N}$ 
will remain undetected, that is,
$\Machine^{\psi}$ on input $\vec u$
will behave the same way, and in particular
still terminate after $N$ steps.
This establishes continuity of $t_\Machine(\cdot,\vec u)$.
By compactness of $\dom(\calF)$,
the following maxima thus exist:
\[ t_\Machine(\vec u) \;:=\; \max \big\{t_\Machine(\varphi,\vec u)\::\:\varphi\in\calF\big\}
\qquad
t_\Machine(n)\;:=\;\max\big\{ t_\Machine(\vec u)\::\: 
\vec u\in\{\sdzero,\sdone\}^{n}\big\} \]
\item[b)]
As mentioned in the proof of (a), 
$\Machine^{\varphi}$ and $\Machine^{\psi}$ 
will behave identically on all inputs $\vec u\in\{\sdzero,\sdone\}^n$ 
for $\varphi,\psi\in\dom(\calF)$ 
with $\dD(\varphi,\psi)\leq2^{-t(n)}$:
Meaning $\calF(\varphi)$ and $\calF(\psi)$
have distance $\leq 2^{-n}$.
\item[c)]
By hypothesis, $\calF\big(\varphi\big)(\vec u)\in\{\sdzero,\sdone\}$
depends only on the restriction $\displaystyle\varphi\big|_{\{\sdzero,\sdone\}^{<m}}$
for $m:=\mu(n)$ and $n:=|\vec u|$. Thus 
\[ \psi\big(\vec u,\varphi(),\varphi(\sdzero),\varphi(\sdone),\varphi(\sdzero\sdzero),\ldots\varphi(\sdone^{m-1})\big)
\;:=\;\calF\big(\varphi\big)(\vec u) \]
is well-defined an oracle.
And, for given $\vec u$ and $\varphi$,
making this query of length $\calO(n+2^m)$
recovers the value $\calF\big(\varphi\big)(\vec u)$.
%\item[d)] ZZZ
%\item[e)]
%\item[f)]
%|item[g)]
\qed\end{enumerate}\end{proof}
%

%%%%%%%%%%%%%%%%%%%%%%%%%%%%%%%%%%%%%%%%%
\subsection{Representation Theory of Compact Metric Spaces}
\label{ss:Future}

Both stream and hyper representations 
translate (notions of computability and bit-complexity) to their co-domains 
from `universal' \cite{Benyamini}
compact structures (which are already naturally equipped with formal conceptions of computing,
namely) $\Cantor$ and $\Cantor'$, respectively.
Matthias Schr\"{o}der [personal communication 2017] has suggested a third candidate domain
of generalized representations: the Hilbert Cube $\Hilbert=\prod_{j\geq0} [0;1]$ 
equipped with metric $\dH(\bar x,\bar y)=\sup_j |x_j-y_j|/2^j$; recall Example~\ref{x:Entropy}a).
This parallels earlier developments in continuous computability theory
considering \emph{equilogical} \cite{DBLP:journals/tcs/BauerBS04}
and \emph{quotients of countably-based topological} (=QCB) 
spaces \cite{DBLP:conf/cie/Schroder06}.
And its suggests the following generalization:

\begin{definition}
\label{d:Representation}
Fix a compact metric space $(\Universe,D)$ with entropy $\Theta$.
\begin{enumerate}
\item[a)]
A \emph{$\Universe$-representation} of another compact metric space $(X,d)$
is a surjective partial mapping 
$\xi:\subseteq\Universe\twoheadrightarrow X$.
\item[b)]
Let $\eta$ denote the entropy of $(X,d)$. \\
Call $\xi$ \emph{linearly admissible} if
it has a modulus of continuity $\kappa$ such that \\
(i) $\Theta\circ\kappa\leq\calO\cala(\eta)$, i.e.,
$\exists C\in\IN \; \forall n\in\IN: \; \big(\Theta\circ\kappa\big)(n)\leq C+C\cdot \eta(n+C)$  \\
and (ii) for every uniformly continuous partial surjection
$\zeta:\subseteq\Universe\twoheadrightarrow X$
it holds $\zeta\preccurlyeqO\xi$, meaning: 
There exists a map $F:\subseteq\dom(\zeta)\to\dom(\xi)$
such that $\zeta=\xi\circ F$ and,
for every modulus of continuity $\nu$ of $\zeta$,
$F$ has a modulus of continuity $\mu$
satisfying $\mu\circ\kappa\leq\calo(\nu)$.
\item[c)]
Call $\xi$ \emph{polynomially admissible} if
it has a modulus of continuity $\kappa$ such that \\
(i) $\Theta\circ\kappa\leq\calP\calo(\eta)$, and \\
(ii) for every uniformly continuous partial surjection
$\zeta:\subseteq\Universe\twoheadrightarrow X$
it holds $\zeta\preccurlyeqP\xi$, meaning: 
There exists a map $F:\subseteq\dom(\zeta)\to\dom(\xi)$
such that $\zeta=\xi\circ F$ and,
for every modulus of continuity $\nu$ of $\zeta$,
$F$ has a modulus of continuity $\mu$
satisfying $\mu\circ\kappa\leq\calp(\nu)$.
\item[d)]
$(\Universe,D)$ is linearly/polynomially \emph{universal}
if every compact metric space $(X,d)$
admits a linearly/polynomially admissible $\Universe$-representation.
\end{enumerate}
\end{definition}
Note how (b) and (c) boil down to Definition~\ref{d:Admissible}d+e) 
in case $(\Universe,D)=(\Cantor,\dC)$ of linear entropy $\Theta=\id$. 
And Theorem~\ref{t:Linear} now means that 
Cantor space is linearly universal.
It seems worthwhile to identify and classify linearly/polynomially 
universal compact metric spaces.

%%%%%%%%%%%%%%%%%%%%%%%%%%%%%%%%%%%%%%%%%
\bibliographystyle{alpha}

\bibliography{cca,signdigit}

\end{document}